\documentclass[11pt,draftcls,onecolumn]{IEEEtran}
\usepackage{cite}

\ifCLASSINFOpdf
\else
\fi

\usepackage{amsmath,graphicx,mystyle,epstopdf,hyperref,url,cite,booktabs,amsfonts}
\usepackage{cite}
\usepackage[ruled]{algorithm2e}
\usepackage{mathtools}

\newcommand{\pp}{\mathcal{P}}
\newcommand{\MM}{\beta}
\newcommand{\mm}{\alpha}

\newcommand{\gt}{*}
\newcommand{\mmx}{\mm_{x^\gt}}

\newcommand{\eg}{e_g}
\newcommand{\nup}{\nu_p}
\newcommand{\nug}{\nu_g}
\newcommand{\nut}{\mathbf{e}}

\newcommand{\nablaa}{\widetilde{\nabla}}
\newcommand{\g}{\widetilde g}
\newcommand{\n}{\widetilde n}

\newcommand{\vertiii}[1]{{\left\vert\kern-0.25ex\left\vert\kern-0.25ex\left\vert #1 
		\right\vert\kern-0.25ex\right\vert\kern-0.25ex\right\vert}}

\newlength \myfigwidth
\newlength \myfigwidthll
\newlength \myfigwidthllI

\newlength \descriptionwidth

\if@twocolumn
\else
\fi
\if@twocolumn
\else
\fi

\if@twocolumn
\else
\fi

\begin{document}

\title{Inexact Gradient Projection 
	and Fast Data Driven Compressed Sensing}
%
%
%

\author{Mohammad~Golbabaee,~Mike~E.~Davies,~\IEEEmembership{Fellow,~IEEE,}

\thanks{MG and MED are with the Institute for Digital Communications (IDCOM), School of Engineering, University of Edinburgh, EH9 3JL, United Kingdom. E-mail: \{m.golbabaee, mike.davies\}@ed.ac.uk. This work is partly funded by the EPSRC grant EP/M019802/1 and the ERC C-SENSE project (ERC-ADG-2015-694888). MED is also supported by the  Royal Society Wolfson Research Merit Award.
	}
}

\maketitle
\begin{abstract}
We study convergence of the iterative projected gradient (IPG) algorithm for arbitrary (possibly non-convex) sets and when both the gradient and projection oracles are computed approximately. We consider different notions of  approximation of which we show that 
the Progressive Fixed Precision (PFP) and the $(1+\epsilon)$-optimal oracles can achieve the same  accuracy as for the exact IPG algorithm. We show that the former scheme is also able to maintain the (linear) rate of convergence of the exact algorithm, under the same embedding assumption. In contrast, the $(1+\epsilon)$-approximate oracle requires a stronger embedding condition, moderate compression ratios and it typically slows down the convergence. 

We apply our results to accelerate solving a class of data driven compressed sensing problems, where we replace iterative 
 exhaustive searches over large datasets by fast approximate nearest neighbour search strategies based on the cover tree data structure. For datasets with low intrinsic dimensions our proposed algorithm achieves a complexity logarithmic  in terms of the dataset population as opposed to the linear complexity of a brute force search. 
By running several numerical experiments we conclude similar observations as predicted by our theoretical analysis.

\end{abstract}


%
\IEEEpeerreviewmaketitle

\section{Introduction}  
Signal inference under limited and noisy observations is a major line of research in signal processing, machine learning and statistics and it has a wide application ranging from biomedical imaging, astrophysics, remote sensing  to data mining. 
Incorporating the structure of signals is proven to significantly help with an accurate inference since natural datasets often have limited degrees of freedom as compared to their original ambient dimensionality. This fact has been invoked in Compressed sensing (CS)  literature by adopting efficient signal models to achieve accurate signal reconstruction given near-minimal number of measurements i.e. much smaller than the signal ambient dimension (see \cite{DonohoCS, CRT:CS,Tropp:SOMP1,Candes:EMC,BW:manifold,modelbasedCS} and e.g. \cite{RichCSreview} for an overview on different CS models). 
CS consists of a linear sampling protocol:
\eql{
	\label{eq:CSsampling}
	y \approx Ax^\gt,
}
where a linear mapping $A$ samples a $m$-dimensional vector $y$ of noisy measurements from a ground truth signal $x^\gt$ which typically lives in a high ambient dimension $n\gg m$. Natural signals often have efficient compact representations using nonlinear models such as low dimensional smooth manifolds, low-rank matrices or the Union of Subspaces (UoS) that itself includes sparse (unstructured) or structured sparse (e.g. group, tree or analysis sparsity) representations in properly chosen orthobases or redundant dictionaries~\cite{RichCSreview}.
CS reconstruction algorithms for estimating $x^\gt$ from $y$ are in general more computationally complex (as opposed to the simple linear acquisition of CS) as they typically require solving a nonlinear optimization problem based around a prior signal model.  
 A proper model should be carefully chosen in order to efficiently promote the low-dimensional structure of signal meanwhile not bringing a huge computational burden to the reconstruction algorithm.
 
Consider the following constrained least square problem for CS reconstruction:
\eql{\label{eq:p1}
	\min_{x\in \Cc} \{f(x):= \frac{1}{2}\norm{y-Ax}^2\},
}
where, the constraint set $\Cc\in \RR^n$ represents the signal model. First order algorithms in the form of projected Landweber iteration a.k.a. iterative projected gradient (IPG) descent or Forward-Backward are very popular for solving \eqref{eq:p1}. 
Interesting features of IPG include flexibility of handling various and often complicated signal models, e.g. $\Cc$ might be convex, nonconvex or/and semi-algebraic such as sparsity or rank constraints (these last models result in challenging combinatorial optimization problems but with tractable projection operators).  
Also IPG (and more generally the proximal-gradient methods) has been considered to be particularly useful for big data applications \cite{Volkan:bigdata}. It is memory efficient due to using only first order local oracles e.g., the gradient and the projection onto $\Cc$, it can be implemented in a distributed/parallel fashion, and it is also robust to using cheap statistical estimates e.g. in stochastic descend methods~\cite{Bottou:SGD} to shortcut heavy gradient computations.

In this regard a major challenge that IPG may encounter is the computational burden of performing an \emph{exact} projection step onto certain complex models (or equivalently, performing an exact but complicated gradient step). In many interesting inverse problems the model projection amounts to solving another optimization within each iteration of IPG. This includes important cases in practice such as the total variation penalized least squares \cite{Chambolle2011,TVprojGabirel}, low-rank matrix completion \cite{Ma2011} or tree sparse model-based CS \cite{modelbasedCS}. Another example is the convex inclusion constraints $\Cc=\bigcap_i \Cc_i$, appearing in multi constrained problems e.g.~\cite{SPCA,LRJS}, where one might be required to perform  a Djkstra type feasibility algorithm at each iteration~\cite{Dykstra,Dykstra2}. Also, for data driven signal models the projection will typically involve some form of search through potentially large datasets. 
In all these cases accessing an exact oracle could be either computationally inefficient or even not possible (e.g. in analysis sparse recovery~\cite{GiryIPGaprox} or  tensor rank minimization~\cite{Holger:tensor} where the exact projection is NP hard), and therefore a natural line of thought is to carry those steps with cheaper \emph{approximate} oracles.

\subsection{Contributions}
In this paper we feature an important property of the IPG algorithm; that \emph{it is robust against deterministic  errors in calculation of the projection and gradient steps.} 
We cover different types of oracles: i) A \emph{fixed precision} (FP) oracle which compared to the exact one has an additive bounded approximation error.  ii) A \emph{progressive fixed precision} (PFP) oracle which allows for larger (additive) approximations in the earlier iterations and refines the precision as the algorithm  progresses. iii) A $(1+\epsilon)$-approximate oracle which introduces a notion of relatively optimal approximation with a multiplicative error (as compared to the exact oracle). 

Our analysis uses a notion of model restricted bi-Lipschitz \emph{embedding} similar to e.g. \cite{Blumen}, however in a more local form and with an improved conditioning (we discuss this in more details in Section~\ref{sec:main}). 
With that respect, our analysis differs from the previous related works in the convex settings as the embedding  enables us for instance to prove a globally optimal recovery result for nonconvex models, as well as establishing linear rate of convergences for the inexact IPG applied for solving CS problems (e.g. results of \cite{BachinexactIPG} on linear convergence of the inexact IPG assumes strong convexity which does not hold in solving underdetermined least squares such as CS). 

In summary, we show that the FP type oracles restrict the final accuracy of the main reconstruction problem. 
This limitation can be overcome by increasing the precision at an appropriate rate using the PFP type oracles where one could achieve the same solution accuracy as for the exact IPG algorithm under the same embedding assumptions (and even with the convergence rate). We show that the $(1+\epsilon)$-approximate projection can also achieve the accuracy of exact IPG however under a stronger embedding assumption, moderate compression ratios and using possibly more iterations (since using this type of oracle typically decreases the rate of convergence). In all the cases above we study conditions that provide us with linear convergence results. 

Finally we apply this theory to a stylized data driven compressed sensing application that requires a nearest neighbour search order to calculate the model projection. We shortcut the computations involved,  (iteratively) performing exhaustive searches over large datasets, by using approximate nearest neighbour search strategies corresponding to the aforementioned oracles and motivated by the cover tree structure introduced in \cite{beygelzimer2006cover}. Our proposed algorithm achieves a complexity logarithmic  in terms of the dataset population (as opposed to the linear complexity of a brute force search). 
By running several numerical experiments on different  datasets we conclude similar observations as predicted by our theoretical results.
\subsection{Paper organization}
The rest of this paper is organized as follows: In Section~\ref{sec:SOA} we review and compare our results to the previous related works on inexact IPG. In Section~\ref{sec:prelim} we define the inexact IPG algorithm for three types of approximate oracles. Section~\ref{sec:main} includes our main theoretical results on robustness and linear convergence of the inexact IPG for solving CS reconstruction problem. 
 In  Section~\ref{sec:datadrivenCS} we discuss an application of the proposed inexact algorithms to accelerate solving data driven CS problems. We also briefly discuss the cover tree data structure and the associated exact/approximate search strategies.  Section~\ref{sec:expe} is dedicated to the numerical experiments on using inexact IPG for data driven CS. 
And finally we discuss and conclude our results in Section~\ref{sec:conclusion}.

\section{Related works}\label{sec:SOA}

Inexact proximal-gradient methods (in particular IPG)  and their Nesterov accelerated variants have been the subject of a substantial amount of work in convex optimization~\cite{Nesterov-inexact,dasper-inexact,villa:inexactFISTA,BachinexactIPG,charles:inexactFISTA}. Here we review some highlights and refer the reader for a comprehensive literature review to \cite{BachinexactIPG}.
Fixed precision approximates have been studied 
for the gradient step e.g. for using the smoothing techniques where the gradient is not explicit and requires solving an auxiliary optimization, see for more details \cite{Nesterov-inexact} and \cite{dasper-inexact} for a semi-definite programming example in solving sparse PCA. A similar approach has been extensively applied for carrying out the projection (or the proximity operator) approximately in cases where it does not have an analytical expression and requires solving another optimization within each iteration  e.g. in total variation constrained inverse problems \cite{TVprojGabirel,Chambolle2011} or the overlapping group Lasso problem \cite{FoGLasso}. Fortunately in convex settings 
one can stop the inner optimization when its duality gap falls below a certain threshold and achieve a fixed precision approximate projection. In this case the solution accuracy of the main problem is proportional to the approximation level introduced within each iteration. Recently \cite{BachinexactIPG} studied the progressive fixed precision (PFP) type approximations for solving convex problems, e.g. a sparse CUR type factorization, via gradient-proximal (and its accelerated variant) methods. The authors show that IPG can afford  larger approximation errors (in both gradient and projection/proximal steps) in the earlier stages of the algorithm and by increasing the approximation accuracy at an appropriate rate one can attain a similar convergence rate and accuracy as for the exact IPG however with significantly less computation.

A part of our results draws a similar conclusion for solving nonconvex constrained least squares that appears in CS problems by using inexact IPG. Note that an FP type approximation has been previously considered for the nonconvex IPG e.g. for the UoS \cite{Blumen} or the manifold \cite{MIP} CS models, however these works only assume inexact projections whereas our result covers  an approximate gradient step as well. Of more importance and to the best of our  knowledge, our results on incorporating the PFP type approximation in nonconvex IPGs and analysing the associated rate of (global) convergence is the first result of its kind. We also note that the results in \cite{BachinexactIPG} are mainly about minimizing convex cost functions (i.e. not necessarily recovery guarantees)
and that the corresponding   
linear convergence results 
only hold for uniformly strong convex objectives. We instead  
cover cases with local (and not uniform) strong convexity and establish the linear convergence of IPG for solving underdetermined inverse problems such as CS.



Using relative $(1+\epsilon)$-approximate projections (described in Section \ref{sec:relative}) for  CS recovery has been subject of more recent research activities (and mainly for nonconvex models). 
In \cite{GiryIPGaprox} the authors studied this type of approximation for an  IPG sparse recovery algorithm under the UoS model in redundant dictionaries. Our result encompasses this as a particular choice of model and additionally allows for inexactness in the gradient step. The work of \cite{StoIHT} studied similar inexact oracles for a stochastic gradient-projection type algorithm customized for sparse UoS and low-rank models (see also \cite{MatrixAlpsapprox} for low-rank CS using an accelerated variant of IPG). We share a similar conclusion with those works; for a large $\epsilon$ more measurements are required for CS recovery and the convergence becomes slow. 
Hegde et al. \cite{HegdeISIT} proposed such projection oracle for tree-sparse signals and use it for the related model-based CS problem using a CoSamp type algorithm (see also~\cite{daven:analysiscosamp,Giri:analysiscosamp} for related works on inexact CoSamp type algorithms).  
In a later work~\cite{Hegde15} the authors consider a modified variant of IPG with $(1+\epsilon)$-approximate projections for application to structured sparse reconstruction problems (specifically tree-sparse and earthmover distance CS models). For this scenario they are able to introduce a modified gradient estimate (called the Head approximation oracle) to strengthen the recovery guarantees by removing the dependency on $\epsilon$, albeit with a more conservative Restricted Isometry Property. Unfortunately, this technique does not immediately generalize to arbitrary signal models and we therefore do not pursue this line of research here.


\section{Preliminaries}\label{sec:prelim}
Iterative projected gradient iterates between calculating the gradient and projection onto the model 
i.e. for positive integers $k$ the \emph{exact} form of IPG follows:
\eql{\label{eq:IP}
	x^{k} = \pp_\Cc\left(x^{k-1}-\mu \nabla f(x^{k-1})\right) }
where, $\mu$ is the step size, $\nabla f(x)=A^T(Ax-y)$ and $\pp_\Cc$ denote the exact gradient and the Euclidean projection oracles, respectively.  
The exact IPG requires   
the constraint set $\Cc$ 
to have a well defined, not necessarily unique but  computationally tractable Euclidean projection $\pp_{\Cc}:\RR^n\rightarrow \Cc$ 
\eq{
	\pp_{\Cc}(x)\in \argmin_{u\in\Cc}\norm{u-x}.}
Throughout we use $\norm{.}$ as a shorthand for  the Euclidean norm $\norm{.}_{\ell_2}$. 

In the following we define three types of approximate oracles which frequently appear in the literature and could be incorporated within the IPG iterations. We also briefly discuss their applications. Each of these approximations are applicable to the data driven problem we will consider in Section~\ref{sec:datadrivenCS}.
\subsection{Fixed Precision (FP) approximate oracles}
\label{sec:FP}
We first consider 
approximate oracles with \emph{additive} bounded-norm 
errors, namely the fixed precision gradient oracle $\nablaa^{\nug} f(x):\RR^n\rightarrow \RR^n$ where:
\begin{align}
\norm{\nablaa^{\nug} f(x) -\nabla f(x)}\leq \nug, \label{eq:grad} 
\end{align}
and the fixed precision projection oracle $\pp_\Cc^{\nup}:\RR^n\rightarrow\Cc$ where:
\begin{align}
\pp_\Cc^{\nup}(x) \in \Big\{ u\in \Cc :\,	\norm{u-x}^2 \leq \inf_{u'\in \Cc}\norm{u'-x}^2 +\nup^2  \Big\}.\label{eq:proj1}
\end{align} 
The values of $\nug,\nup$ denote the levels of inaccuracy in calculating the gradient and projection steps respectively.\footnote{Note that our fixed precision projection \eqref{eq:proj1} is defined on the squared norms in a same way as in e.g. \cite{Blumen,MIP}. 
	} 
The corresponding \emph{inexact} IPG
iterates as follows:
\eql{\label{eq:inIP} x^k = \pp^{\nup^k}_{\Cc}\left(x^{k-1}
	-\mu \nablaa^{\nug^k} f(x^{k-1})\right).}
Note that, unlike \cite{Blumen,MIP} in this formulation we allow for variations in the inexactness levels at different stages of IPG. 
The case where the accuracy levels are bounded by a constant threshold $\nup^k=\nup$ and $\nug^k=\nug$, $\forall k$, refers to an inexact IPG algorithm with \emph{fixed precision} (FP) approximate oracles.

\subsubsection*{Examples}
Such errors may occur for instance in distributed network
optimizations where the gradient calculations could be noisy during the communication on the network, or in CS under certain UoS models with infinite subspaces~\cite{Blumen}  where an exact projection might not exist by definition (e.g. when $\Cc$ is an open set) however a FP type approximation could be achievable. It also has application in finite (discrete) super resolution~\cite{recht:discretize}, source localization and separation~\cite{TASL14,SCOOP} and data driven CS problems e.g., in Magnetic Resonance Fingerprinting~\cite{MRF,BLIPsiam},  where typically a continuous manifold is discretized and approximated  by a large dictionary for e.g. sparse recovery tasks.

\subsection{Progressive Fixed Precision (PFP) approximate oracles}\label{sec:PFP}
One obtains a \emph{Progressive Fixed Precision} (PFP) approximate IPG by refining the FP type precisions thorough the course of iterations. Therefore any FP gradient or projection oracle which has control on tuning the accuracy parameter could be used in this setting and follows \eqref{eq:inIP} with decaying sequences $\nup^k,\nug^k$.
\subsubsection*{Examples}
For instance this case includes projection schemes that require iteratively solving an auxiliary optimization (e.g. the total variation ball \cite{TVprojGabirel}, sparse CUR factorization \cite{BachinexactIPG} or the multi-constraint inclusions~\cite{Dykstra,Dykstra2}, etc) and can progress (at an appropriate rate) on the accuracy of their solutions by adding more subiterations. We also discuss in Section~\ref{sec:covertree} another example 
of this form which is customized for fast approximate nearest neighbour searches with  
application to the data driven CS framework.

\subsection{$(1+\epsilon)$-approximate projection}\label{sec:epsproj}
Obtaining a fixed precision (and thus PFP) accuracy in  projections onto certain constraints might be still computationally exhaustive, whereas a notion of relative optimality 
could be more efficient to implement. The $(1+\epsilon)$-approximate projection is defined as follows: for a given $\epsilon\geq0$,
\eql{\label{eq:eproj}
	\pp_\Cc^{\epsilon}(x) \in \Big\{ u\in \Cc :\,	\norm{u-x} \leq (1+\epsilon)\inf_{u'\in \Cc}\norm{u'-x}  \Big\}. 
}
We note that $\pp_\Cc^{\epsilon}(x)$ might not be unique. In this regard, the inexact IPG algorithm with a $(1+\epsilon)$-approximate projection takes the following form:
\eql{\label{eq:inIP2}
	x^k = \pp_\Cc^{\epsilon}\left(x^{k-1}
	-\mu \nablaa^{\nug^k} f(x^{k-1})\right).}
Note that we still assume a fixed precision gradient oracle with flexible accuracies $\nug^k$.  
One could also consider a $(1+\epsilon)$-approximate gradient oracle and in combination with those aforementioned inexact projections however for brevity and since the analysis would be quite  similar to the relative approximate projection we decide to skip more details on this case. 
\subsubsection*{Examples}
The tree $s$-sparse projection in $\RR^n$ and in the exact form requires solving a dynamical programming with $O(ns)$ running time \cite{DRthompson} whereas solving this problem approximately with $1+\epsilon$ accuracy requires the time complexity $O(n\log(n))$ \cite{HegdeISIT} which better suits  imaging problems in practice with a typical Wavelet sparsity level $s=\Omega(\log(n))$. Also \cite{StoIHT,MatrixAlpsapprox} show that one can reduce the cost of low-rank matrix completion problem by using randomized linear algebra methods, e.g. see \cite{Deshpande2006,HalkoTropp}, and carry out fast low-rank factorizations with a $1+\epsilon$ type approximation. In a recent work on low-rank tensor CS~\cite{Holger:tensor} authors consider using a $1+\epsilon$ approximation (within an IPG algorithm) for low-rank tensor factorization since the exact problem is generally NP hard~\cite{NPtensor}. 
Also in Section~\ref{sec:covertree} we discuss a data driven CS recovery algorithm which uses $1+\epsilon$ approximate nearest neighbour searches in order to break down the complexity of the projections from $O(d)$ (using an exhaustive search) to 
$O(\log (d))$, 
for  $d$ denoting a large number of data points with low intrinsic dimensionality.  

%



\section{Main results}
\label{sec:main}
\subsection{Uniform linear embeddings}
The success of CS paradigm heavily relies  on 
the embedding property of certain random sampling matrices which preserves signal information for low dimensional but often complicated/combinatorial models. It has been shown that IPG can stably predict the true signal $x^\gt$ from noisy CS measurements provided that $A$ satisfies the so called Restricted Isometry Property (RIP): 
\eql{\label{eq:RIP}(1-\theta)\norm{x-x'}^2 \leq \norm{A(x-x')}^2\leq (1+\theta)\norm{x-x'}^2, \quad \forall x,x' \in \Cc
}
for a small constant  $0<\theta<1$.
This has been shown for models such as sparse, low-rank and low-dimensional smooth manifold signals and by using 
IPG type reconstruction 
algorithms which in the nonconvex settings are also known as Iterative Hard Thresholding \cite{IHTCS, Ma2011, AIHT,MIP, modelbasedCS}. Interestingly these results indicate that under the RIP condition (and without any assumption on the initialization) the first order IPG algorithms with cheap local oracles can globally solve
nonconvex optimization problems. 

For instance random orthoprojectors and i.i.d. subgaussian matrices $A$ satisfy RIP when the number of measurements $m$ is proportional to the intrinsic dimension of the model (i.e. signal sparsity level, rank of a data matrix or the dimension of a smooth signal manifold, see e.g. \cite{RichCSreview} for a review on comparing different CS models and their measurement complexities) and sublinearly scales with the ambient dimension $n$. 

A more recent work generalizes the theory of IPG to arbitrary \emph{bi-Lipschitz embeddable} models \cite{Blumen}, that is for given $\Cc$ and $A$ it holds
\eq{\mm \norm{x-x'}^2\leq \norm{A(x-x')}^2\leq \MM \norm{x-x'}^2 \quad \forall x,x'\in \Cc.}
for some constants $\mm, \MM >0$. Similar to the RIP these constants are defined \emph{uniformly} over the constraint set i.e. $\forall x,x'\in \Cc$. There Blumensath shows that if \eq{\MM<1.5\mm,} then IPG robustly solves the corresponding noisy CS reconstruction problem \emph{for all} $x^\gt\in \Cc$. This result also relaxes the RIP requirement to a nonsymmetric and unnormalised notion of linear embedding whose implication in deriving sharper recovery bounds is previously studied by \cite{JaredJeff}. 

\subsection{Hybrid (local-uniform) linear embeddings}
Similarly the notion of restricted embedding plays a key role in our analysis. However we adopt a more local form of embedding and show that it is still able to guarantee stable CS reconstruction. 
{\ass{ \label{def:Lip}
		Given $(x_0\in \Cc, \Cc, A)$ there exists constants  $\MM,\mm_{x_0}>0$ for which the following inequalities hold:
		\begin{itemize}
			\item Uniform Upper Lipschitz Embedding (ULE)
			\begin{align*}
			\norm{A(x-x')}^2\leq \MM \norm{x-x'}^2 \quad \forall x,x'\in \Cc
			\end{align*}
			\item Local Lower Lipschitz Embedding (LLE)
			\begin{align*}
			\norm{A(x-x_0)}^2 \geq \mm_{x_0} \norm{x-x_0}^2 \quad \forall x\in \Cc
			\end{align*}
		\end{itemize}	
		Upon existence, $\MM$ and $\mm_{x_0}$ denote  respectively the smallest and largest constants for which the inequalities above hold.		
	}}
	\newline

This is a weaker assumption compared to RIP or the uniform bi-Lipschitz embedding. Note that for any $x_0\in \Cc$ we have: \eq{
\mm\leq\mm_{x_0}\leq\MM\leq \vertiii{A}^2
} 
(where $\vertiii{.}$ denotes the matrix spectral norm i.e. the largest singular value). 
However with such an assumption one has to sacrifice the \emph{universality} of the RIP-dependent results for a signal $x^\gt$ dependent analysis. Depending on the study, local analysis could be very useful to avoid e.g. worst-case scenarios that might unnecessarily restrict the recovery analysis~\cite{me:modelselecion}. Similar local assumptions in the convex settings are shown   to improve the measurement bound and the speed of convergence up to very sharp constants~\cite{recht:GW, Oymak:tradeoff}. 

Unfortunately we are currently unable to make the analysis fully local as we require the uniform ULE constraint. Nonetheless, one can always plug the stronger bi-Liptchitz assumption into our results throughout (i.e. replacing $\mmx$ with $\mm$) and regain the universality.




\subsection{Linear convergence of (P)FP inexact IPG for CS recovery}
In this section we show that IPG is robust against deterministic (worst case) errors. Moreover, we show that for certain decaying approximation errors, the IPG solution maintain  the same accuracy as for the approximation-free  algorithm. 

{\thm{\label{th:inexactLS1} Assume $(x^\gt\in \Cc, \Cc,A)$ satisfy the main Lipschitz assumption  with constants $\MM< 2\mmx$. Set the step size $(2 \mmx )^{-1}<\mu\leq\MM^{-1}$. The sequence generated by Algorithm \eqref{eq:inIP} obeys the following bound:
		\eql{\label{eq:errbound}
			\norm{x^{k}-x^\gt}\leq  \rho^k \left(\norm{x^\gt}+\sum_{i=1}^k \rho^{-i} \nut^i \right)+ \frac{2\sqrt{\MM}}{\mmx(1-\rho)}w
		}
		where 
		\begin{align*}
		\rho=\sqrt{\frac{1}{\mu \mmx} -1} \qandq
		\nut^i=
		\frac{2\nug^i}{\mmx} + \frac{\nup^i}{\sqrt{\mu \mmx}},			
		\end{align*} 
		and $w=\norm{y-Ax^\gt}$.
	}} 

{\rem{Theorem \ref{th:inexactLS1} implications for the exact IPG (i.e. $\nup^k=\nug^k=0$) and inexact FP approximate IPG (i.e. $\nup^k=\nup,\nug^k=\nug, \forall k$)  improve  \cite[Theorem~2]{Blumen} in three ways: first by relaxing the uniform lower Lipschitz constant $\mm$ to a local form $\mmx\geq \mm$ with the possibility of conducting  a local recovery/convergence analysis. Second, by improving the embedding condition for CS stable recovery to 
\eql{\label{eq:cond}\MM < 2\mmx,
	} 
or $\MM < 2\mm$ for a uniform recovery $\forall x^\gt\in\Cc$. And third, by improving 
the rate $\rho$ of convergence.}}

The following corollary is an immediate consequence of the linear convergence result established in Theorem~\ref{th:inexactLS1} for which we do not provide a proof:
{\cor{\label{cor:FP}With assumptions in Theorem \ref{th:inexactLS1} the IPG algorithm with FP approximate oracles achieves the solution accuracy
\eq{\norm{x^K-x^\gt}\leq 
		 	\frac{1}{1-\rho}\left(\frac{2\nug}{\mmx} + \frac{\nup}{\sqrt{\mu \mmx}}+ \frac{2\sqrt{\MM}}{\mmx}w\right) +\tau}
for any $\tau>0$ and in a finite number of iterations
\eq{
	K=\left\lceil\frac{1}{\log(\rho^{-1})} \log\left( \frac{ \norm{x^\gt}}{\tau}\right)\right\rceil
	}			
}}
As it turns out in our experiments and aligned with the result of Corollary~\ref{cor:FP}, the solution accuracy of IPG can not exceed the precision level introduced by a PF oracle. In this sense Corollary~\ref{cor:FP} is tight as a trivial converse example would be that IPG starts from the optimal solution $x^\gt$ but an adversarial FP scheme projects it to another point within a fixed distance. 

Interestingly one can deduce another implication from  Theorem~\ref{th:inexactLS1} and overcome such limitation by  using a PFP type oracle.
Remarkably 
one achieves a linear convergence to a solution with the same accuracy as for the exact IPG, 
as long as $\nut^k$ geometrically decays. 	
The following corollary makes this statement explicit:
	
	
{\cor{\label{cor:decay}Assume $\nut^k\leq Cr^k$ for some error decay rate $0<r<1$ and a constant $C$. Under the assumptions of Theorem~\ref{th:inexactLS1} the solution updates $\norm{x^{k}-x^*}$ of the IPG algorithm with PFP approximate oracles is bounded above by:
		\begin{align*}
			&\max(\rho,r)^k \left(\norm{x^\gt}+\frac{C}{1-\frac{\min(\rho,r)}{\max(\rho,r)}}\right) +
			\frac{2\sqrt{\MM}}{\mm_{x^\gt}(1-\rho)} w,  &r\neq \rho \\
			&\rho^k \Big(\norm{x^\gt}+Ck\Big) +
			\frac{2\sqrt{\MM}}{\mm_{x^\gt}(1-\rho)} w, &r=\rho 
		\end{align*}
		Which implies a linear convergence at rate 
		\begin{align*}
		\bar \rho = \choice{\max(\rho,r)\quad r\neq\rho \\
			\rho+\xi\qquad\quad r=\rho}
		\end{align*}
		for an arbitrary small $\xi>0$.	
	}
}
{\rem{Similar to Corollary \ref{cor:FP} 
one can increase the final solution precision of the FPF type  IPG with logarithmically more iterations i.e. in a finite number $K=O(\log(\tau^{-1}))$ of iterations one achieves $\norm{x^{K}-x^*}\leq O(w)+\tau$. Therefore in contrast with the FP oracles one achieves an accuracy within the noise level $O(w)$ that is the precision of an approximation-free IPG.
}}

{\rem{Using the PFP type oracles can also maintain the rate of linear convergence identical as for the exact IPG. For this the approximation errors suffice to follow a geometric decaying rate of  $r<\rho$. 
}}

{\rem{The embedding condition~\eqref{eq:cond} sufficient to guarantee our stability results is invariant to the  precisions of the FP/PFP oracles and it is the same as for an exact IPG.}}

\subsection{Linear convergence of inexact IPG with $(1+\epsilon)$-approximate projection for CS recovery}
\label{sec:relative}
In this part we focus on the inexact  algorithm~\eqref{eq:inIP2} with a $(1+\epsilon)$-approximate projection. As it turns out by the following theorem we require a stronger embedding condition to guarantee the CS stability compared to the previous algorithms.

{\thm{\label{th:inexactLS2}  Assume $(x^\gt\in \Cc, \Cc,A)$ satisfy the main Lipschitz assumption and that
		\eq{\sqrt{2\epsilon+\epsilon^2}\leq \delta\frac{\sqrt{\mmx}}{\vertiii{A}} \qandq \MM < (2-2\delta+\delta^2) \mmx} 	
		for $\epsilon\geq 0$ and some constant $\delta \in [0,1)$.
		Set the step size $\left((2-2\delta+\delta^2) \mmx\right)^{-1}<\mu\leq\MM^{-1}$. The sequence generated by Algorithm \eqref{eq:inIP2} obeys the following bound:
		\eq{
			\norm{x^{k}-x^\gt}\leq  \rho^k \left(\norm{x^\gt}+\kappa_g \sum_{i=1}^k \rho^{-i} \nug^i \right)+ 
			\frac{\kappa_w}{1-\rho}w
		}
		where 
		\begin{align*}
		&\rho=\sqrt{\frac{1}{\mu \mmx} -1}+ \delta, \quad
		\kappa_g = \frac{2}{\mmx}+\frac{\sqrt\mu}{\vertiii{A}}\delta, \\		
		&\kappa_w= 2\frac{\sqrt{\MM}}{\mmx}+\sqrt\mu\delta, \qandq w=\norm{y-Ax^\gt}.
		\end{align*}		
		
	}} 	
{\rem{Similar conclusions follow as in Corollaries~\ref{cor:FP} and \ref{cor:decay} on the linear convergence, logarithmic number of iterations vs. final level of accuracy (depending whether the gradient oracle is exact or FP/PFP) however with a stronger requirement than \eqref{eq:cond} on the embedding; increasing $\epsilon$ i.e. consequently $\delta$, limits the recovery guarantee and slows down the convergence (compare $\rho$ in Theorems~\ref{th:inexactLS1} and \ref{th:inexactLS2}). Also approximations of this type result in amplifying distortions (i.e. constants $\kappa_w, \kappa_g$) due to the measurement noise and gradient errors. For example for an exact or a geometrically decaying PFP gradient updates and for a fixed $\epsilon$ chosen according to the Theorem~\ref{th:inexactLS2}  assumptions, Algorithm~\eqref{eq:inIP2} achieves a full precision accuracy $\norm{x^{K}-x^*}\leq O(w)+\tau$ (similar to the exact IPG) in a finite number $K=O(\log(\tau^{-1}))$ of iterations.
		}}



{\rem{\label{rem:stringe}The assumptions of Theorem~\ref{th:inexactLS2} impose a stringent requirement on the scaling of the approximation parameter i.e. $\epsilon=O\left(\sqrt{\frac{\mmx}{\vertiii{A}}}\right)$ which is not purely dependent on the model-restricted embedding condition but also on the spectral norm  $\vertiii{A}$. In this sense since  $\vertiii{A}$ ignores the structure $\Cc$ of the problem it might scale very differently than  the corresponding embedding constants $\mmx,\mm$ and $\MM$. For instance a $m\times n$  i.i.d. Gaussian matrix has w.h.p. $\vertiii{A}=\Theta(n)$ (when $m\ll n$) whereas, e.g. for  sparse signals, the embedding constants $\mm,\MM$ w.h.p. scale as $O(m)$. A similar gap exists for other low dimensional signal models and for other compressed sensing matrices e.g. random orthoprojectors. 
This indicates that the $1+\epsilon$ oracles may be
sensitive to the CS sampling ratio i.e. for $m\ll n $
we may be limited to use very small approximations  $\epsilon = O(\sqrt{ \frac{m}{n}} )$. 


In the following we show by a deterministic example that this requirement is indeed tight. We also empirically observe in Section \ref{sec:expe} that such a limitation indeed holds in randomized settings (e.g. i.i.d. Gaussian $A$) and on average. 
Although it would be desirable to modify the IPG algorithm to avoid such restriction, as was done in~\cite{Hegde15} for specific structured sparse models, we note that this is the same term that appears due to 'noise folding' when the signal model is not exact or when there is noise in the signal domain (see the discussion in Section~\ref{sec:inexactmodel}). As such most practical CS systems will inevitably have to avoid regimes of extreme undersampling.
}}


\subsubsection*{A converse example}
Consider a noiseless CS recovery problem where $n =2$, $m=1$ and the sampling matrix (i.e. here a row vector) is
\eq{A=[\cos(\gamma)\quad -\sin(\gamma)]}
for some parameter $0\leq\gamma<\pi/2$. Consider the following one-dimensional signal model along the first coordinate:
\eq{
\Cc = \{x\in \RR^2: x(1)\in \RR,\, x(2)=0\}.
}
We have indeed $\vertiii{A}=1$. It is easy to verify that both of the embedding constants w.r.t. to $\Cc$ are
\eq{ \mmx=\MM = \cos(\gamma)^2.}
Therefore one can tune $\gamma\rightarrow \pi/2$ to obtain arbitrary small ratios for $\sqrt{\frac{\mmx}{\vertiii{A}}}=\cos(\gamma)$. 

Assume the true signal, the corresponding CS measurement,  and the initialization point are
\eq{
x^\gt=[1 \quad 0]^T,\quad y=Ax^\gt=\cos(\gamma) \qandq x^0=[0 \quad 0]^T.
}
Consider an adversarial $(1+\epsilon)$-approximate projection oracle which performs the following step for any given $x\in \RR^2$:
\eq{
		\pp_\Cc^{\epsilon}(x) := [x(1)+\epsilon x(2) \quad 0]^T.
	}
For simplicity we assume no errors on the gradient step. 	
By setting $\mu=1/\cos(\gamma)^2$, the corresponding inexact IPG updates as 
\eq{	
	x^k(1)= 1+\epsilon\tan(\gamma) \left(x^{k-1}(1)-1\right)  
}
and only along the first dimension (we note that due to the choice of oracle $x^k(2)=0, \forall k$). Therefore we have
\eq{	
	x^k(1)= 1-\left(\epsilon\tan(\gamma)\right)^k.
}
which requires $\epsilon<\tan^{-1}(\gamma)=O(\cos(\gamma))$ for convergence, and it diverges otherwise. As we can see for $\gamma\rightarrow \pi/2$  (i.e. where $A$ becomes extremely unstable w.r.t. sampling the first dimension) the range of admissible $\epsilon$ shrinks, regardless of the fact that the restricted embedding $\mmx=\MM$ exhibits a  perfect isometry; which is an ideal situation for solving a noiseless CS (i.e. in this case an exact IPG takes only one iteration to converge).
\subsection{When the projection is not onto the signal model}\label{sec:inexactmodel}

 One can also make a distinction between  the projection set $\Cc$ and the signal model here denoted as $\Cc'$ (i.e. $x^\gt\in\Cc'$) by modifying our earlier definitions \eqref{eq:proj1} and \eqref{eq:eproj} in the following ways:
an approximate FP projection reads, 
\begin{align}
\pp_\Cc^{\nup}(x) \in \Big\{ u\in \Cc :\,	\norm{u-x}^2 \leq \inf_{u'\in \Cc'}\norm{u'-x}^2 +\nup^2  \Big\},\label{eq:proj1-1}
\end{align} 
and a $(1+\epsilon)$-approximate projection reads
\eql{\label{eq:eproj-1}
	\pp_\Cc^{\epsilon}(x) \in \Big\{ u\in \Cc :\,	\norm{u-x} \leq (1+\epsilon)\inf_{u'\in \Cc'}\norm{u'-x}  \Big\}. 
}
With respect to such a distinction,  Theorems~\ref{th:inexactLS1} and \ref{th:inexactLS2} still hold (with the same embedding assumption/constants on the projection set $\Cc$), conditioned that $x^\gt\in \Cc$.
Indeed this can be verified by following identical steps as in the proof of both theorems.
This allows throughout the flexibility of considering an approximate projection onto a possibly larger set $\Cc$ including the original signal model $\Cc'$ i.e. $x^\gt \in \Cc'\subseteq \Cc$, which for instance finds application in fast tree-sparse signal or low-rank matrix CS recovery, see \cite{HegdeISIT,MatrixAlpsapprox}. Such an inclusion is also important to derive a uniform recovery result \emph{for all} $x^*\in\Cc'$. 

The case where $x^\gt \notin \Cc$ can also be bounded in a similar fashion as in \cite{Blumen}. We first consider a proximity point in $\Cc$ i.e. $x^{o}:=\argmin_{u\in\Cc} \norm{x^\gt-u}$  and update the noise term  to $w:=\norm{y-Ax^{o}}\leq \norm{y-Ax^\gt}+\norm{A(x^\gt-x^{o})} $. We then use Theorems~\ref{th:inexactLS1} and \ref{th:inexactLS2} to derive an error bound, here on $\norm{x^k-x^{o}}$. For this we assume the embedding condition \emph{uniformly} holds over the projection set $\Cc$ (which includes $x^{o}$). As a result we get a bound on the error $\norm{x^k-x^\gt} \leq \norm{x^\gt-x^{o}}+\norm{x^k-x^{o}}$ which includes a bias term with respect to the distance of $x^\gt$ to $\Cc$. Note that since here $w$ also includes a signal (and not only measurement) noise term introduced by  $\norm{A(x^\gt-x^{o})}$, the results are subjected to \emph{noise folding} i.e. a noise amplification 
with a similar unfavourable scaling (when $m \ll n$) to our discussion in Remark~\ref{rem:stringe} (for more details on CS noise folding see e.g.~\cite{eldar:noisefolding,daven:noisefolding}).

\section{Proofs}
\subsection{Proof of Theorem \ref{th:inexactLS1}}
We start from a similar argument as in \cite[proof of Therorem~2]{Blumen}. 
Set $g := 2\nabla f(x^{k-1})=2A^T(Ax^{k-1}-y)$ and $\g:=2\nablaa^{\nug} f(x^{k-1})= g+2\eg^k$ for some vector $\eg^k$ which by definition~\eqref{eq:grad} is bounded $\norm{\eg^k}\leq \nug^k$. It follows that
\ifCLASSOPTIONtwocolumn
\begin{align*} 
&\norm{y-Ax^k}^2-\norm{y-Ax^{k-1}}^2	\\
&= \langle x^k-x^{k-1},g \rangle +\norm{A(x^k-x^{k-1})}^2 \\
&\leq \langle x^k-x^{k-1},g \rangle + \MM \norm{x^k-x^{k-1}}^2, 
\end{align*}
\else
\begin{align*} 
\norm{y-Ax^k}^2-\norm{y-Ax^{k-1}}^2	&= \langle x^k-x^{k-1},g \rangle +\norm{A(x^k-x^{k-1})}^2 \\
&\leq \langle x^k-x^{k-1},g \rangle + \MM \norm{x^k-x^{k-1}}^2, 
\end{align*}
\fi
where the last inequality follows from the ULE property in Definition \ref{def:Lip}. Assuming $\MM \leq 1/\mu$, we have
\ifCLASSOPTIONtwocolumn
\begin{align*}
&\langle x^k-x^{k-1},g \rangle + \MM \norm{x^k-x^{k-1}}^2 \\
& \leq \langle x^k-x^{k-1},g \rangle + \frac{1}{\mu} \norm{x^k-x^{k-1}}^2\\
&= \langle x^k-x^{k-1},\g \rangle + \frac{1}{\mu} \norm{x^k-x^{k-1}}^2 - \langle x^k-x^{k-1},2\eg^k \rangle\\
& = \frac{1}{\mu} \norm{x^k-x^{k-1}+\frac{\mu}{2} \g }^2 - \frac{\mu}{4} \norm{\g}^2 - \langle x^k-x^{k-1},2\eg^k \rangle.
\end{align*}
\else
\begin{align*}
\langle x^k-x^{k-1},g \rangle + \MM \norm{x^k-x^{k-1}}^2 
& \leq \langle x^k-x^{k-1},g \rangle + \frac{1}{\mu} \norm{x^k-x^{k-1}}^2\\
&= \langle x^k-x^{k-1},\g \rangle + \frac{1}{\mu} \norm{x^k-x^{k-1}}^2 - \langle x^k-x^{k-1},2\eg^k \rangle\\
& = \frac{1}{\mu} \norm{x^k-x^{k-1}+\frac{\mu}{2} \g }^2 - \frac{\mu}{4} \norm{\g}^2 - \langle x^k-x^{k-1},2\eg^k \rangle.
\end{align*}
\fi
Due to the update rule of Algorithm \eqref{eq:inIP} and the inexact (fixed-precision) projection step, we have
\ifCLASSOPTIONtwocolumn
\begin{align*}
	&\norm{x^k-x^{k-1}+\frac{\mu}{2} \g }^2 \\
	&\leq  \norm{\pp_{\Cc}(x^{k-1}-\frac{\mu}{2} \g)-x^{k-1}+\frac{\mu}{2} \g }^2 +(\nup^k)^2\\
	&\leq \norm{x^\gt-x^{k-1}+\frac{\mu}{2} \g }^2 +(\nup^k)^2.
\end{align*}
\else
\begin{align*}
\norm{x^k-x^{k-1}+\frac{\mu}{2} \g }^2 
&\leq  \norm{\pp_{\Cc}(x^{k-1}-\frac{\mu}{2} \g)-x^{k-1}+\frac{\mu}{2} \g }^2 +(\nup^k)^2\\
&\leq \norm{x^\gt-x^{k-1}+\frac{\mu}{2} \g }^2 +(\nup^k)^2.
\end{align*}
\fi
The last inequality holds for any member of $\Cc$ and thus here for $x^\gt$. Therefore we can write
\ifCLASSOPTIONtwocolumn
\begin{align}
&\norm{y-Ax^k}^2-\norm{y-Ax^{k-1}}^2 \nonumber	\\
&\leq \frac{1}{\mu} \norm{x^\gt-x^{k-1}+\frac{\mu}{2} \g }^2 - \frac{\mu}{4} \norm{\g}^2 \nonumber\\
&\qquad - \langle x^k-x^{k-1},2\eg^k \rangle +(\frac{\nup^k}{\sqrt\mu})^2 \nonumber \\
&= \langle x^\gt-x^{k-1},\g \rangle + \frac{1}{\mu} \norm{x^\gt-x^{k-1}}^2 \nonumber\\
&\qquad - \langle x^k-x^{k-1},2\eg^k \rangle 
 +(\frac{\nup^k}{\sqrt\mu})^2\nonumber\\
&\leq \langle x^\gt-x^{k-1},g \rangle + \frac{1}{\mu} \norm{x^\gt-x^{k-1}}^2 \nonumber\\
&\qquad+2\nug^k\norm{x^k-x^*} +(\frac{\nup^k}{\sqrt\mu})^2. \label{eq:p1b2}
\end{align}
\else
\begin{align}
\norm{y-Ax^k}^2-\norm{y-Ax^{k-1}}^2 \nonumber	
&\leq \frac{1}{\mu} \norm{x^\gt-x^{k-1}+\frac{\mu}{2} \g }^2 - \frac{\mu}{4} \norm{\g}^2 
 - \langle x^k-x^{k-1},2\eg^k \rangle +(\frac{\nup^k}{\sqrt\mu})^2 \nonumber \\
&= \langle x^\gt-x^{k-1},\g \rangle + \frac{1}{\mu} \norm{x^\gt-x^{k-1}}^2 
 - \langle x^k-x^{k-1},2\eg^k \rangle 
+(\frac{\nup^k}{\sqrt\mu})^2\nonumber\\
&\leq \langle x^\gt-x^{k-1},g \rangle + \frac{1}{\mu} \norm{x^\gt-x^{k-1}}^2 
+2\nug^k\norm{x^k-x^*} +(\frac{\nup^k}{\sqrt\mu})^2. \label{eq:p1b2}
\end{align}
\fi
The last line replaces $\g= g+2\eg^k$ and uses the Cauchy-Schwartz inequality.

Similarly we use the LLE property in Definition \ref{def:Lip} to obtain an upper bound on $ \langle x^\gt-x^{k-1},g \rangle$:
\begin{align*} 
\langle x^\gt-x^{k-1},g \rangle 	&= w^2-\norm{y-Ax^{k-1}}^2 +\norm{A(x_0-x^{k-1})}^2 \\
&\leq w^2 -\norm{y-Ax^{k-1}}^2 +\mmx\norm{x^\gt-x^{k-1}}^2,
\end{align*}
where $w=\norm{ y-Ax^\gt}$. Replacing this bound in \eqref{eq:p1b2} and cancelling $-\norm{y-Ax^{k-1}}^2$ from both sides of the inequality yields
\ifCLASSOPTIONtwocolumn
\begin{align}
&\norm{y-Ax^k}^2- 2\nug^k\norm{x^k-x^\gt}\nonumber \\ 
&\leq \left(\frac{1}{\mu}-\mmx \right)\norm{x^{k-1}-x^\gt}^2 + (\frac{\nup^k}{\sqrt\mu})^2+w^2. \label{eq:p1b3}
\end{align}
\else
\begin{align}
\norm{y-Ax^k}^2- 2\nug^k\norm{x^k-x^\gt}
\leq \left(\frac{1}{\mu}-\mmx \right)\norm{x^{k-1}-x^\gt}^2 + (\frac{\nup^k}{\sqrt\mu})^2+w^2. \label{eq:p1b3}
\end{align}
\fi
We continue to lower bound the left-hand side of this inequality:
\ifCLASSOPTIONtwocolumn
\begin{align*}
&\norm{y-Ax^k}^2- 2\nug^k\norm{x^k-x^\gt}\\
&= \norm{A(x^k-x^\gt)}^2+w^2-2\langle y-Ax^\gt, A(x^k-x^\gt)\rangle\\
&- 2\nug^k\norm{x^k-x^\gt} \\
&\geq \norm{A(x^k-x^\gt)}^2+w^2-2w \norm{A(x^k-x^\gt)}\\
&- 2\nug^k\norm{x^k-x^\gt} \\
& \geq \mmx\norm{x^k-x^\gt}^2+w^2-2(w \sqrt{\MM}+\nug^k)\norm{x^k-x^\gt}\\
&= \left(\sqrt{\mmx}\norm{x^k-x^\gt}-\frac{\nug^k}{\sqrt{\mmx}}- \sqrt{\frac{\MM}{\mmx}}w\right)^2 \\
&- (\frac{\nug^k}{\sqrt{\mmx}})^2 -(\frac{\MM}{\mmx}-1)w^2.
\end{align*}
\else
\begin{align*}
\norm{y-Ax^k}^2- 2\nug^k\norm{x^k-x^\gt}
&= \norm{A(x^k-x^\gt)}^2+w^2-2\langle y-Ax^\gt, A(x^k-x^\gt)\rangle- 2\nug^k\norm{x^k-x^\gt} \\
&\geq \norm{A(x^k-x^\gt)}^2+w^2-2w \norm{A(x^k-x^\gt)}
- 2\nug^k\norm{x^k-x^\gt} \\
& \geq \mmx\norm{x^k-x^\gt}^2+w^2-2(w \sqrt{\MM}+\nug^k)\norm{x^k-x^\gt}\\
&= \left(\sqrt{\mmx}\norm{x^k-x^\gt}-\frac{\nug^k}{\sqrt{\mmx}}- \sqrt{\frac{\MM}{\mmx}}w\right)^2 - (\frac{\nug^k}{\sqrt{\mmx}})^2 -(\frac{\MM}{\mmx}-1)w^2.
\end{align*}
\fi
The first inequality uses the Cauchy-Schwartz's and the second inequality follows from the ULE and LLE properties. Using this bound together with \eqref{eq:p1b3} we get
\ifCLASSOPTIONtwocolumn
\begin{align*}
&\left(\sqrt{\mmx}\norm{x^k-x^\gt}-\frac{\nug^k}{\sqrt{\mmx}}- \sqrt{\frac{\MM}{\mmx}}w\right)^2\\
&\leq \left(\frac{1}{\mu}-\mmx \right)\norm{x^{k-1}-x^\gt}^2 + (\frac{\nug^k}{\sqrt{\mmx}})^2+ (\frac{\nup^k}{\sqrt\mu})^2+\frac{\MM}{\mmx}w^2 \\
&\leq \left(\sqrt{\frac{1}{\mu}-\mmx} \norm{x^{k-1}-x^\gt} + \frac{\nug^k}{\sqrt{\mmx}}+ \frac{\nup^k}{\sqrt\mu}+\sqrt{\frac{\MM}{\mmx}}w \right)^2.
\end{align*}
\else
\begin{align*}
\left(\sqrt{\mmx}\norm{x^k-x^\gt}-\frac{\nug^k}{\sqrt{\mmx}}- \sqrt{\frac{\MM}{\mmx}}w\right)^2
&\leq \left(\frac{1}{\mu}-\mmx \right)\norm{x^{k-1}-x^\gt}^2 + (\frac{\nug^k}{\sqrt{\mmx}})^2+ (\frac{\nup^k}{\sqrt\mu})^2+\frac{\MM}{\mmx}w^2 \\
&\leq \left(\sqrt{\frac{1}{\mu}-\mmx} \norm{x^{k-1}-x^\gt} + \frac{\nug^k}{\sqrt{\mmx}}+ \frac{\nup^k}{\sqrt\mu}+\sqrt{\frac{\MM}{\mmx}}w \right)^2.
\end{align*}
\fi
The last inequality assumes $\mu\leq \mmx^{-1}$ which holds since we previously assumed $\mu\leq \MM^{-1}$. As a result we deduce that
\begin{align}
\norm{x^k-x^\gt}\leq \rho \norm{x^{k-1}-x^\gt} + \nut^k + 2\frac{\sqrt{\MM}}{\mmx}w \label{eq:p1b4}
\end{align}
for $\rho$ and $\nut^k$ defined in Theorem \ref{th:inexactLS1}. Applying this bound recursively (and setting $x^0=0$) completes the proof:
\begin{align*}
\norm{x^k-x^\gt}\leq \rho^k \norm{x^\gt} + \sum_{i=1}^k \rho^{k-i} \nut^i + \frac{2\sqrt{\MM}}{\mmx(1-\rho)}w.
\end{align*} 
Note that for convergence we require $\rho<1$ and therefore, a lower bound on the step size which is $\mu> (2\mmx)^{-1}$. 

\subsection{Proof of Corollary~\ref{cor:decay}}
Following the error bound \eqref{eq:errbound} derived in   Theorem~\ref{th:inexactLS1} and by setting $\nut^k\leq C r^k$ we obtain:
		\eq{
			\norm{x^{k}-x^\gt}\leq  \rho^k \left(\norm{x^\gt}+C\sum_{i=1}^k (r/\rho)^{i}  \right)+ \frac{2\sqrt{\MM}}{\mmx(1-\rho)}w,			
		}
which for $r<\rho$ it implies 		
		\eq{
			\norm{x^{k}-x^\gt}\leq 
			 \rho^k \left(\norm{x^\gt}+\frac{C}{1-r/\rho}\right)+ \frac{2\sqrt{\MM}}{\mmx(1-\rho)}w,
		}
and for $r>\rho$ implies 
\begin{align*}
\norm{x^{k}-x^\gt}&\leq  \rho^k \norm{x^\gt}+C r^k \sum_{i=1}^k (\rho/r)^{k-i}  + \frac{2\sqrt{\MM}}{\mmx(1-\rho)}w\\
&\leq r^k \left(\norm{x^\gt}+\frac{C}{1-\rho/r}\right)+ \frac{2\sqrt{\MM}}{\mmx(1-\rho)}w,	
\end{align*}
and for $r=\rho$ we immediately get
\eq{
\norm{x^{k}-x^\gt}\leq  \rho^k \norm{x^\gt}+C k \rho^k + \frac{2\sqrt{\MM}}{\mmx(1-\rho)}w.	
}
Note that there exists a constant $c$ such that for an arbitrary small $\xi>0$ it holds $k\rho^k\leq c(\rho+\xi)^k$. Therefore we also achieve a linear convergence for the case $r=\rho$.
\subsection{Proof of Theorem \ref{th:inexactLS2}}
As before set $g= 2A^T(Ax^{k-1}-y)$ and $\g= g+2\eg^k$ for some bounded gradient error vector $\eg^k$ i.e. $\norm{\eg^k}\leq \nug^k$. Note that 
here the update rule of Algorithm \eqref{eq:inIP2} uses the  $(1+\epsilon)$-approximate projection  which by definition \eqref{eq:eproj} implies
\ifCLASSOPTIONtwocolumn
\begin{align*}
&\norm{x^k-x^{k-1}+\frac{\mu}{2} \g }^2 =  \norm{\pp^{\epsilon}_{\Cc}(x^{k-1}-\frac{\mu}{2} \g)-x^{k-1}+\frac{\mu}{2} \g }^2\\
&\leq  (1+\epsilon)^2\norm{\pp_{\Cc}(x^{k-1}-\frac{\mu}{2} \g)-x^{k-1}+\frac{\mu}{2} \g }^2\\
&\leq \norm{x^\gt-x^{k-1}+\frac{\mu}{2} \g }^2 + \phi(\epsilon)^2\frac{\mu^2}{4}\norm{\g}^2
\end{align*}
\else
\begin{align*}
\norm{x^k-x^{k-1}+\frac{\mu}{2} \g }^2 &=  \norm{\pp^{\epsilon}_{\Cc}(x^{k-1}-\frac{\mu}{2} \g)-x^{k-1}+\frac{\mu}{2} \g }^2\\
&\leq  (1+\epsilon)^2\norm{\pp_{\Cc}(x^{k-1}-\frac{\mu}{2} \g)-x^{k-1}+\frac{\mu}{2} \g }^2\\
&\leq \norm{x^\gt-x^{k-1}+\frac{\mu}{2} \g }^2 + \phi(\epsilon)^2\frac{\mu^2}{4}\norm{\g}^2
\end{align*}
\fi
where $\phi(\epsilon):=\sqrt{2\epsilon+\epsilon^2}$. For the last inequality we replace $\pp_{\Cc}(x^{k-1}-\frac{\mu}{2} \g)$ with two feasible points $x^\gt,x^{k-1}\in \Cc$. 

As a result by only replacing $\nug^k$ with $\mu\phi(\epsilon)\norm{\g}/2$, we can follow identical steps as for the proof of Theorem \ref{th:inexactLS1} up to \eqref{eq:p1b4}, revise the definition of $\nut^k:={2\nug^k}/{\mmx} + {\sqrt{\mu}\phi(\epsilon)\norm{\g}}/(2\sqrt{{\mmx}})$ and write
\ifCLASSOPTIONtwocolumn
\begin{align*}
\norm{x^k-x^\gt}\leq& \sqrt{\frac{1}{\mu\mmx}-1} \norm{x^{k-1}-x^\gt} \\
&+ \frac{2\nug^k}{\mmx} +\frac{\phi(\epsilon)}{2} \sqrt{\frac{\mu}{\mmx}}\norm{\g} + 2\frac{\sqrt{\MM}}{\mmx}w. 
\end{align*}
\else
\begin{align*}
\norm{x^k-x^\gt}\leq \sqrt{\frac{1}{\mu\mmx}-1} \norm{x^{k-1}-x^\gt} 
+ \frac{2\nug^k}{\mmx} +\frac{\phi(\epsilon)}{2} \sqrt{\frac{\mu}{\mmx}}\norm{\g} + 2\frac{\sqrt{\MM}}{\mmx}w. 
\end{align*}
\fi
Note that so far we only assumed $\mu\leq \MM^{-1}$. 

On the other hand by triangle inequality we have
\begin{align*}
	\norm{\g}&\leq \norm{g}+2\nug^k\\
	&\leq 2\norm{A^TA(x^{k-1}-x^\gt)}+2\norm{A^T(y-Ax^\gt)}+2\nug^k \\
	&\leq 2\sqrt \MM\vertiii{A}\norm{(x^{k-1}-x^\gt)}+2\vertiii{A}w+2\nug^k\\
	&\leq 2\sqrt{ 1/\mu}\vertiii{A}\norm{(x^{k-1}-x^\gt)}+2\vertiii{A}w+2\nug^k.
\end{align*}
The third inequality uses the ULE property and the last one holds since $\mu\leq \MM^{-1}$.
Therefore, we get
\ifCLASSOPTIONtwocolumn
\begin{align*}
&\norm{x^k-x^\gt}\leq
\left(\sqrt{\frac{1}{\mu\mmx}-1}+ \phi(\epsilon)\frac{\vertiii{A}}{\sqrt{\mmx}}\right) \norm{x^{k-1}-x^\gt} \\
&+ \left( \frac{2}{\mmx} +\phi(\epsilon){\sqrt{\frac{\mu}{\mmx}}}\right) \nug^k 
+ \left( 2\frac{\sqrt{\MM}}{\mmx}+ \phi(\epsilon)\sqrt{\frac{\mu}{\mmx}} \vertiii{A} \right)w. 
\end{align*}
\else
\begin{align*}
\norm{x^k-x^\gt}\leq&
\left(\sqrt{\frac{1}{\mu\mmx}-1}+ \phi(\epsilon)\frac{\vertiii{A}}{\sqrt{\mmx}}\right) \norm{x^{k-1}-x^\gt} \\
&+ \left( \frac{2}{\mmx} +\phi(\epsilon){\sqrt{\frac{\mu}{\mmx}}}\right) \nug^k 
+ \left( 2\frac{\sqrt{\MM}}{\mmx}+ \phi(\epsilon)\sqrt{\frac{\mu}{\mmx}} \vertiii{A} \right)w. 
\end{align*}
\fi
Based on assumption $\phi(\epsilon)\frac{\vertiii{A}}{\sqrt{\mmx}}\leq \delta$ of the theorem  we can deduce
\ifCLASSOPTIONtwocolumn
\begin{align*}
\norm{x^k-x^\gt}\leq&
\rho \norm{x^{k-1}-x^\gt} + \left( \frac{2}{\mmx} +\frac{\sqrt \mu}{\vertiii{A}} \delta\right) \nug^k \\
&
+ \left( 2\frac{\sqrt{\MM}}{\mmx}+\sqrt{\mu}\delta \right)w 
\end{align*}
\else
\begin{align*}
\norm{x^k-x^\gt}\leq
\rho \norm{x^{k-1}-x^\gt} + \left( \frac{2}{\mmx} +\frac{\sqrt \mu}{\vertiii{A}} \delta\right) \nug^k 
+ \left( 2\frac{\sqrt{\MM}}{\mmx}+\sqrt{\mu}\delta \right)w 
\end{align*}
\fi
where $\rho=\sqrt{\frac{1}{\mu\mmx}-1}+\delta$.

Applying this bound recursively (and setting $x^0=0$) completes the proof:
\eq{
\norm{x^{k}-x^\gt}\leq  \rho^k \norm{x^\gt}+\kappa_g \sum_{i=1}^k \rho^{k-i} \nug^i+ \frac{\kappa_w}{1-\rho}w
}
for $\kappa_g, \kappa_w$ defined in Theorem \ref{th:inexactLS2}. The condition for convergence is $\rho<1$ which implies $\delta<1$ and a lower bound on the step size which is $\mu> (\mmx+(1-\delta)^2\mmx)^{-1}$. 
		
\section{Application in Data driven compressed sensing}
\label{sec:datadrivenCS}
Many CS reconstruction programs resort to signal models promoted by certain (semi) algebraic functions $h(x):\RR^n\rightarrow \RR_+\cup\{ 0\}$. For example we can have
\eq{\Cc := \big\{x\in \RR^n : h(x)\leq \zeta\big\},
	}
where $h(x)$ may be chosen as the $\ell_{0}$ or $\ell_{0-2}$ semi-norms or as the $\rank(x)$  which promotes sparse, group-sparse or low-rank (for matrix spaces) solutions, respectively. One might also replace those penalties with their corresponding  
convex relaxations namely, the $\ell_1$, $\ell_{1-2}$ norms or the nuclear norm. 

\emph{Data driven} compressed sensing however corresponds to cases where in the absence of an algebraic physical model one resorts to collecting a large number of data samples in a dictionary and use it as a \emph{point cloud} model for CS reconstruction~\cite{RichCSreview}.  
Data driven CS finds numerous applications e.g. in Hyperspectral imagery \cite{TIPHSI}, Mass spectroscopy (MALDI imaging) \cite{Kobarg2014}, Raman Imaging~\cite{ramanCS} and Magnetic Resonance Fingerprinting (MRF)~\cite{MRF,BLIPsiam} just to name a few. For instance the USGS Hyperspectral library \footnote{\url{http://speclab.cr.usgs.gov}} contains the spectral signatures (reflectance) of thousands of substances measured across a few hundred frequency bands. This side information is shown to be useful for CS reconstruction and classification in both convex and nonconvex settings (see e.g. \cite{TIPHSI} for more details and relations to sparse approximation in redundant dictionaries).  
Data driven CS may also apply to algebraic models with non trivial projections. For example in the MRF reconstruction problem one first constructs a huge  dictionary of fingerprints i.e. the magnetization responses (across the readout times) for many $T1,T2$ relaxation values (i.e. spin-spin and spin-echo) presented in normal tissues~\cite{MRF}. This corresponds to sampling a two-dimensional manifold associated with the solutions of the \emph{Bloch dynamic  equations}~\cite{BLIPsiam}, which in the MRF settings neither the response nor the projection has an analytic closed-form solution.

\subsection{A data driven CS in product space}
To explore how our theoretical results can be used to accelerate CS reconstruction we consider a stylized data driven application for which we explain how one can obtain each of the aforementioned approximate projections. Consider a multi-dimensional image such as HSI, MALDI or MRF that can be represented by a $\n\times J$ matrix $X$, where $n=\n J$ is the total number of spatio-spectral pixels, $J$ is the spatial resolution and $\n$ is the number of spectral bands 
e.g. $\n=3$ for an RGB image, $\n\approx 400$ for an HSI acquired by NASA's AVIRIS spectrometer, $\n\approx 5000$ for a MALDI image~\cite{Kobarg2014}.
 In the simplest form we assume that 
each spatial pixel corresponds to a certain material with a specific signature, i.e. 
\eq{
	X_{j}\in \widetilde \Cc, \quad \forall j=1,\ldots,J,
} 
where $X_j$ denotes the $j$th column of $X$ and 
\eq{
	\widetilde \Cc:=\bigcup_{i=1}^{d}\{\psi_i\} \in \RR^J}
is the point cloud of a large number $d$ of signatures $\psi_i$ e.g. in a customized spectral library for HSI or MALDI data.

The CS sampling model follows \eqref{eq:CSsampling} by setting $x^*:=X_\text{vec}$, where by $X_\text{vec}\in \RR^n$ we denote the vector-rearranged form of the matrix $X$.
The CS reconstruction reads
\eql{\label{eq:datadrivenCS}
	\min_{\substack{X_j\in \widetilde\Cc,\\ \forall j=1,...,J}} \big\{f(x):= \frac{1}{2}\norm{y-AX_\text{vec}}^2\big\}.
}
or equivalently and similar to problem \eqref{eq:p1} 
\eq{
	\min_{x\in \prod_{j=1}^J \widetilde\Cc} \big\{f(x):= \frac{1}{2}\norm{y-Ax}^2\big\}.
}
The only update w.r.t. problem \eqref{eq:p1} 
is the fact that now the solution(s) $x$ lives in a product space of the same model i.e. 
\eql{\label{eq:prod}\Cc:=\prod_{j=1}^J \widetilde\Cc
} 
(see also~\cite{kronCS} on product/Kronecker space CS however using a sparsity inducing semi-algebraic model). We note that solving directly this problem for a general $A$ (e.g. sampling models which non trivially combine columns/spatial pixels of $X$)  is \emph{exponentially hard} $O(d^J)$ because of the combinatorial nature of the product space constraints. In this regard, a tractable scheme which has been frequently considered for this problem e.g. in \cite{BLIPsiam} would be the application of an IPG type algorithm 
in order to break down the cost into the gradient and projection computations (here the projection requires $O(Jd)$ computations to search the closest signatures  to the current solution) to locally solve the problem at each iteration. 


\subsection{Measurement bound}
The classic Johnson-Lindenstrauss lemma says that one can use random linear transforms to stably embed point clouds into a lower dimension of size $O(\log(\#\widetilde \Cc))$ where $\#$ stands for the set cardinality~\cite{JL}.
{\thm{\label{th:JL} Let $\widetilde \Cc$ be a finite set of points in $\RR^{\n}$. For $A$ drawn at random from the i.i.d. normal distribution and a positive constant $\widetilde \theta$, with very high probability one has
		\eq{(1-\widetilde \theta)\norm{(x-x')} \leq \norm{A(x-x')}\leq (1+\widetilde \theta)\norm{(x-x')}, \quad \forall x,x' \in \widetilde \Cc
			}
			provided $m=O(\log(\#\widetilde \Cc)/\widetilde \theta^2)$.
		}}
		
Note that this definition implies an RIP embedding for the point cloud model according to \eqref{eq:RIP} with a constant $\theta<3\widetilde \theta$ which in turn (and for small enough $\widetilde \theta$) implies the sufficient embedding condition for a stable CS recovery using the exact or approximate IPG. 
This bound considers an arbitrary point cloud and could be improved when data points $\widetilde \Cc\subseteq \Mm$ are derived from a low-dimensional structure, e.g. a smooth manifold $\Mm$ such as the MR Fingerprints, for which one can alternatively use $m=\dim_{E}(\Mm)$ for the corresponding RIP type \emph{embedding dimension} listed e.g. in \cite{RichCSreview}.



%
		
We note that such a measurement bound for a product space model  \eqref{eq:prod}  without considering any structure between spaces turns into 
\eq{
	m=O\left(J\min\left\{\log(d), \dim_E(\Mm)\right\}\right).
	}

\subsection{Cover tree for fast nearest neighbour search}
\label{sec:covertree}
With the data driven CS formalism and discretization of the model  
the projection step of IPG reduces to searching for the nearest signature in each of the product spaces, however in a potentially very large $d$ size dictionary. 
And thus search strategies with linear complexity in $d$ e.g. an exhaustive search, can be a serious bottleneck for solving such problems. 
A very well-established approach to overcome the complexity of an exhaustive nearest neighbour (NN) search on a large dataset consists of hierarchically partitioning the solution space and forming a \emph{tree} whose nodes represents those partitions, and then using branch-and-bound methods on the resulting tree for a fast Approximate NN (ANN) search with $o(d)$ complexity e.g. see~\cite{Navigating,beygelzimer2006cover}.

In this regard, we address the computational shortcoming of the projection step in the exact IPG by preprocessing $\widetilde \Cc$ and form a \emph{cover tree} structure suitable for fast ANN searches \cite{beygelzimer2006cover}. 
A cover tree is a levelled tree whose nodes at different scales form covering nets for data points at multiple resolutions; 
if $\sigma:=\max_{\psi\in\widetilde \Cc}\norm{\psi_\text{root}-\psi}$ corresponds to the maximal coverage by the root, then  
nodes appearing at any finer scale $l>0$ form a $(\sigma 2^{-l})$-covering net for their descendants i.e. as we descend down the tree the covering resolution refines in a dyadic coarse-to-fine fashion.

 We consider three possible search strategies 
 using such a tree structure:
 \begin{itemize}
 	\item \textbf{Exact NN:} which is based on the branch-and-bound algorithm proposed in \cite[Section 3.2]{beygelzimer2006cover}.	Note that we should distinguish between this strategy and performing a brute force search. Although they both perform an exact NN search, the complexity of the proposed algorithm in~\cite{beygelzimer2006cover} is shown to be way less in practical datasets.
 	\item \textbf{$(1+\epsilon)$-ANN:} this search is also based on the branch-and-bound algorithm proposed in \cite[Section 3.2]{beygelzimer2006cover} which has includes an early search termination criteria (depending on the accuracy level $\epsilon$) 
 	for which one obtains an approximate  oracle of type defined by \eqref{eq:eproj}. 
 	Note that the case $\epsilon=0$ refers to the exact tree NN search described above. 
 	
 	\item \textbf{FP-ANN:} that is traversing down the tree up to a scale $l=\lceil \log(\frac{\nup}{\sigma})\rceil$ for which the covering resolution falls below a threshold $\nup$ on the search accuracy. This search results in a fixed precision type approximate oracle as described in Section~\ref{sec:FP} and in a sense it is similar to performing the former search with $\epsilon=0$, however on a truncated (low-resolution) cover tree. 
 	
 \end{itemize}
 All strategies could be applied to accelerate the projection step of an exact or inexact IPG (with variations discussed in Sections~\ref{sec:epsproj} and \ref{sec:FP}) to tackle the data driven CS problem~\eqref{eq:datadrivenCS}.  
 In addition one can iteratively refine the accuracy of the FP-ANN search (e.g. $\nup^k =r^k$ for a certain decay rate $r<1$ and IPG iteration number $k$) and obtain a PFP type approximate IPG discussed in Section~\ref{sec:PFP}.  
 
Note that while the cover tree construction is blind to the explicit structure of the data, several key growth properties such as the tree's explicit depth, the number of children per node, and importantly the overall search complexity are characterized by the intrinsic dimension of the model, called the \emph{doubling dimension},
and defined as follows~\cite{assouad,heinonen}:

{\defn{\label{def:doub} Let $B(q,r)$ denotes a ball of radius $r$ centred at a point $q$ in some metric space. The doubling dimension $\dim_D(\Mm)$ of a set $\Mm$ is the smallest integer such that every ball of $\Mm$ (i.e. $\forall r>0$, $\forall q\in\Mm$, $B(q,2r)\cap \Mm$) can be covered by $2^{\dim_D(\Mm)}$  balls of half radius i.e. $B(q',r)\cap \Mm$, $q'\in \Mm$. 
		}}
		
The doubling dimension has several 
appealing properties e.g. $\dim_D(\RR^n)=\Theta(n)$, $\dim_D(\Mm_1)\leq \dim_D(\Mm_2)$ when $\Mm_1$ is a subspace of $\Mm_2$, and $\dim(\cup_{i=1}^I \Mm_i)\leq \max_i \dim_D(\Mm_i)+\log(I)$~\cite{heinonen,Navigating}. 
Practical datasets are often assumed to have small doubling dimensions e.g. when $\widetilde \Cc \subseteq \Mm$  
 samples a low $K$-dimensional manifold $\Mm$ with certain smoothness and regularity one has $\dim_D(\widetilde \Cc)\leq \dim_D(\Mm)=O(K)$ \cite{dasgupta2008}.\footnote{Although the two notions of embedding $\dim_E$ and doubling $\dim_D$ dimensions scale similarly for certain sets e.g. linear subspaces, UoS, smooth manifolds..., this does not generally hold in $\RR^n$ and one needs to distinguish between them,  
see for more discussions~\cite{Indyk:NNembedding,dasgupta2012}.}

Equipped with such a notion of dimensionality, the following theorem bounds the complexity of a  $(1+\epsilon)$-ANN cover tree search~\cite{Navigating,beygelzimer2006cover}: 

{\thm{\label{thm:NNcomp2}Given a query which might not belong to $\widetilde \Cc$, the approximate $(1+\epsilon)$-ANN search on a cover tree takes at most 
		\eql{
			2^{O(\dim_D(\widetilde \Cc))}\log \Delta+(1/\epsilon)^{O(\dim_D(\widetilde \Cc))} 
		}
		computations in time with $O(\#\widetilde \Cc)$ memory requirement, where  $\Delta$ is the aspect ratio of $\widetilde \Cc$.}	
}

For most applications $\log (\Delta) = O(\log(d))$~\cite{Navigating} and thus for datasets with low dimensional structures i.e. $\dim_D=O(1)$ and by using  moderate approximations one achieves a logarithmic search complexity in $d$, as opposed to the linear complexity of a brute force search.

Note that the complexity of an \emph{exact} cover tree search could be arbitrarily high 
and thus the same applies to the FP and PFP type ANN searches since they are also based on performing exact NN (on a truncated tree). However in the next section we empirically observe  that the complexity of an exact cover tree NN (and also the FP and PFP type ANN) is much lower than performing an exhaustive search.

\section{Numerical experiments}
\label{sec:expe}

We test the performance of the exact/inexact IPG algorithm for our product-space data driven CS reconstruction using the four datasets described in Table \ref{tab:data}. The datasets are uniformly sampled (populated) from 2-dimensional continuous manifolds embedded in a higher ambient dimension, see also  Figure~\ref{fig:datasets}\footnote{The S-manifold, Swiss roll and Oscillating wave are synthetic machine learning  datasets available e.g. in \cite{GMRA12}. The Magnetic Resonance Fingerprints (MRF) is generated by solving the Bloch dynamic equation for a uniform grid of relaxation times $T1,T2$ and for an external magnetic excitation pattern, discussed and implemented in~\cite{MRF}.}. 

To proceed with fast ANN searches within IPG, we separately build a cover tree structure per dataset i.e. a preprocessing step. As illustrated for the MRF manifold in Figure~\ref{fig:CT} the coverage levels 
decrease in a coarse-to-fine manner as we traverse down the tree i.e. increasing the scale.

\ifCLASSOPTIONtwocolumn
\begin{table}[t!]
	\centering
	\scalebox{.91}{
	\begin{tabular}{ccccc}
		\toprule[.2em]
		Dataset & Population ($d$) & Ambient dim. ($\n$)&CT depth&CT res.\\
		\midrule[.1em]
		S-Manifold & 5000 & 200& 14&2.43E-4 \\
		Swiss roll & 5000 & 200 &14&1.70E-4\\
		Oscillating wave & 5000 & 200 &14&1.86E-4\\
		MR Fingerprints & 29760 & 512& 13&3.44E-4\\
		\bottomrule[.2em]
	\end{tabular}}
	\caption{Datasets for data-driven CS evaluations; a cover tree (CT) structure is formed for each dataset. The last two columns respectively report the number of scales and the finest covering resolution of each tree.  }
	\label{tab:data}	
\end{table}
\else
\begin{table}[t!]
	\centering
	\scalebox{1.1}{
		\begin{tabular}{ccccc}
			\toprule[.2em]
			Dataset & Population ($d$) & Ambient dim. ($\n$)&CT depth&CT res.\\
			\midrule[.1em]
			S-Manifold & 5000 & 200& 14&2.43E-4 \\
			Swiss roll & 5000 & 200 &14&1.70E-4\\
			Oscillating wave & 5000 & 200 &14&1.86E-4\\
			MR Fingerprints & 29760 & 512& 13&3.44E-4\\
			\bottomrule[.2em]
		\end{tabular}}
		\caption{Datasets for data-driven CS evaluations; a cover tree (CT) structure is formed for each dataset. The last two columns respectively report the number of scales and the finest covering resolution of each tree.  }
		\label{tab:data}	
	\end{table}
	\fi
\ifCLASSOPTIONtwocolumn
\begin{figure}[t!]
	\centering
	\begin{minipage}{\linewidth}
		\centering
		\subfloat[S-Manifold]{\includegraphics[width=.46\textwidth]{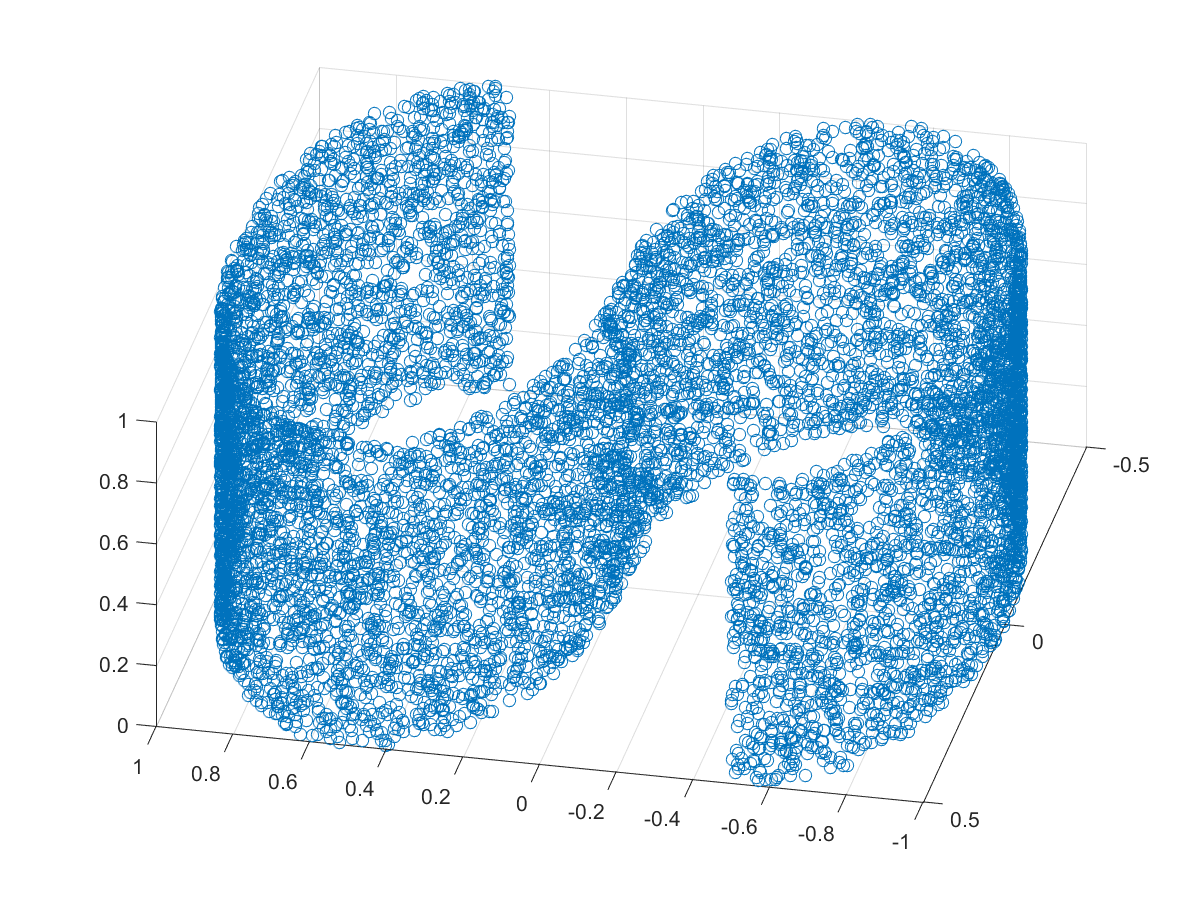} }	
		\quad	
		\subfloat[Swiss roll]{\includegraphics[width=.46\textwidth]{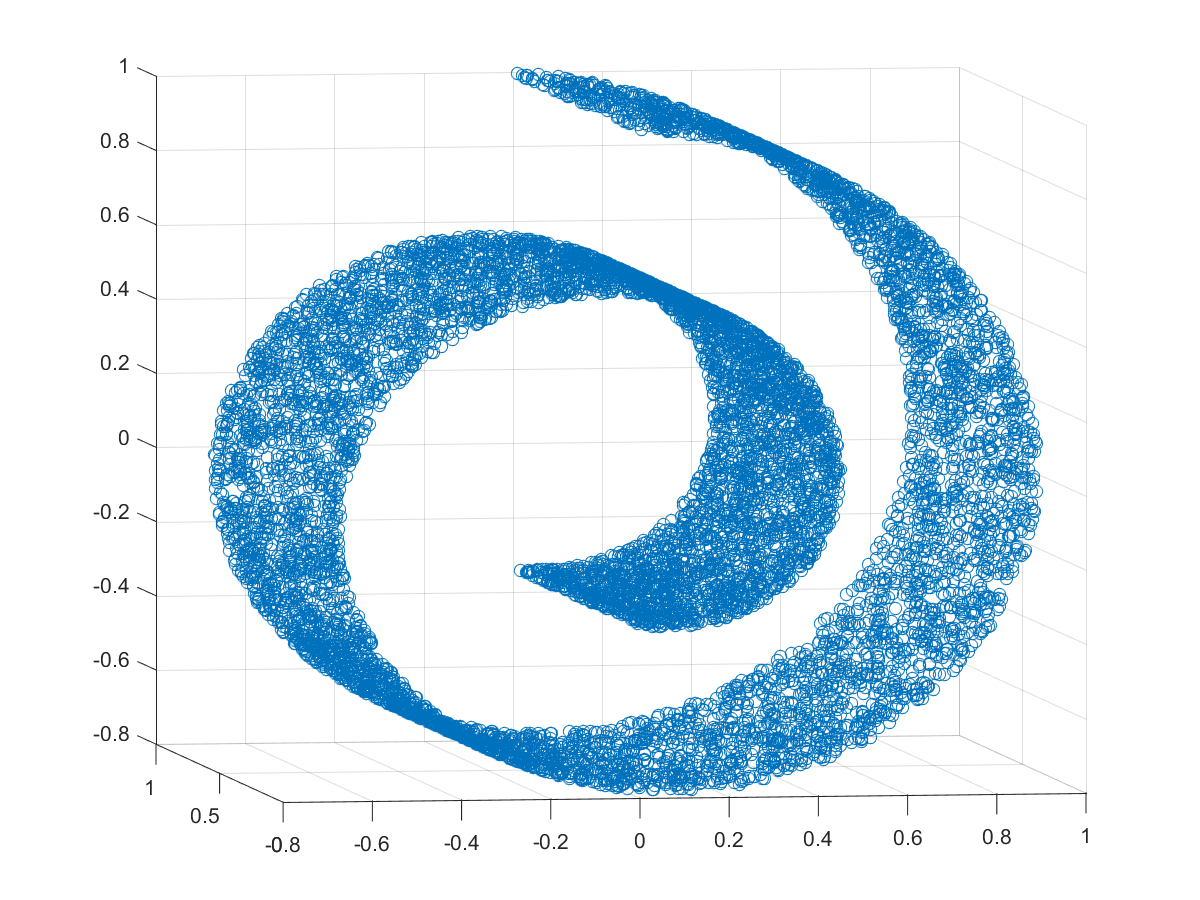} }	
		\quad	
		\subfloat[Oscillating wave]{\includegraphics[width=.46\textwidth]{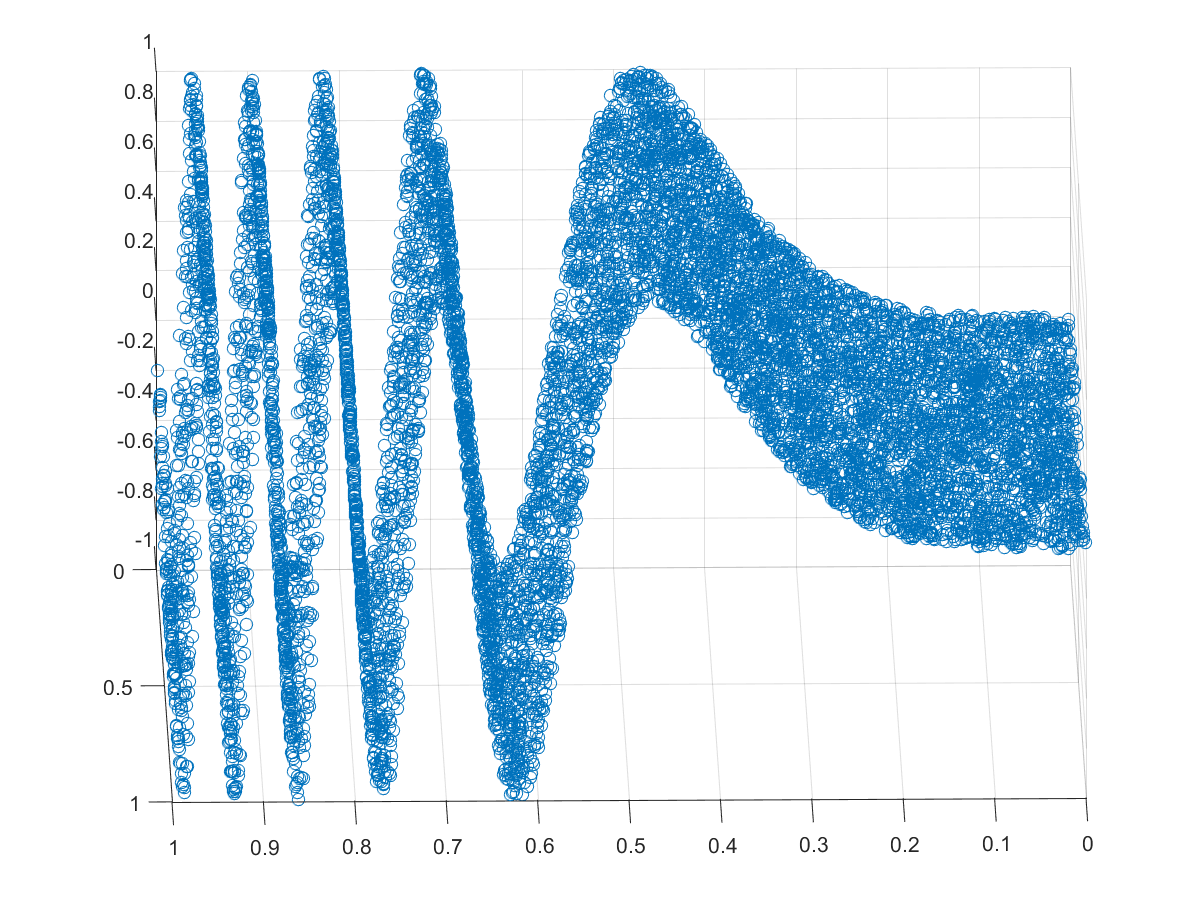} }	
		\quad	
		\subfloat[MR Fingerprints]{\includegraphics[width=.46\textwidth]{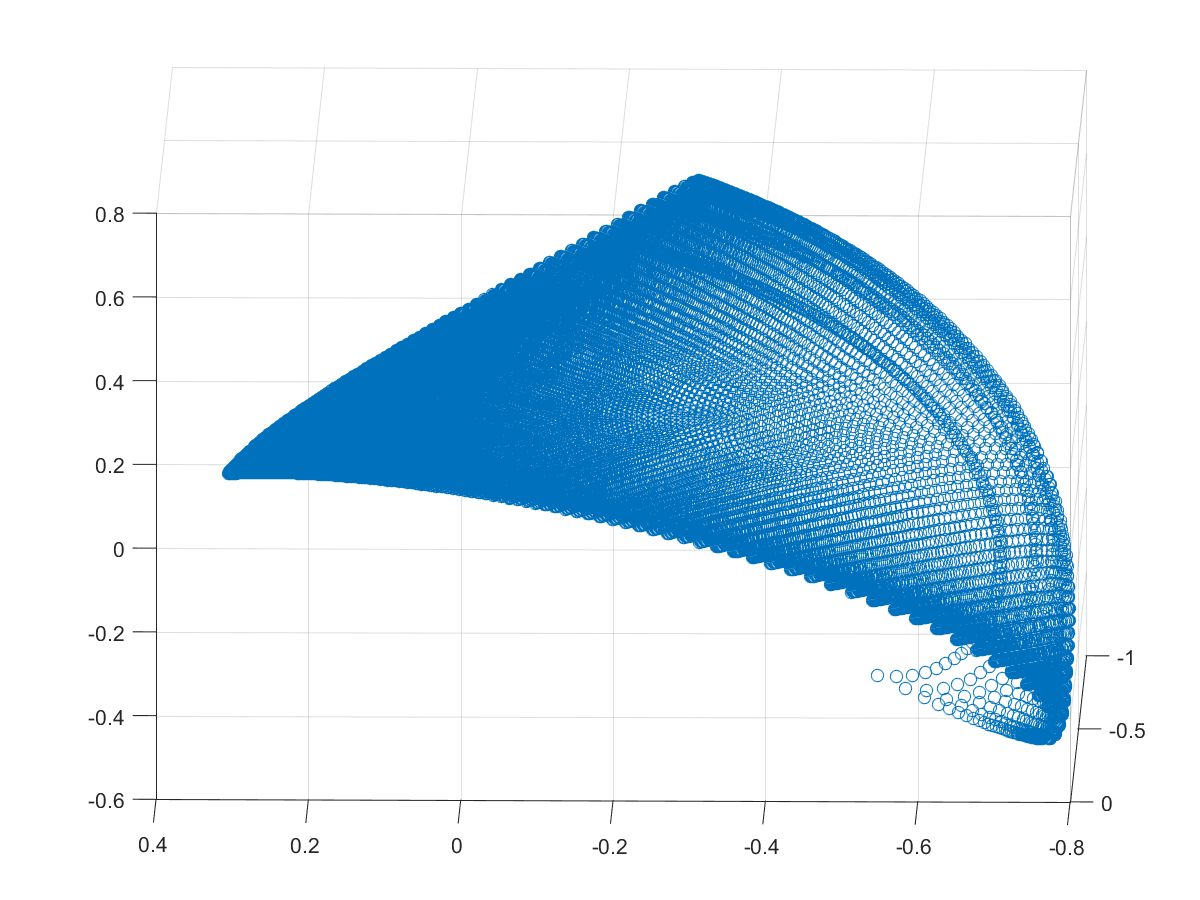} }	
	\caption{Illustration of the low dimensional structures of  datasets presented in Table~\ref{tab:data}. Points are depicted along the first three principal components of each dataset.\label{fig:datasets}}
\end{minipage}
\end{figure}
\else
\begin{figure}[t!]
	\centering
	\begin{minipage}{\linewidth}
		\centering
		\subfloat[S-Manifold]{\includegraphics[width=.35\textwidth]{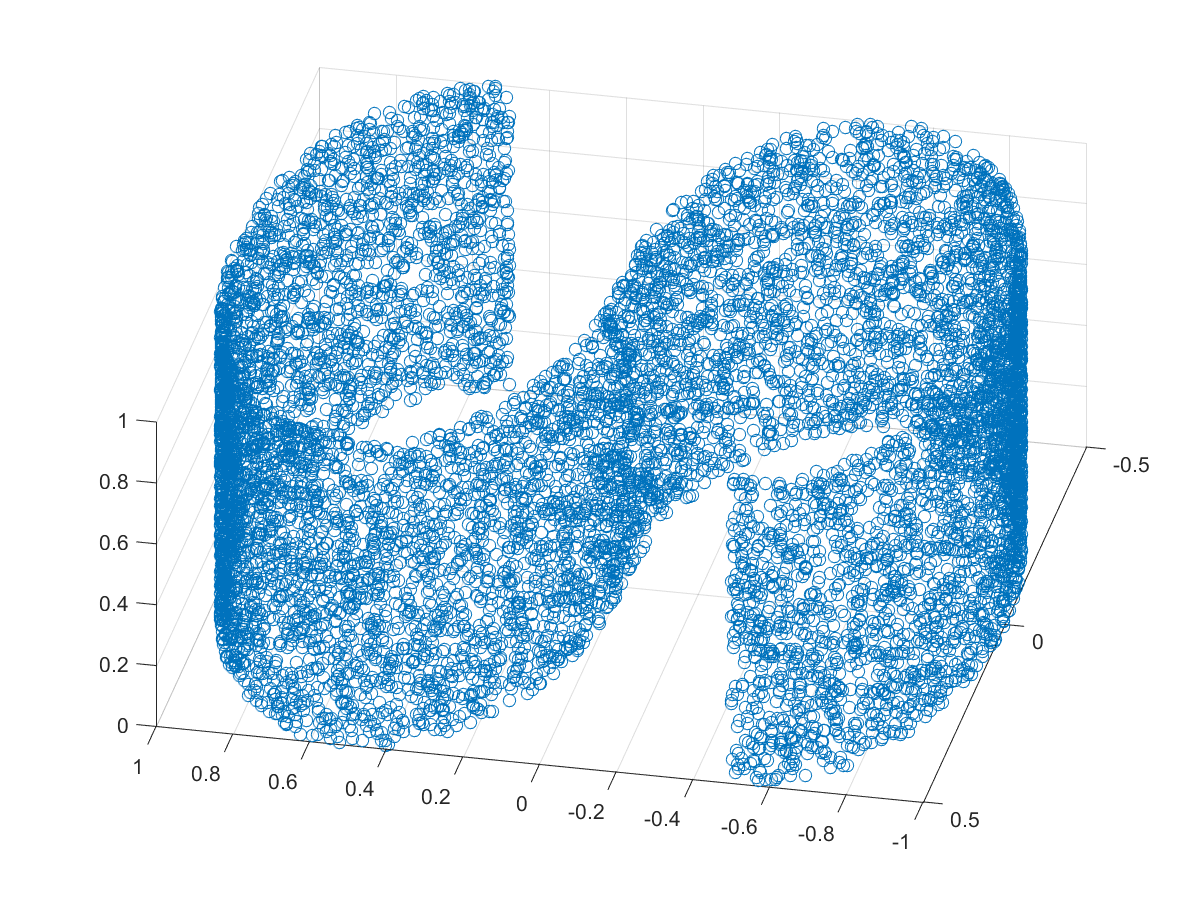} }	
		\quad	
		\subfloat[Swiss roll]{\includegraphics[width=.35\textwidth]{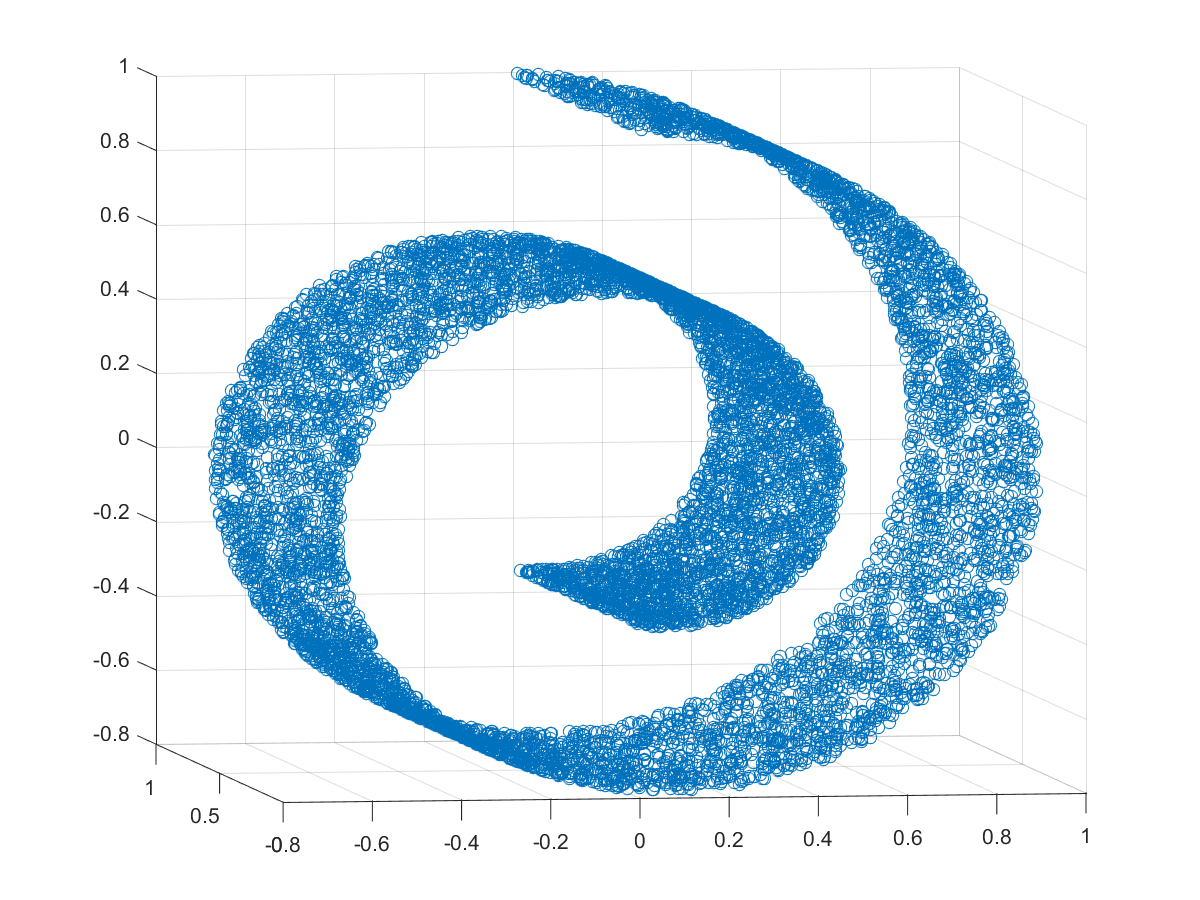} }	
		\quad	
		\subfloat[Oscillating wave]{\includegraphics[width=.35\textwidth]{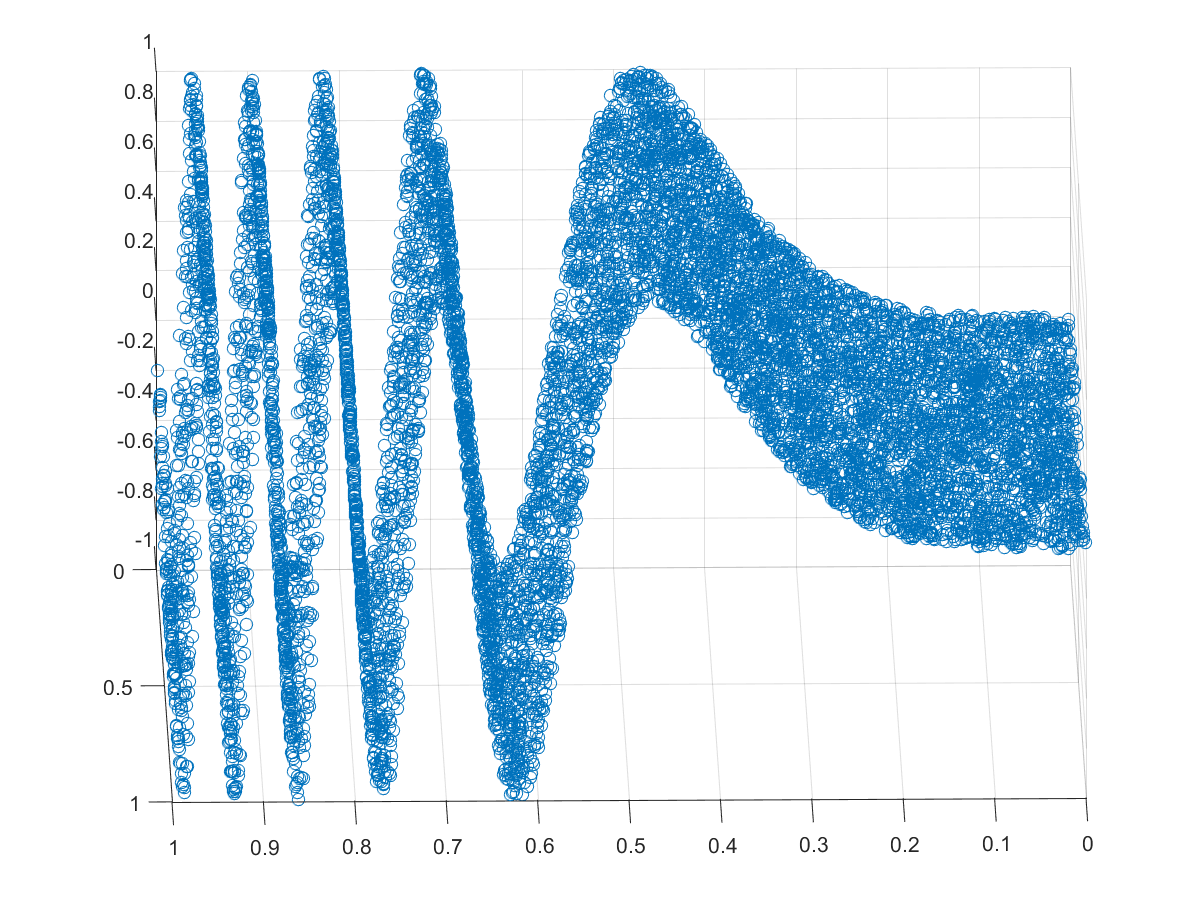} }	
		\quad	
		\subfloat[MR Fingerprints]{\includegraphics[width=.35\textwidth]{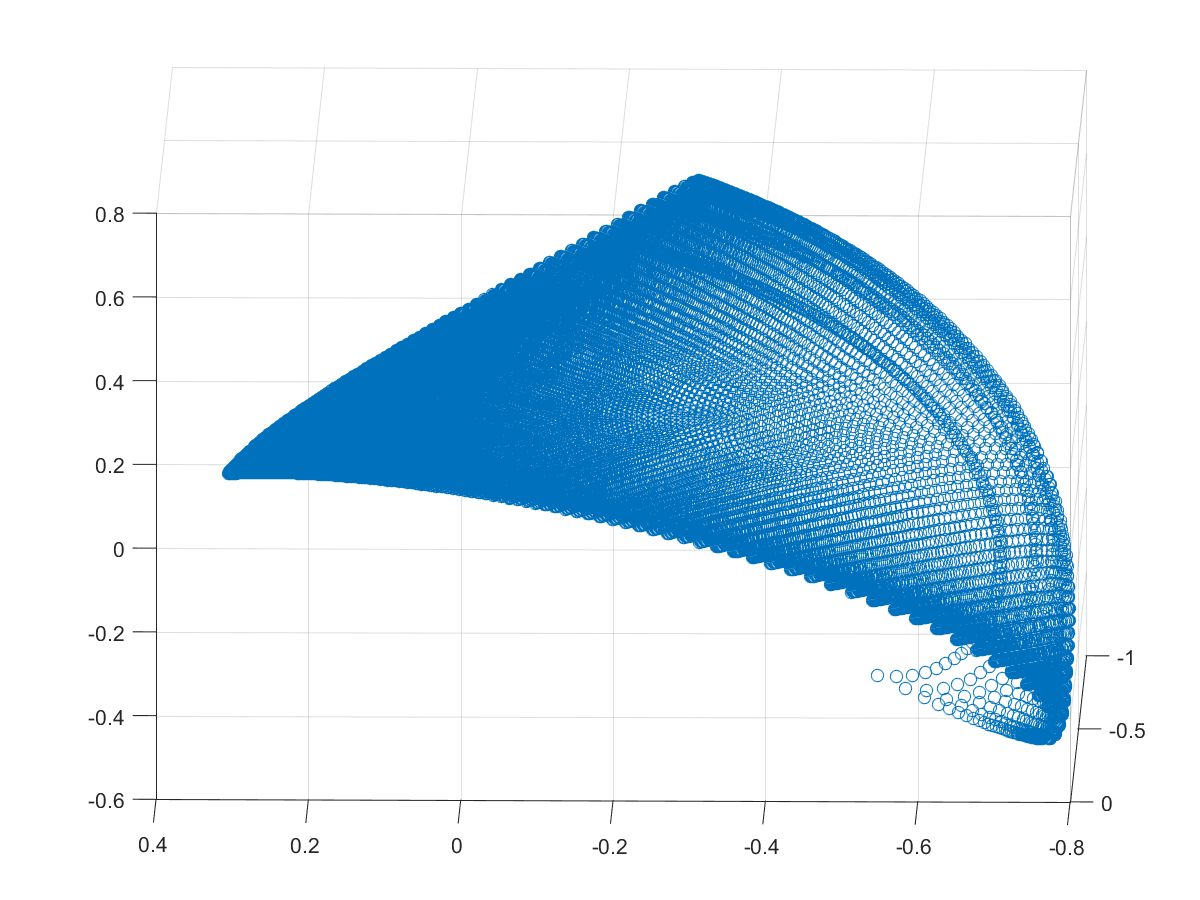} }	
		\caption{Illustration of the low dimensional structures of  datasets presented in Table~\ref{tab:data}. Points are depicted along the first three principal components of each dataset.\label{fig:datasets}}
	\end{minipage}
\end{figure}
\fi

\ifCLASSOPTIONtwocolumn
\begin{figure}[t!]
\centering
\begin{minipage}{\linewidth}
		\subfloat[Scale 2]{\includegraphics[width=.46\textwidth]{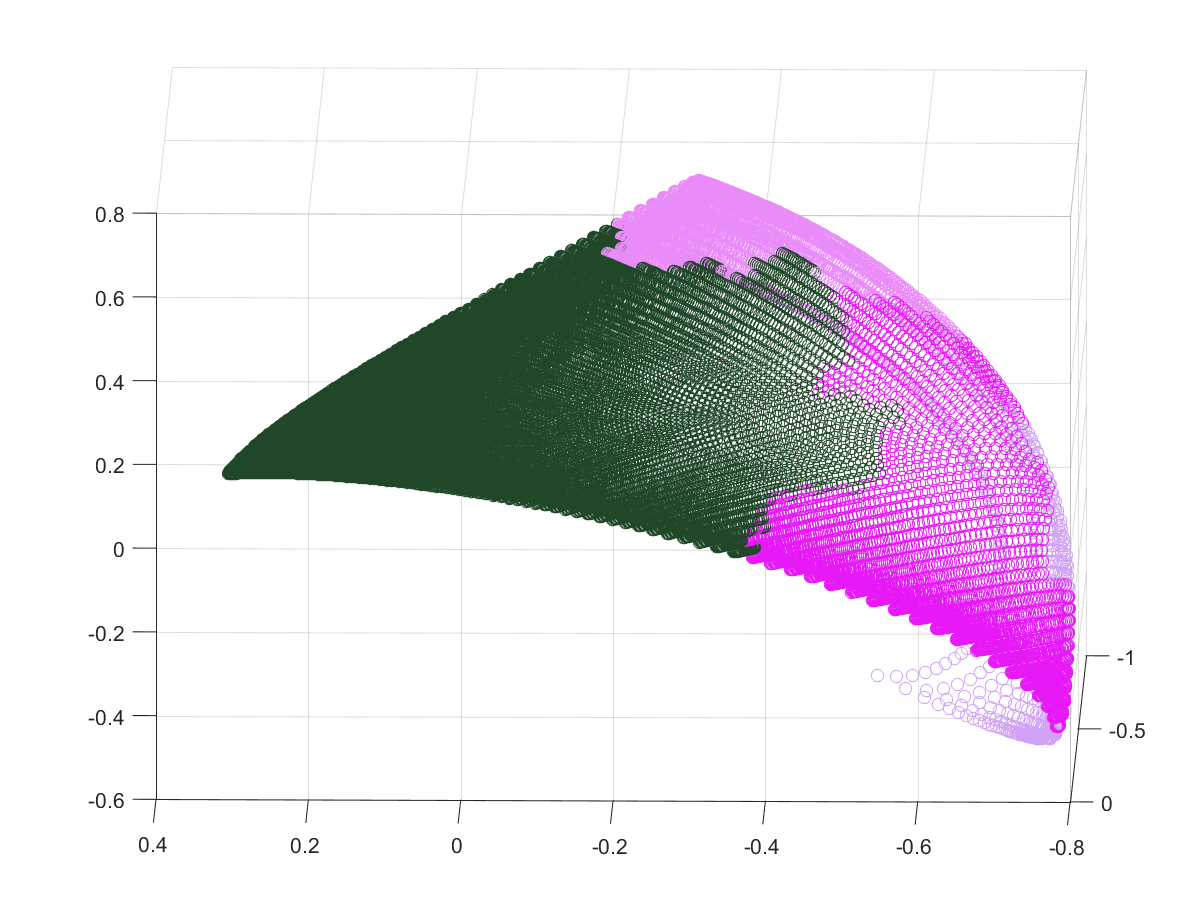} }	
		\quad	
		\subfloat[Scale 3]{\includegraphics[width=.46\textwidth]{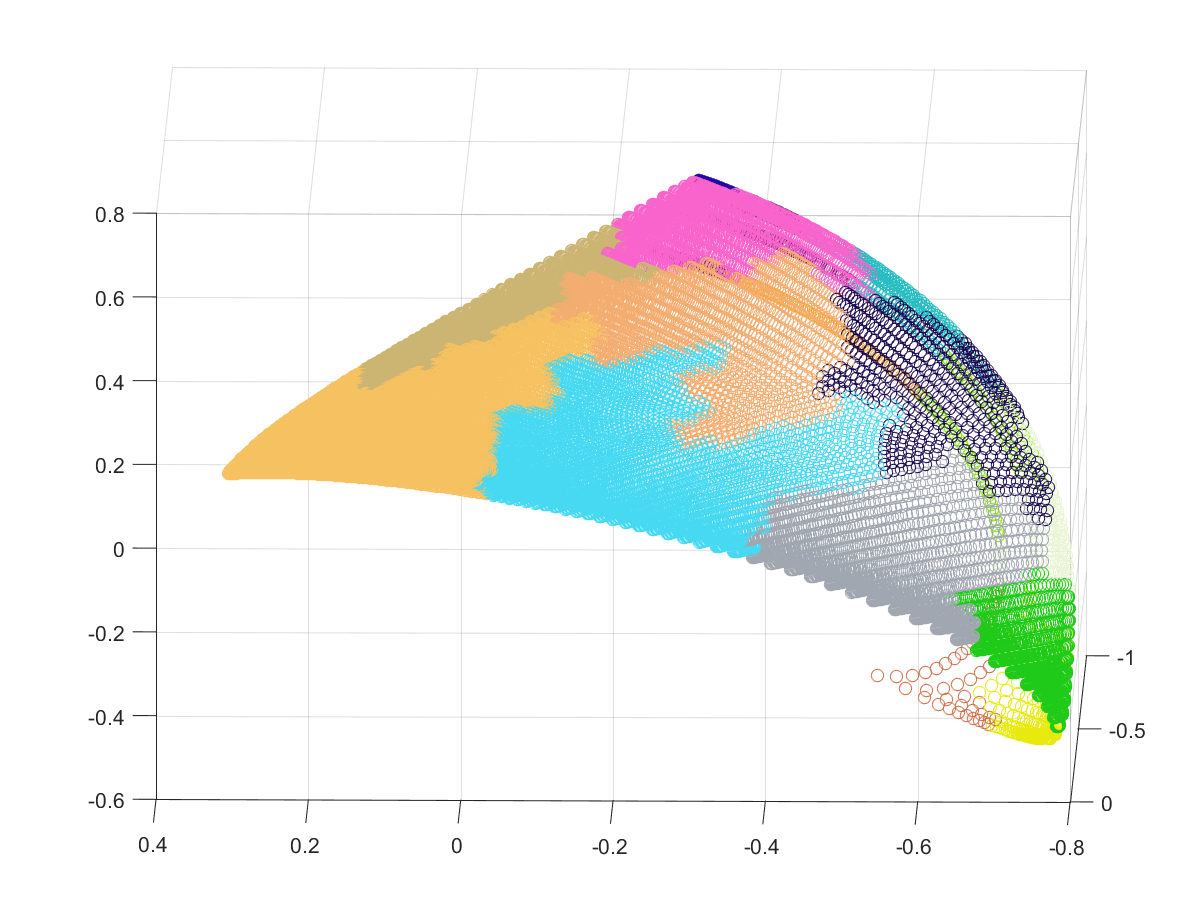} }	
		\\
		\subfloat[Scale 4]{\includegraphics[width=.46\textwidth]{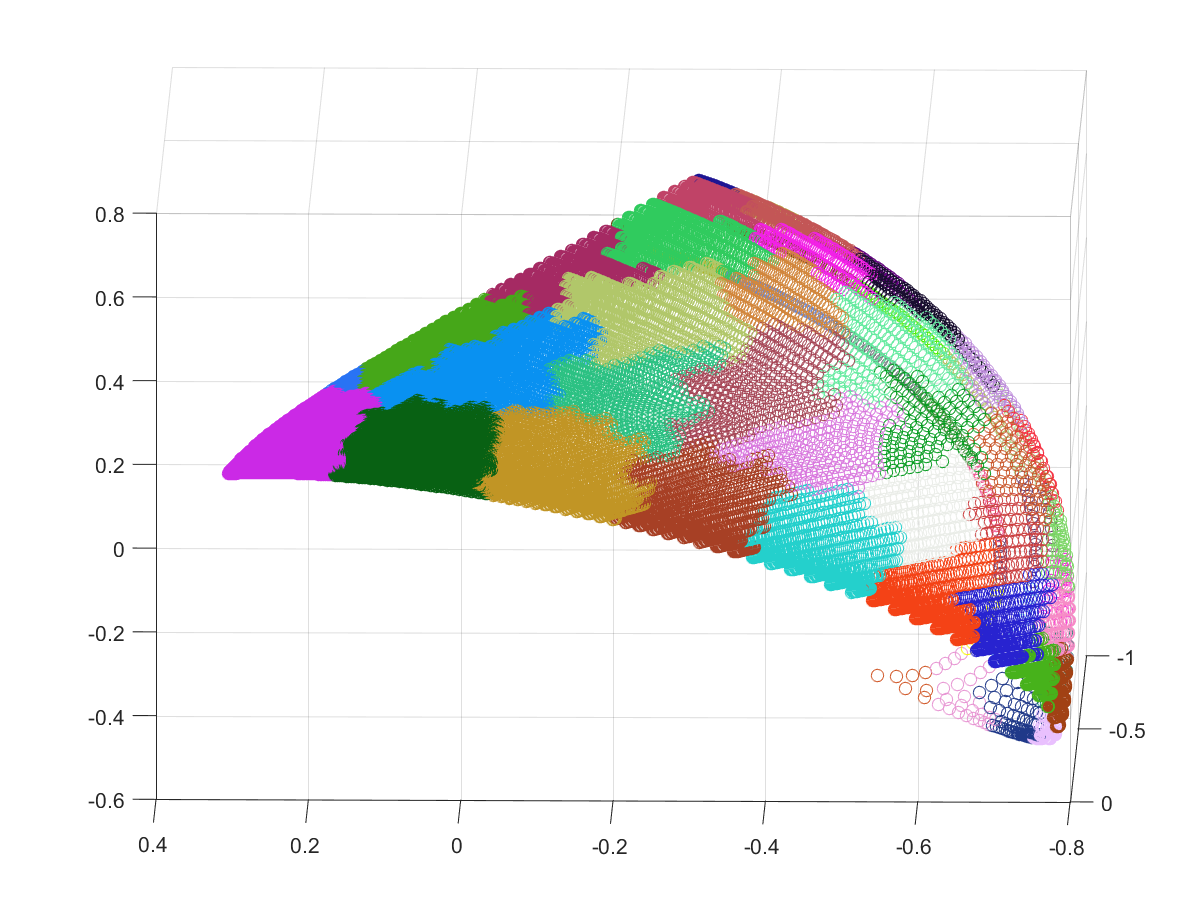} }	
		\quad
		\subfloat[Scale 5]{\includegraphics[width=.46\textwidth]{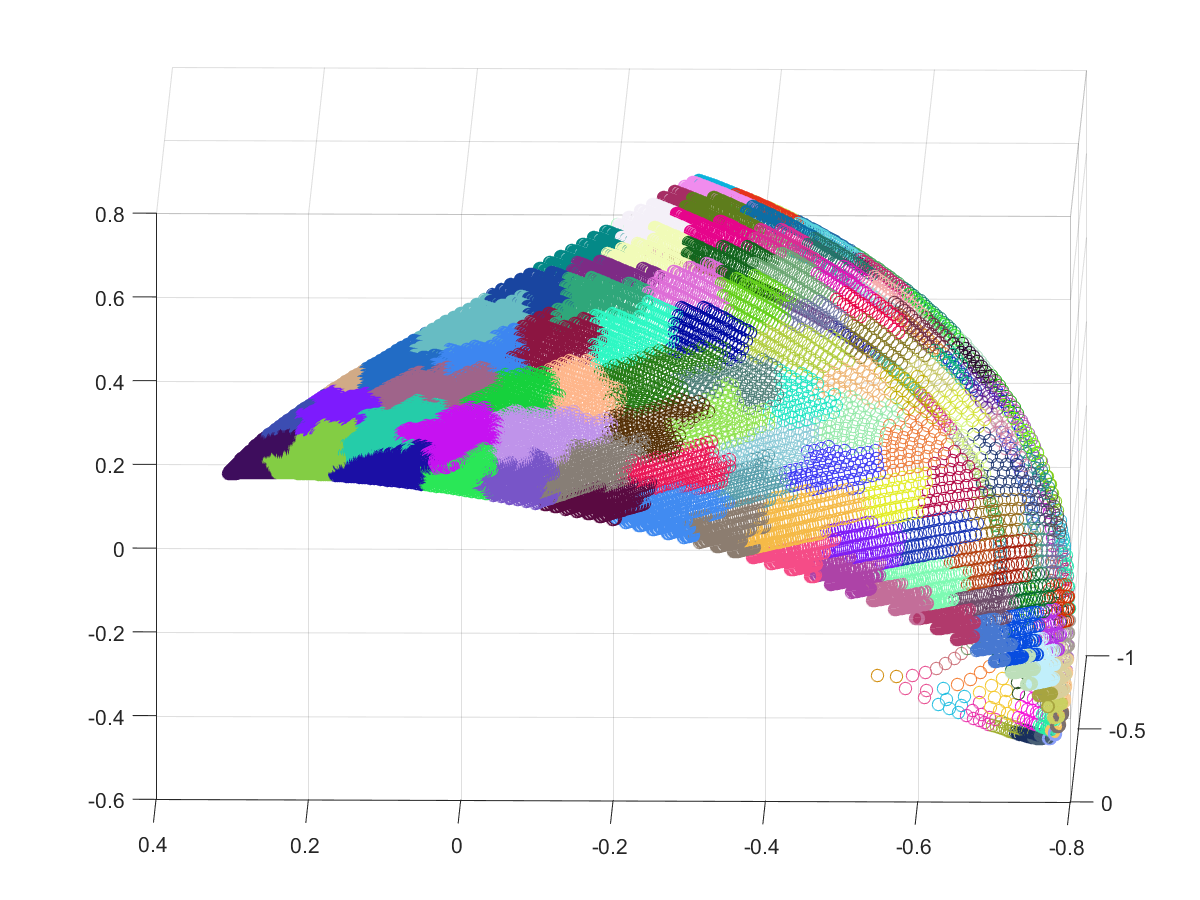} }		
		\caption{A cover tree is built on MR Fingerprints dataset: (a-d) data partitions i.e. descendants covered with parent nodes appearing at scales 2-5
			are highlighted in different colours. The coverage resolution refines 
			by increasing the scale.\label{fig:CT}}
\end{minipage}
\end{figure}
\else
\begin{figure}[t!]
	\centering
	\begin{minipage}{\linewidth}
		\centering
		\subfloat[Scale 2]{\includegraphics[width=.35\textwidth]{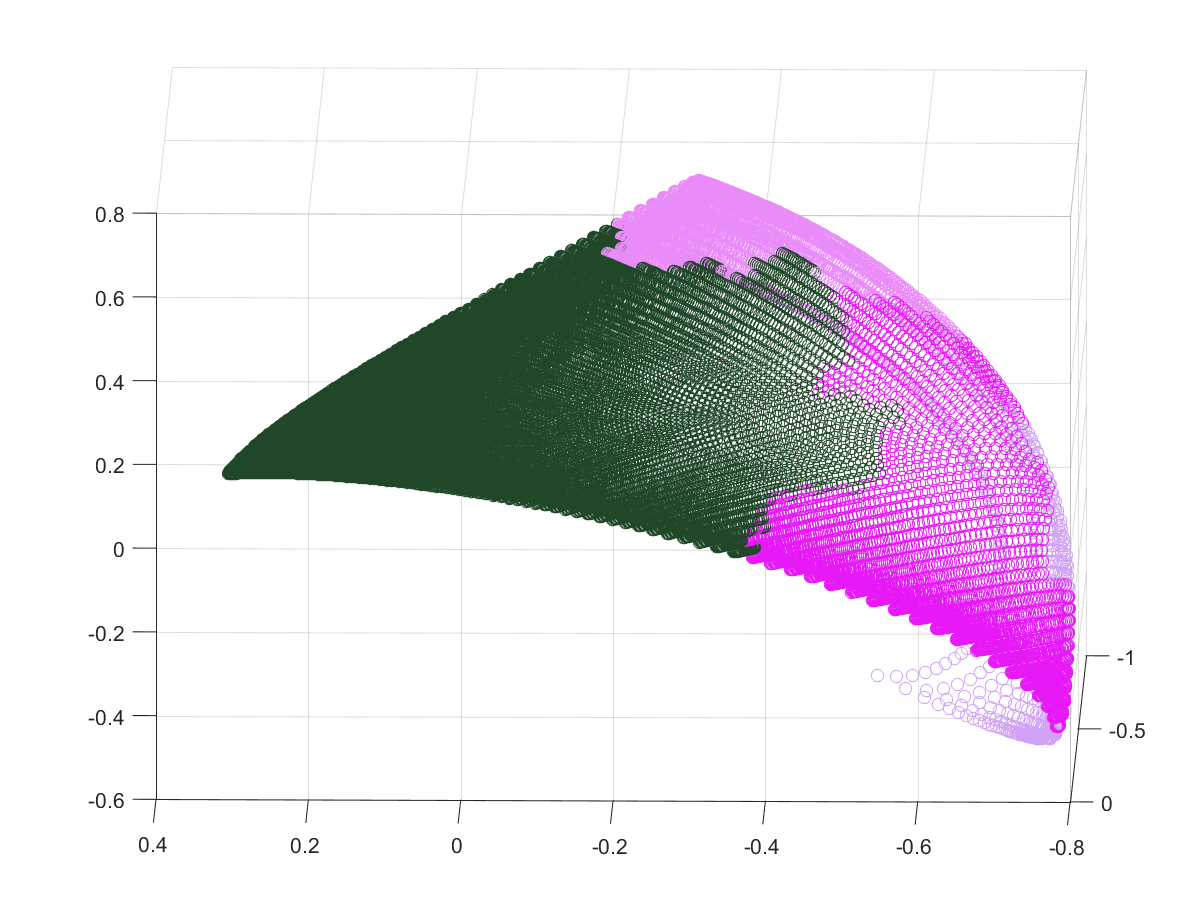} }	
		\quad	
		\subfloat[Scale 3]{\includegraphics[width=.35\textwidth]{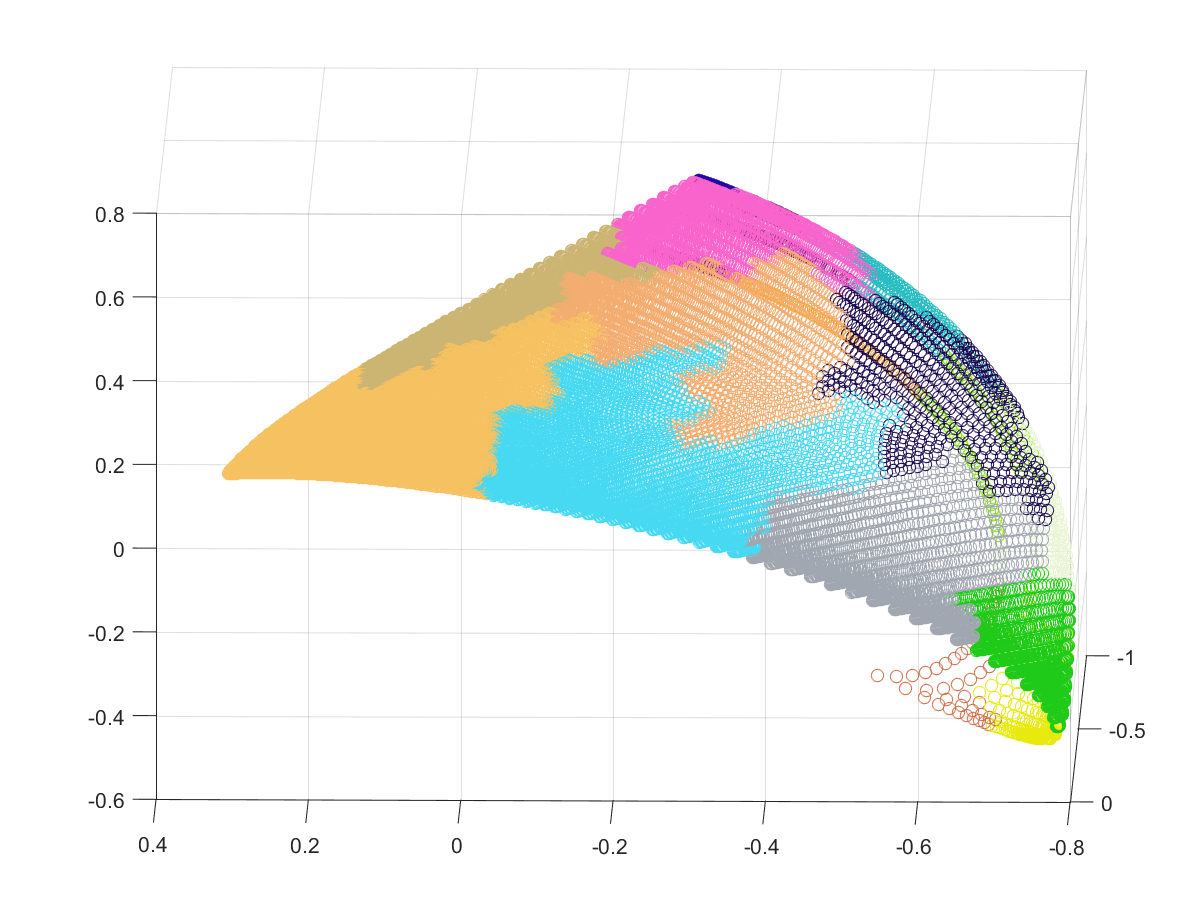} }	
		\\
		\subfloat[Scale 4]{\includegraphics[width=.35\textwidth]{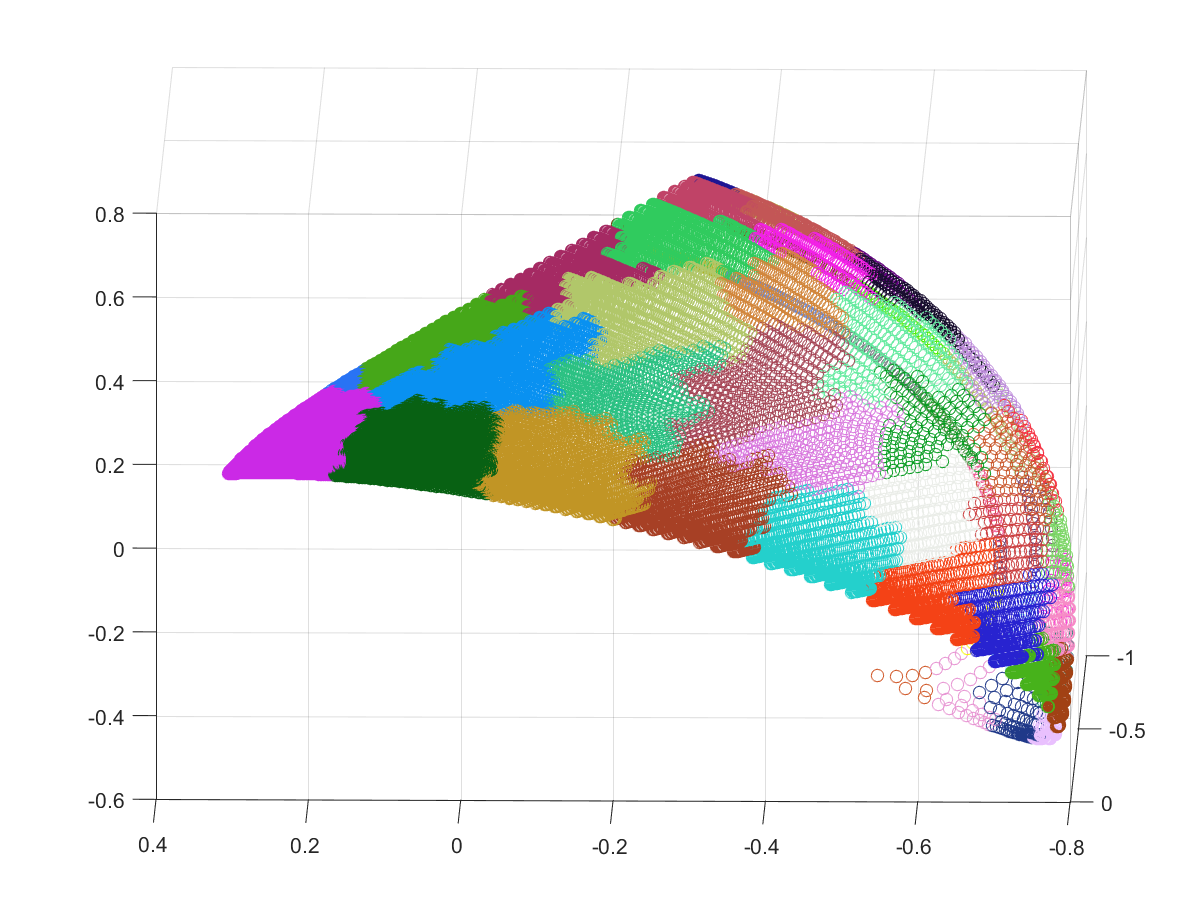} }	
		\quad
		\subfloat[Scale 5]{\includegraphics[width=.35\textwidth]{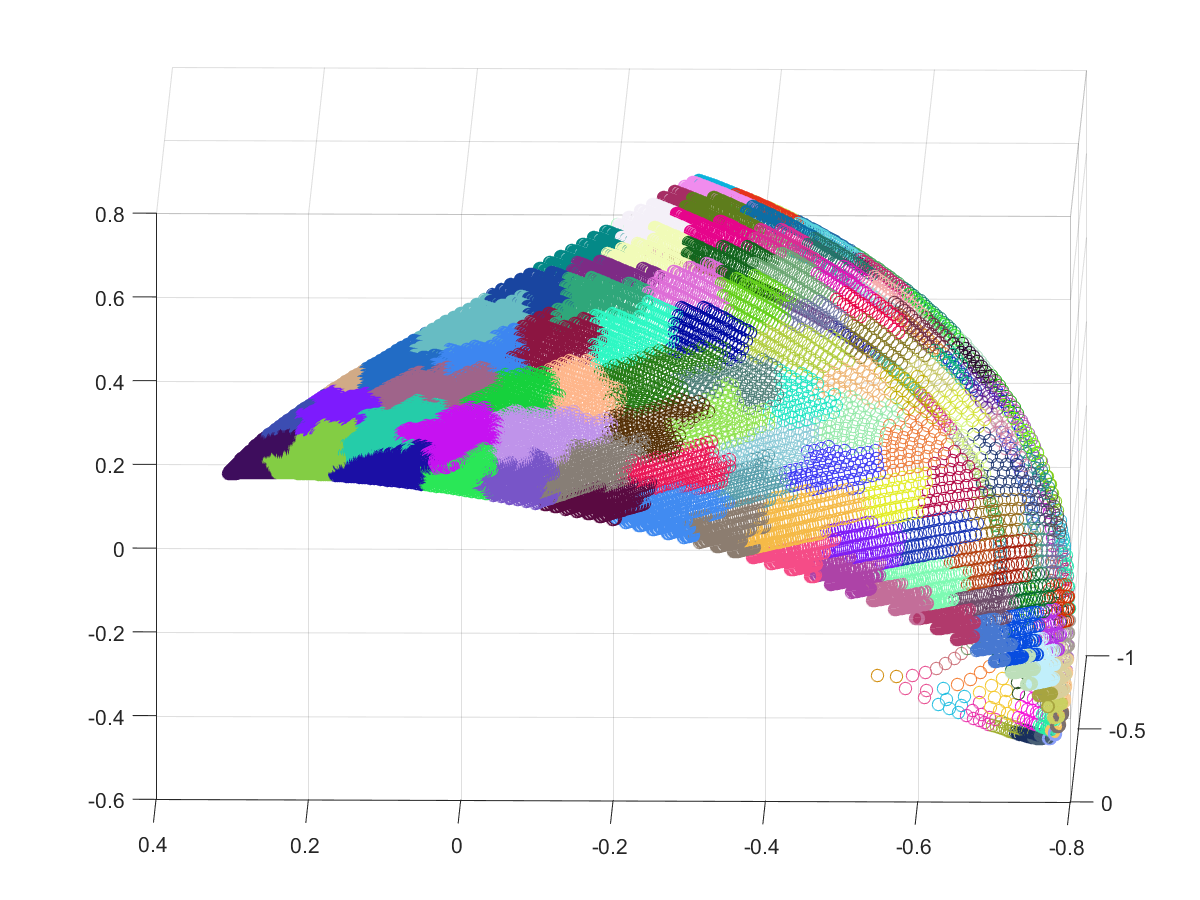} }		
		\caption{A cover tree is built on MR Fingerprints dataset: (a-d) data partitions i.e. descendants covered with parent nodes appearing at scales 2-5
			are highlighted in different colours. The coverage resolution refines 
			by increasing the scale.\label{fig:CT}}
	\end{minipage}
\end{figure}
\fi


\ifCLASSOPTIONtwocolumn
\begin{figure*}[t]
	\centering
	\begin{minipage}{\textwidth}
		\centering
		\subfloat{\includegraphics[width=.225\textwidth]{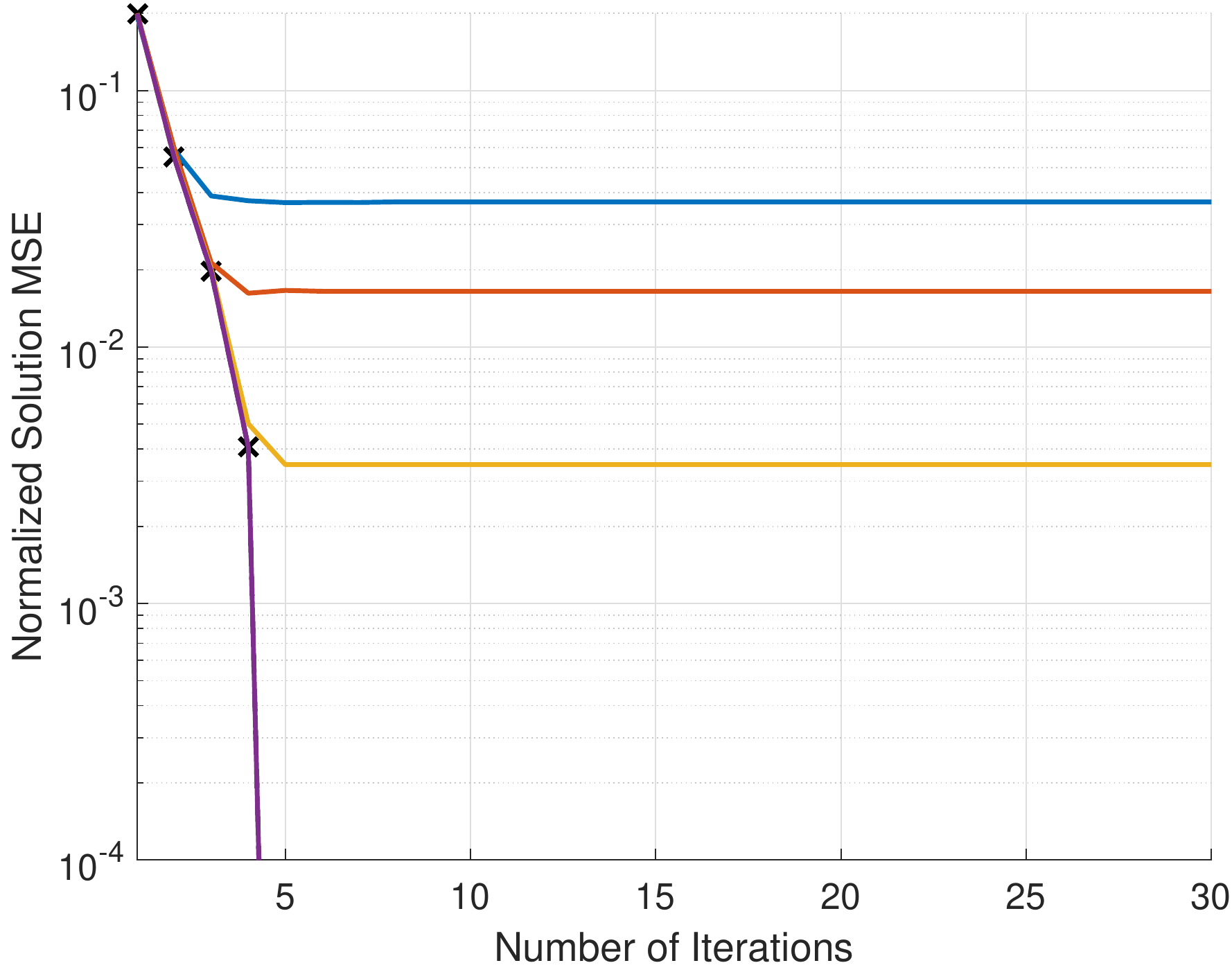} }	
		\quad	
		\subfloat{\includegraphics[width=.225\textwidth]{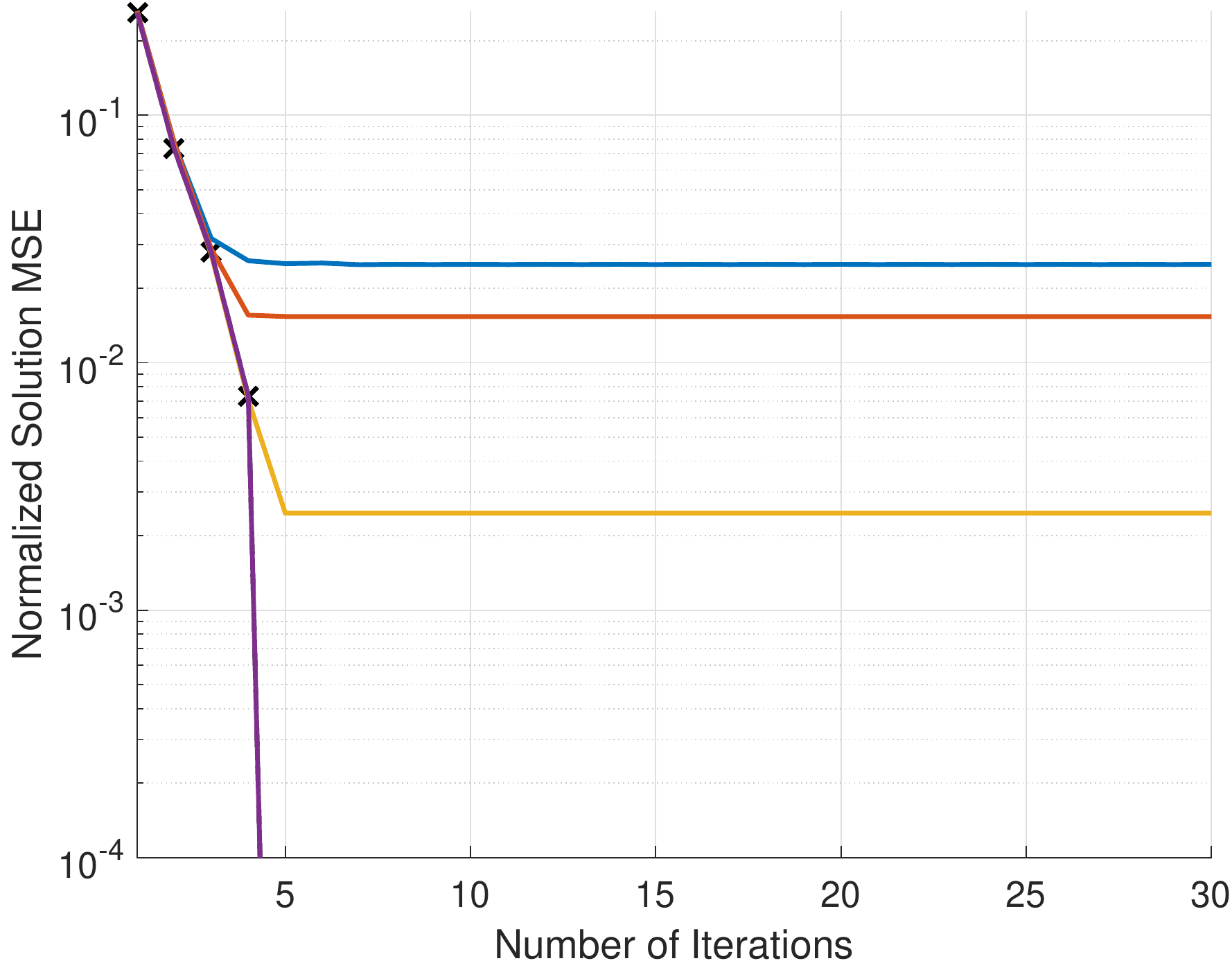} }	
		\quad
		\subfloat{\includegraphics[width=.225\textwidth]{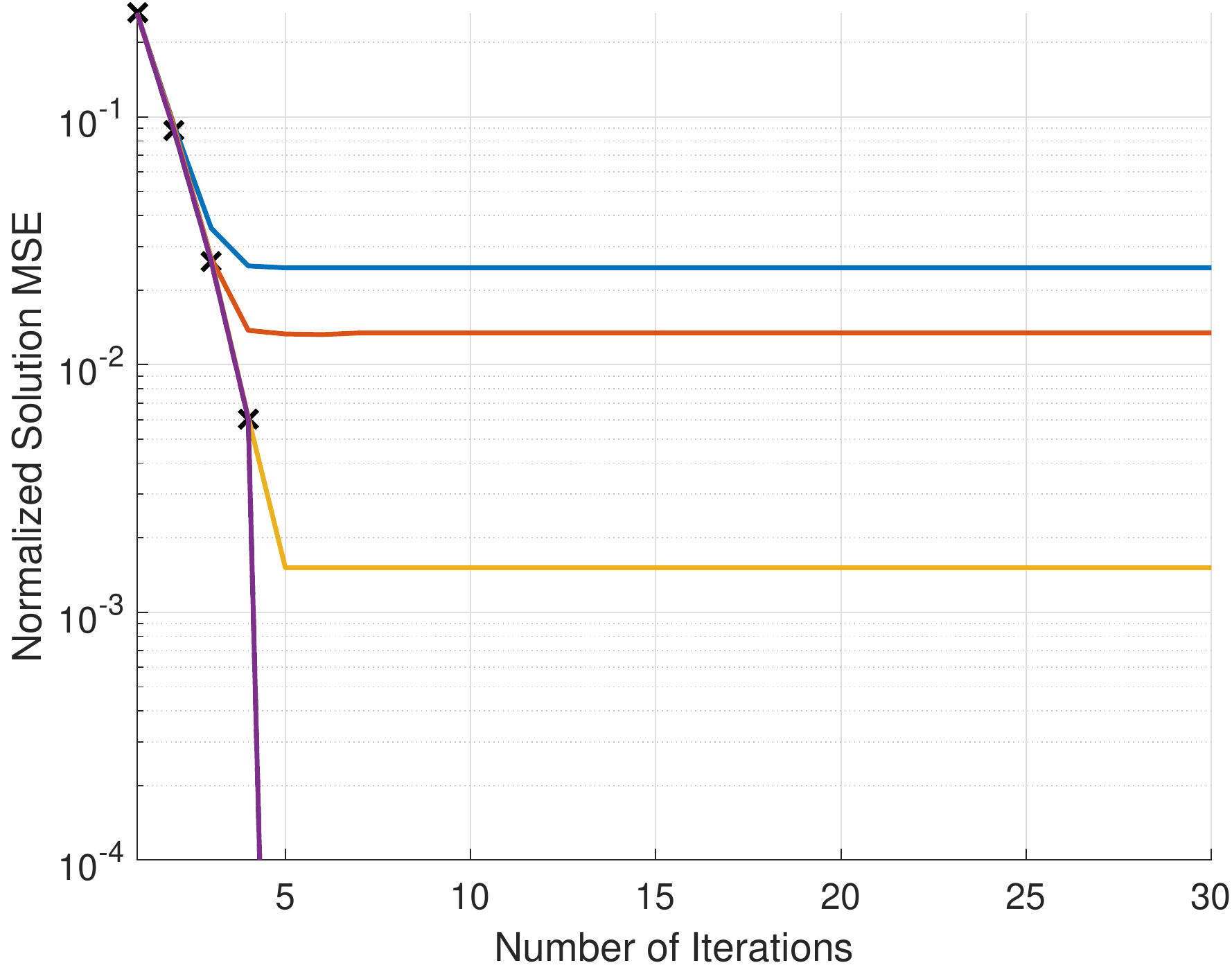} }	
		\quad	
		\subfloat{\includegraphics[width=.225\textwidth]{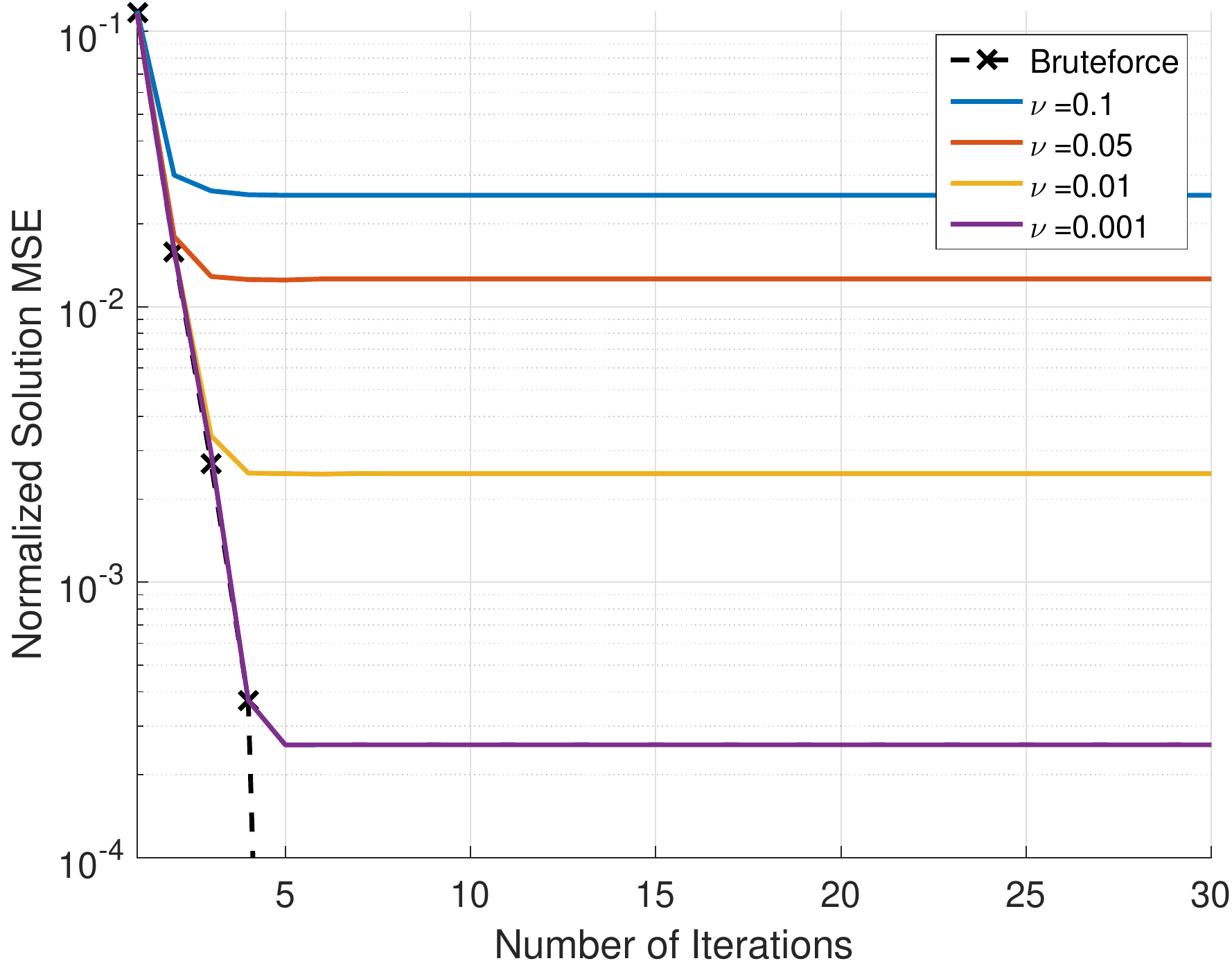} }	
		\quad
		\subfloat{\includegraphics[width=.225\textwidth]{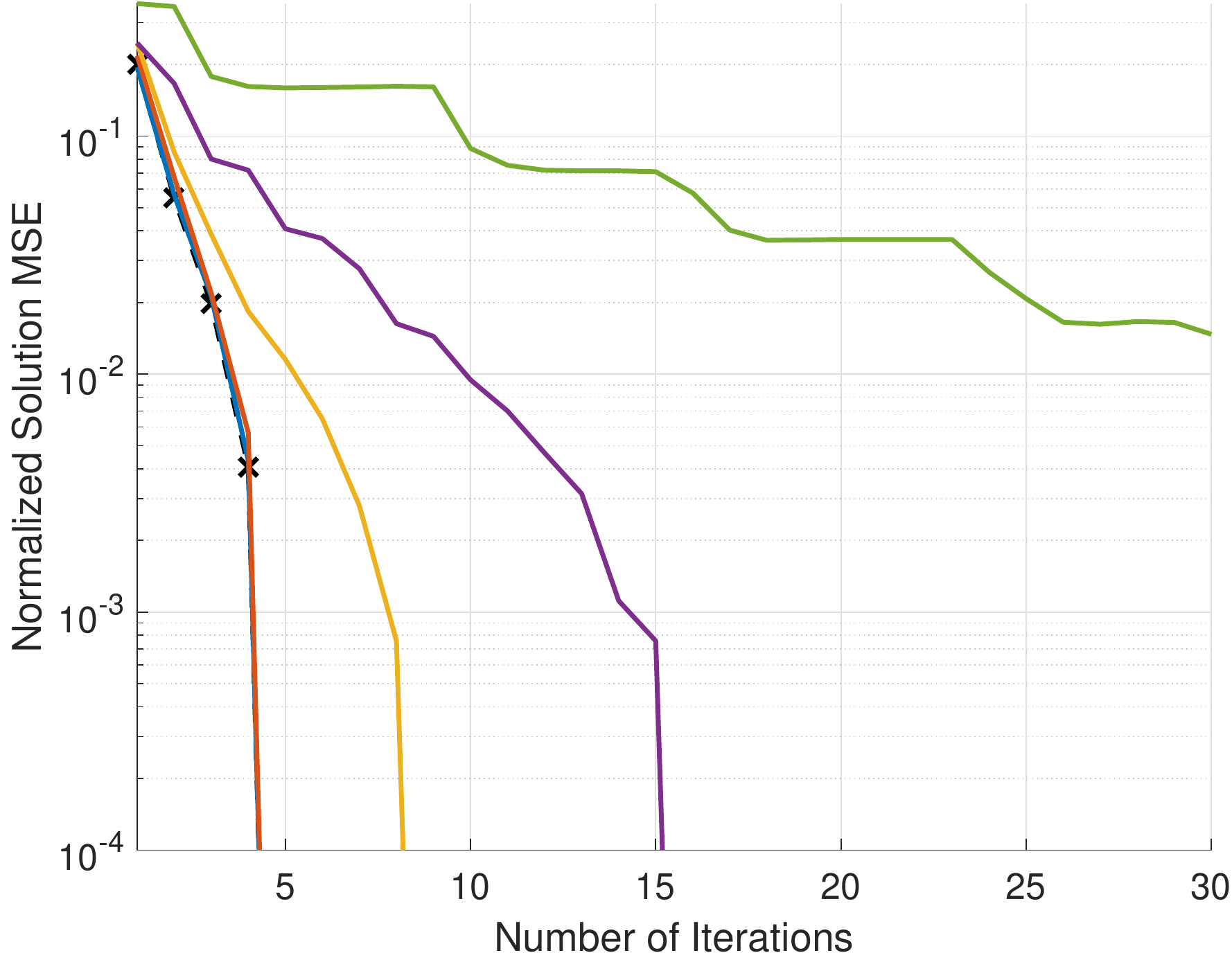} }	
		\quad	
		\subfloat{\includegraphics[width=.225\textwidth]{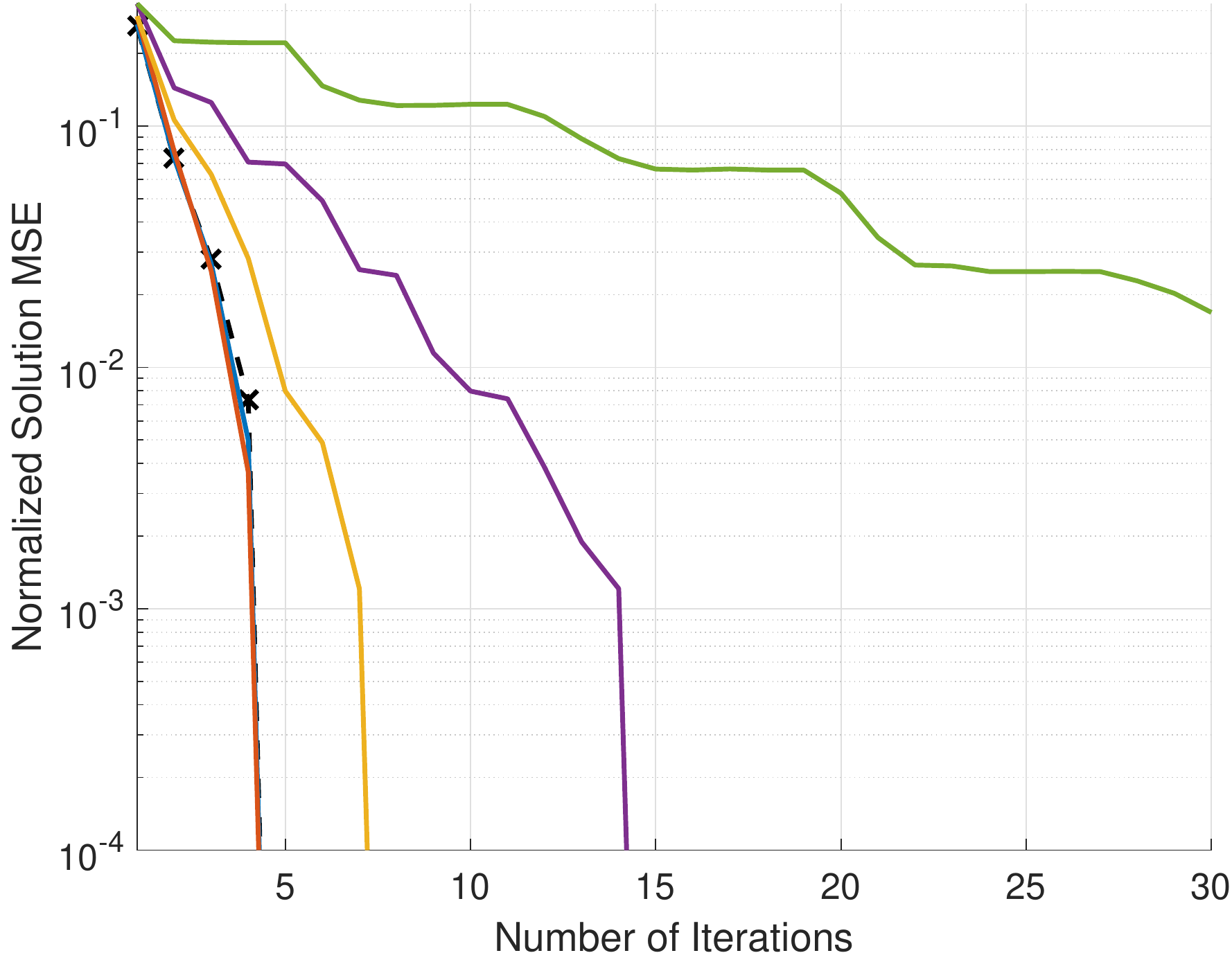} }	
		\quad
		\subfloat{\includegraphics[width=.225\textwidth]{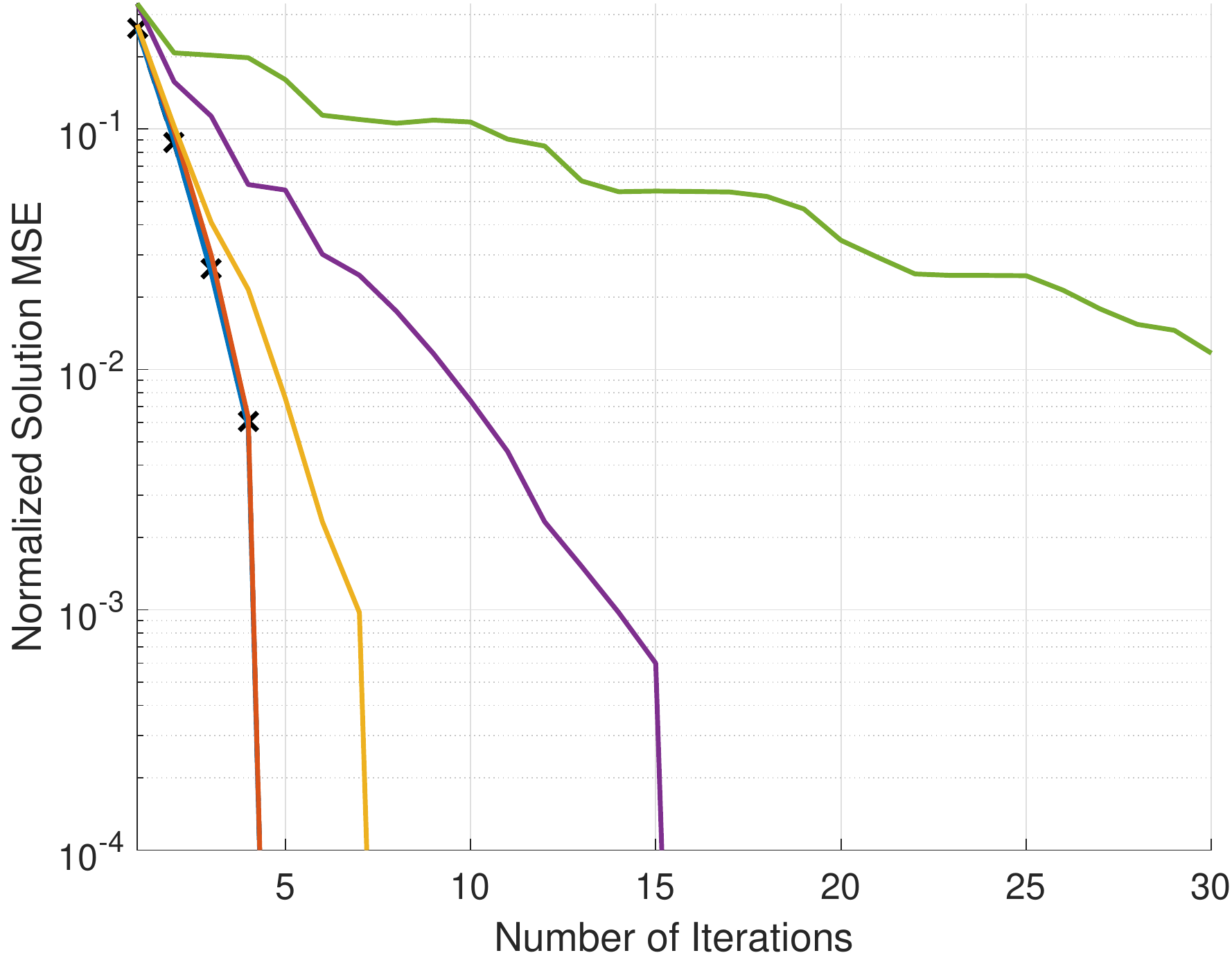} }	
		\quad	
		\subfloat{\includegraphics[width=.225\textwidth]{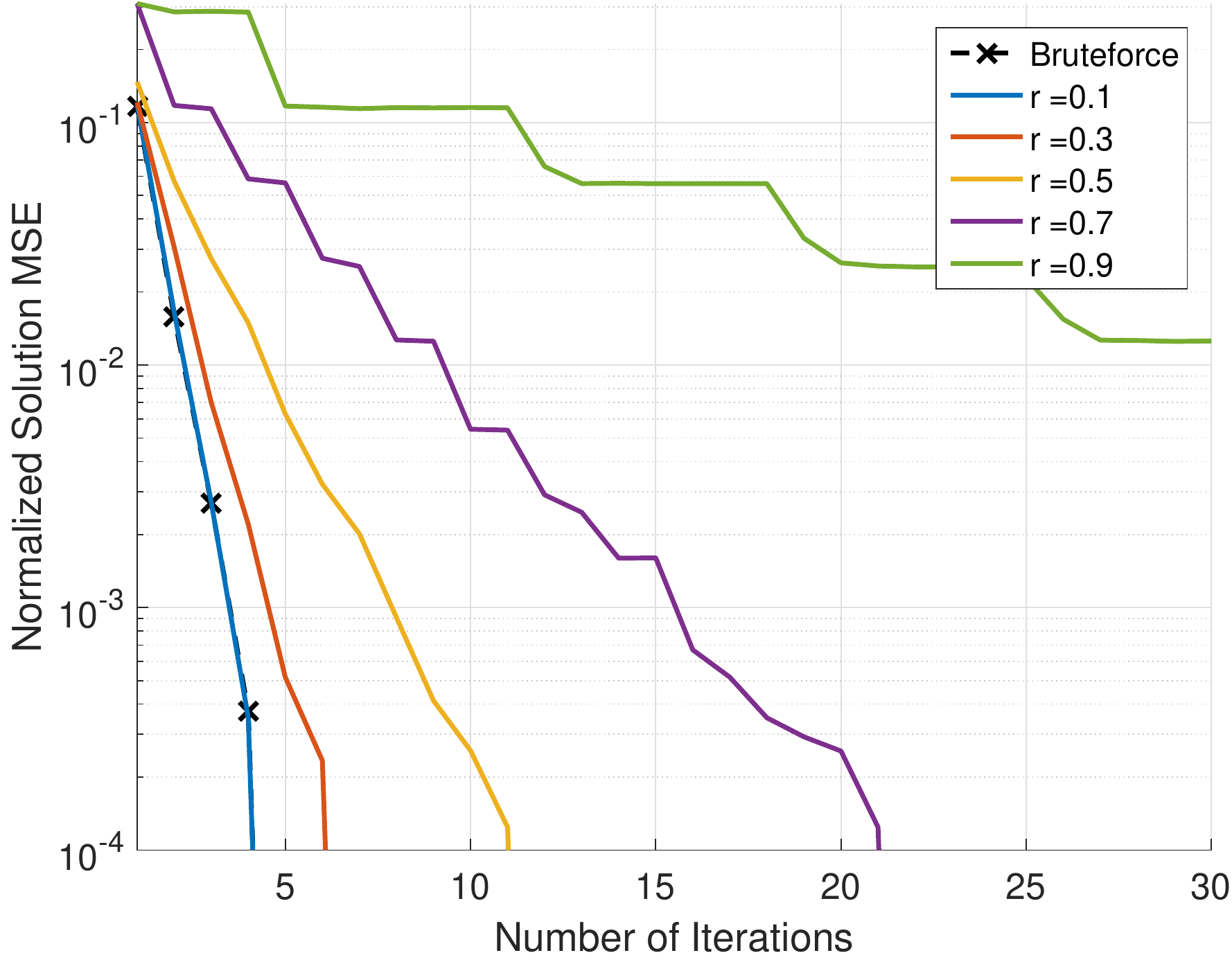} }	
		\quad
		\subfloat{\includegraphics[width=.225\textwidth]{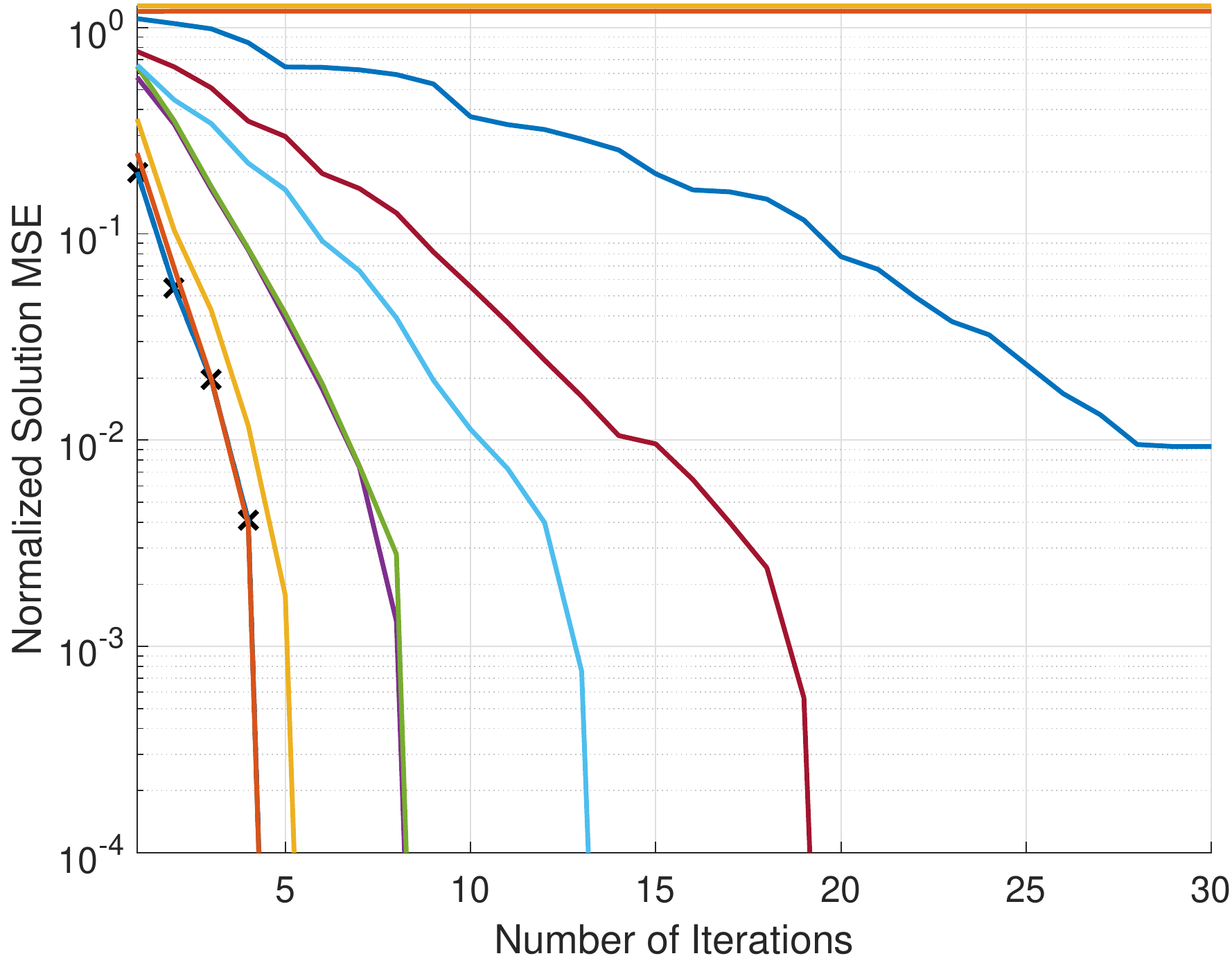} }	
		\quad	
		\subfloat{\includegraphics[width=.225\textwidth]{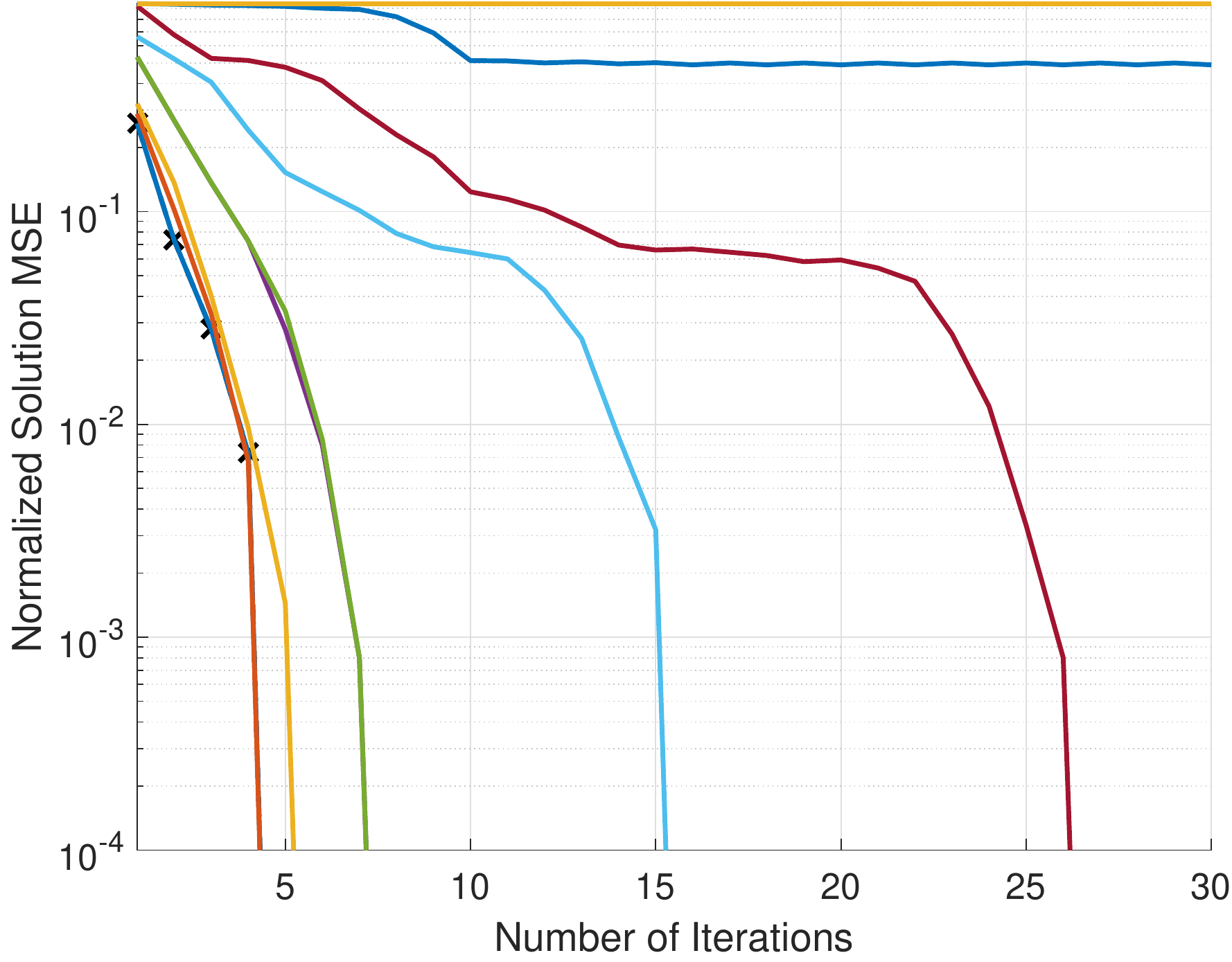} }	
		\quad
		\subfloat{\includegraphics[width=.225\textwidth]{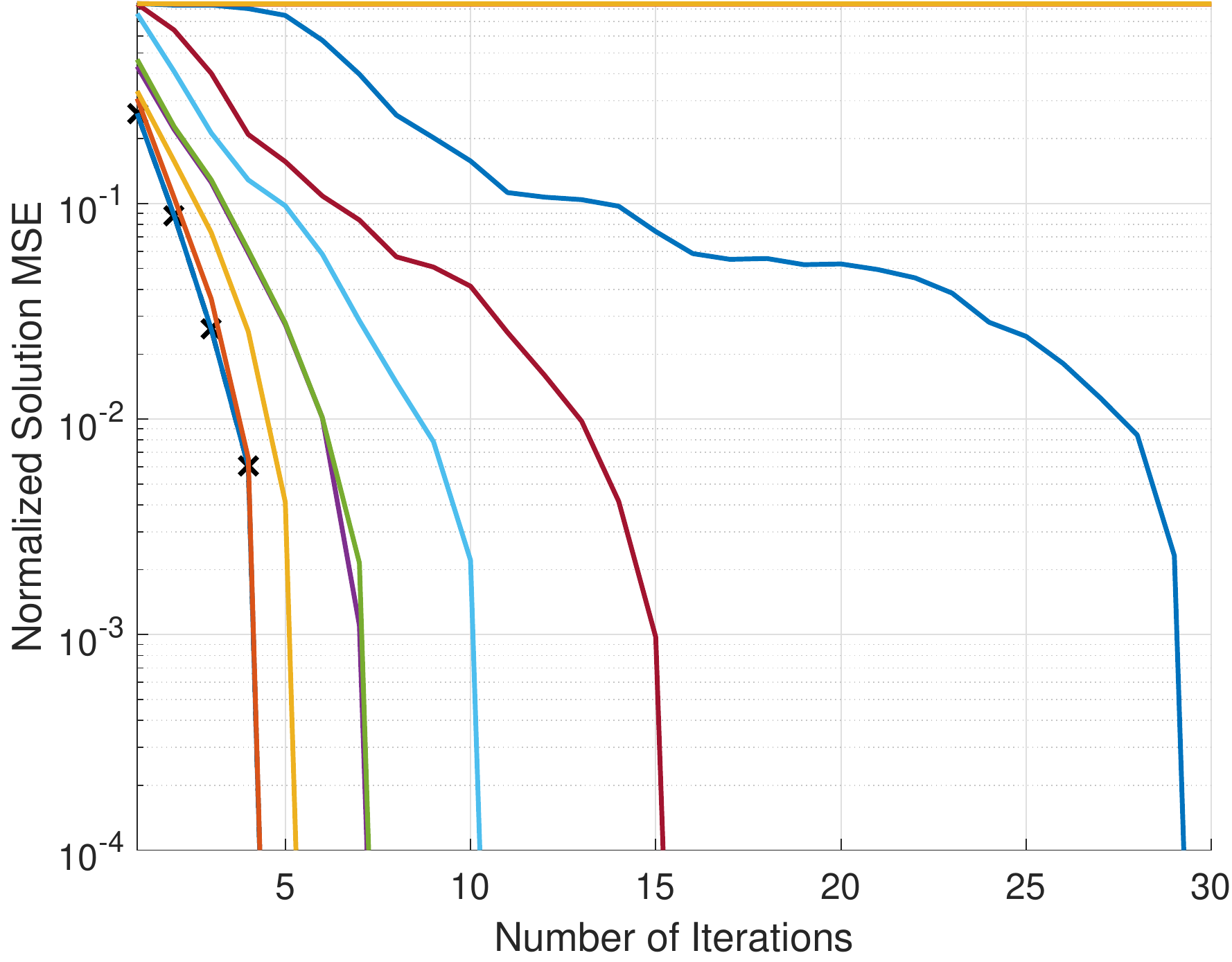} }	
		\quad	
		\subfloat{\includegraphics[width=.225\textwidth]{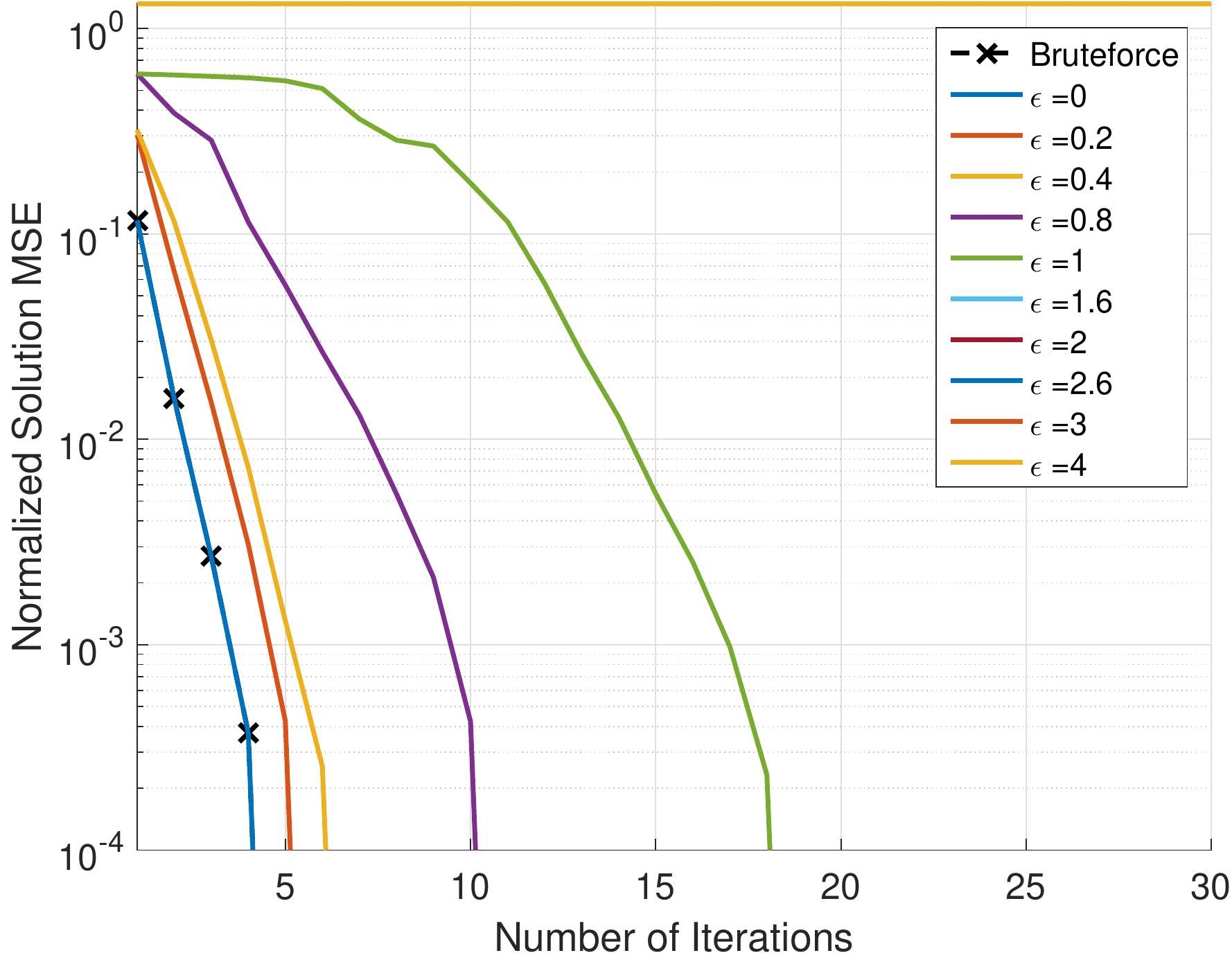} }	
		\caption{Convergence of the exact/inexact IPG for subsampling ratio $\frac{m}{n}=0.2$. 
			Rows from top to bottom correspond to inexact algorithms with FP, PFP and $1+\epsilon$ ANN searches, respectively (legends for the plots in each row are identical and included in the last column). Columns from left to right correspond to  S-Manifold, Swiss roll, Oscillating wave and MR Fingerprints datasets, respectively. \label{fig:Decays}}
	\end{minipage}
\end{figure*}
\else
\begin{figure*}[t]
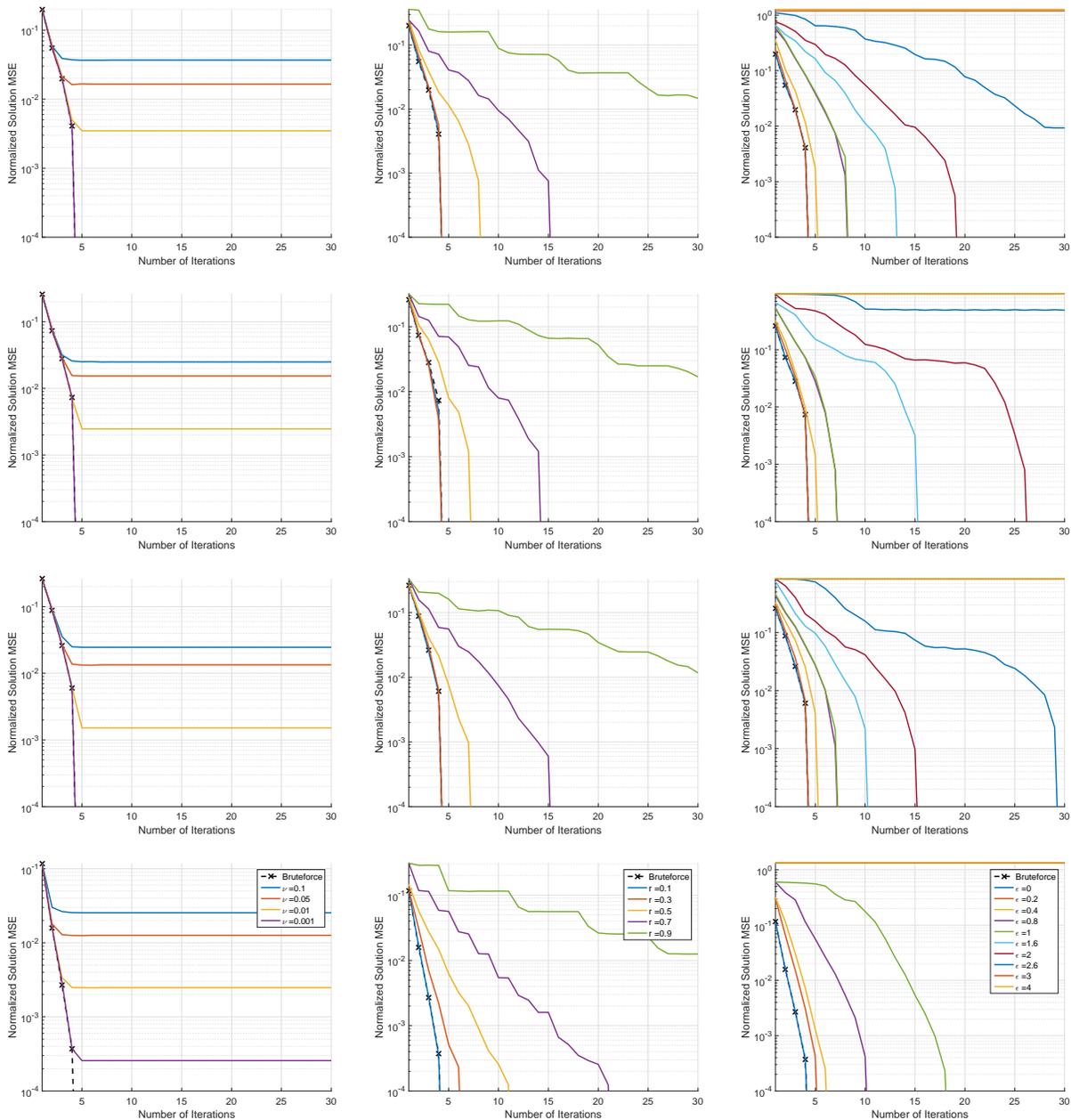

	\centering
	\begin{minipage}{\textwidth}
		\centering
		\subfloat{\includegraphics[width=.3\textwidth]{Sol_iter_data_1_alg_4_compr_4} }	
		\quad
		\subfloat{\includegraphics[width=.3\textwidth]{Sol_iter_data_1_alg_3_compr_4} }	
		\quad
		\subfloat{\includegraphics[width=.3\textwidth]{Sol_iter_data_1_alg_2_compr_4} }
		\quad			
		\subfloat{\includegraphics[width=.3\textwidth]{Sol_iter_data_2_alg_4_compr_4} }	
		\quad	
		\subfloat{\includegraphics[width=.3\textwidth]{Sol_iter_data_2_alg_3_compr_4} }
		\quad	
		\subfloat{\includegraphics[width=.3\textwidth]{Sol_iter_data_2_alg_2_compr_4} }
		\quad
		\subfloat{\includegraphics[width=.3\textwidth]{Sol_iter_data_3_alg_4_compr_4} }	
		\quad
		\subfloat{\includegraphics[width=.3\textwidth]{Sol_iter_data_3_alg_3_compr_4} }
		\quad
		\subfloat{\includegraphics[width=.3\textwidth]{Sol_iter_data_3_alg_2_compr_4} }		
		\quad	
		\subfloat{\includegraphics[width=.3\textwidth]{Sol_iter_data_4_alg_4_compr_4} }		
		\quad	
		\subfloat{\includegraphics[width=.3\textwidth]{Sol_iter_data_4_alg_3_compr_4} }	
		\quad	
		\subfloat{\includegraphics[width=.3\textwidth]{Sol_iter_data_4_alg_2_compr_4} }	
		\caption{Convergence of the exact/inexact IPG for subsampling ratio $\frac{m}{n}=0.2$. 			 
			Columns from left to right correspond to inexact algorithms with FP, PFP and $1+\epsilon$ ANN searches, respectively (legends for the plots in each column are identical and included in the last row). Rows from top to bottom correspond to   S-Manifold, Swiss roll, Oscillating wave and MR Fingerprints datasets, respectively. \label{fig:Decays}}
	\end{minipage}
\end{figure*}
\fi

\ifCLASSOPTIONtwocolumn
\begin{figure*}
	\centering
	\begin{minipage}{\textwidth}
		\centering
		\subfloat
		{\includegraphics[width=.225\textwidth]{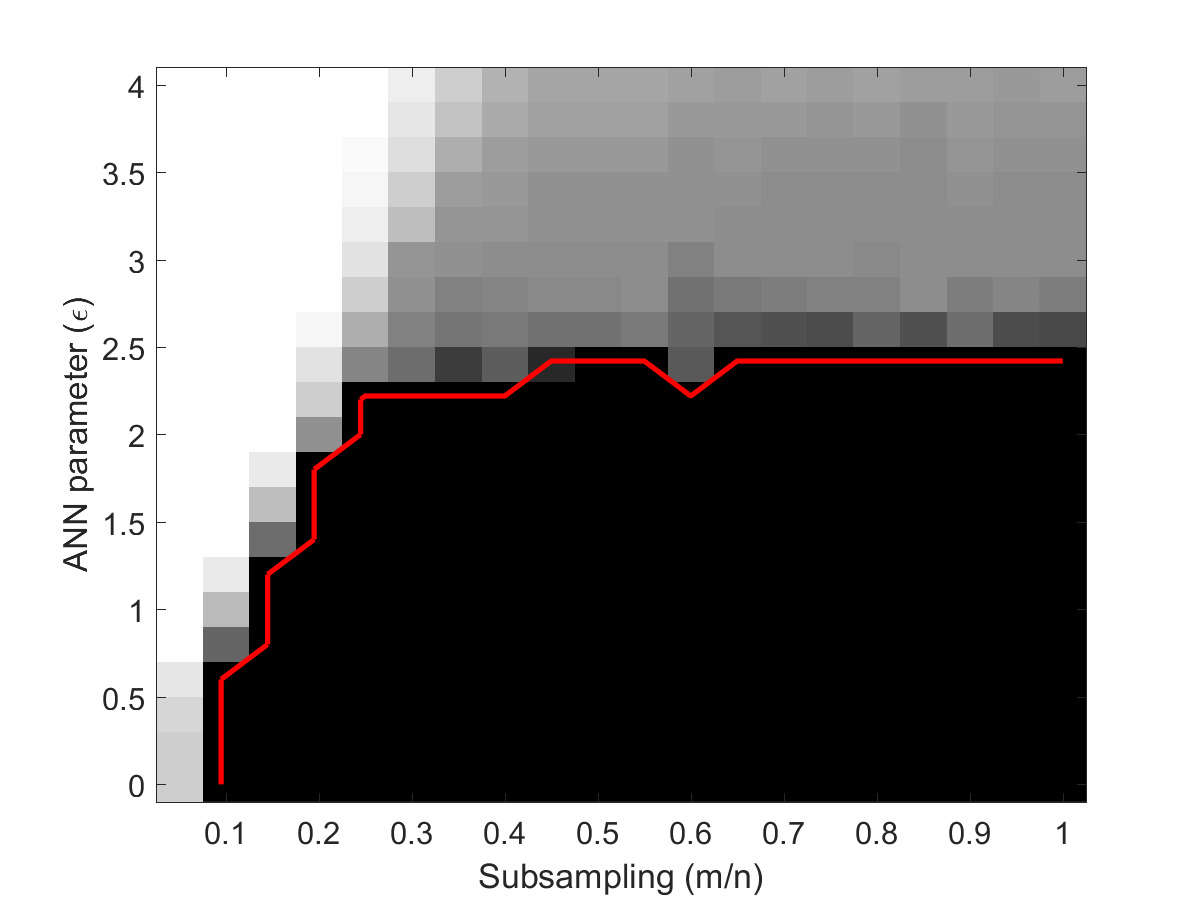} }
		\quad
		\subfloat
		{\includegraphics[width=.225\textwidth]{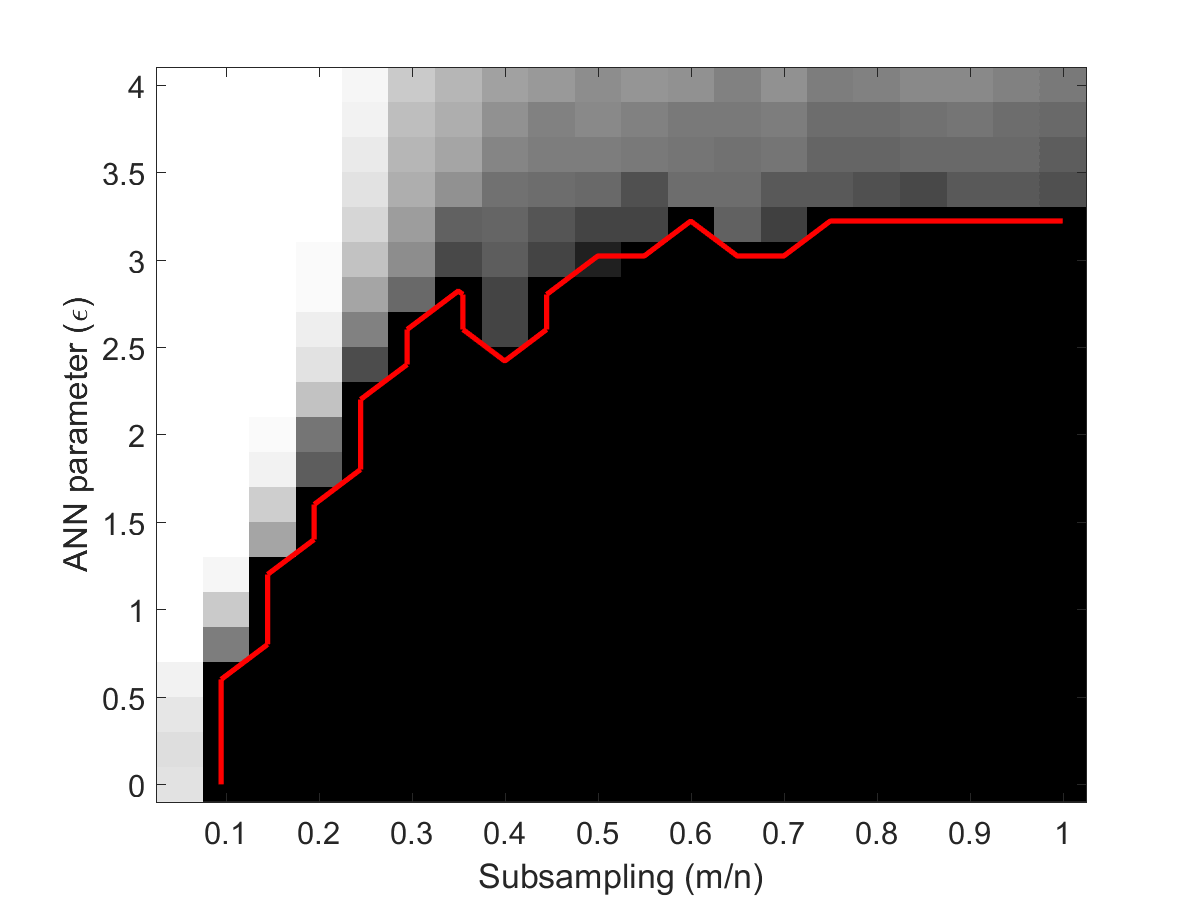} }
		\quad
		\subfloat
		{\includegraphics[width=.225\textwidth]{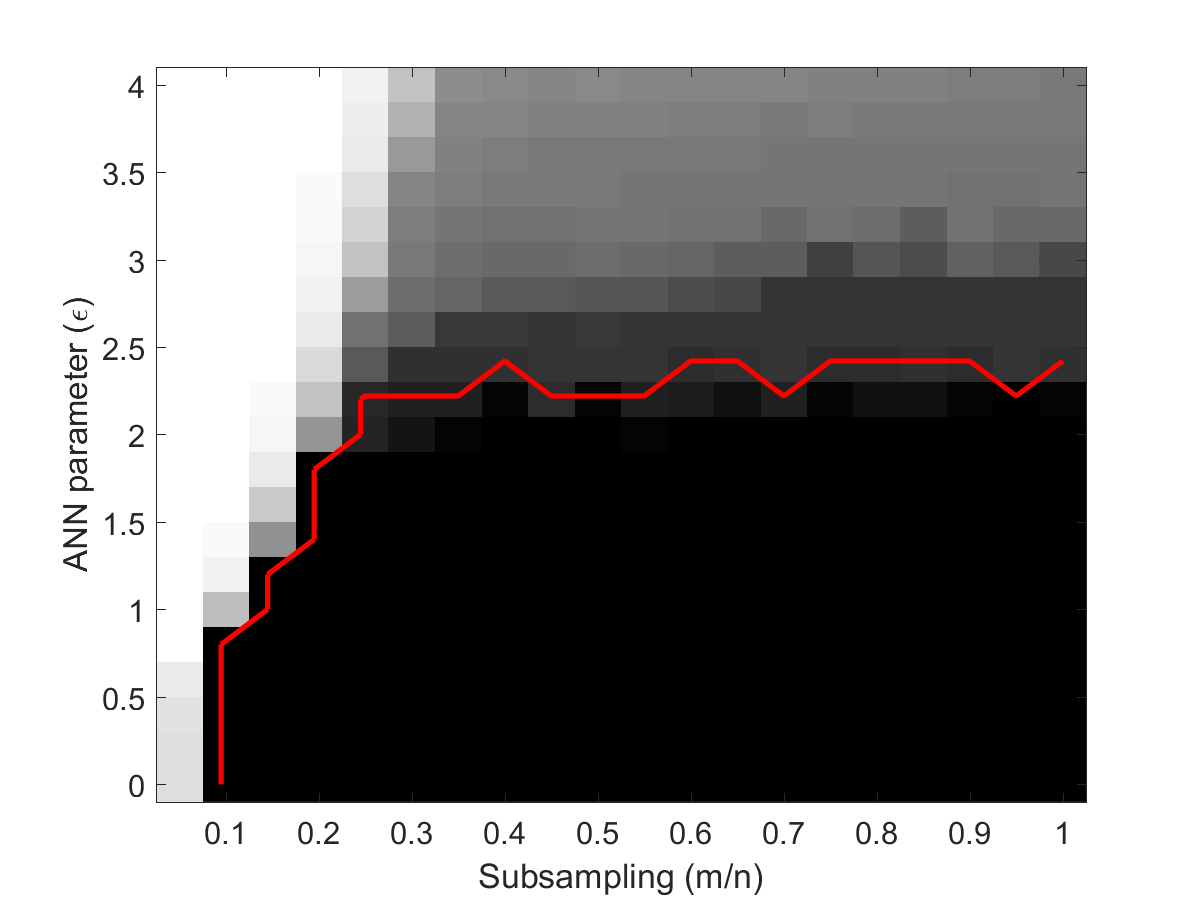} }
		\quad
		\subfloat
		{\includegraphics[width=.225\textwidth]{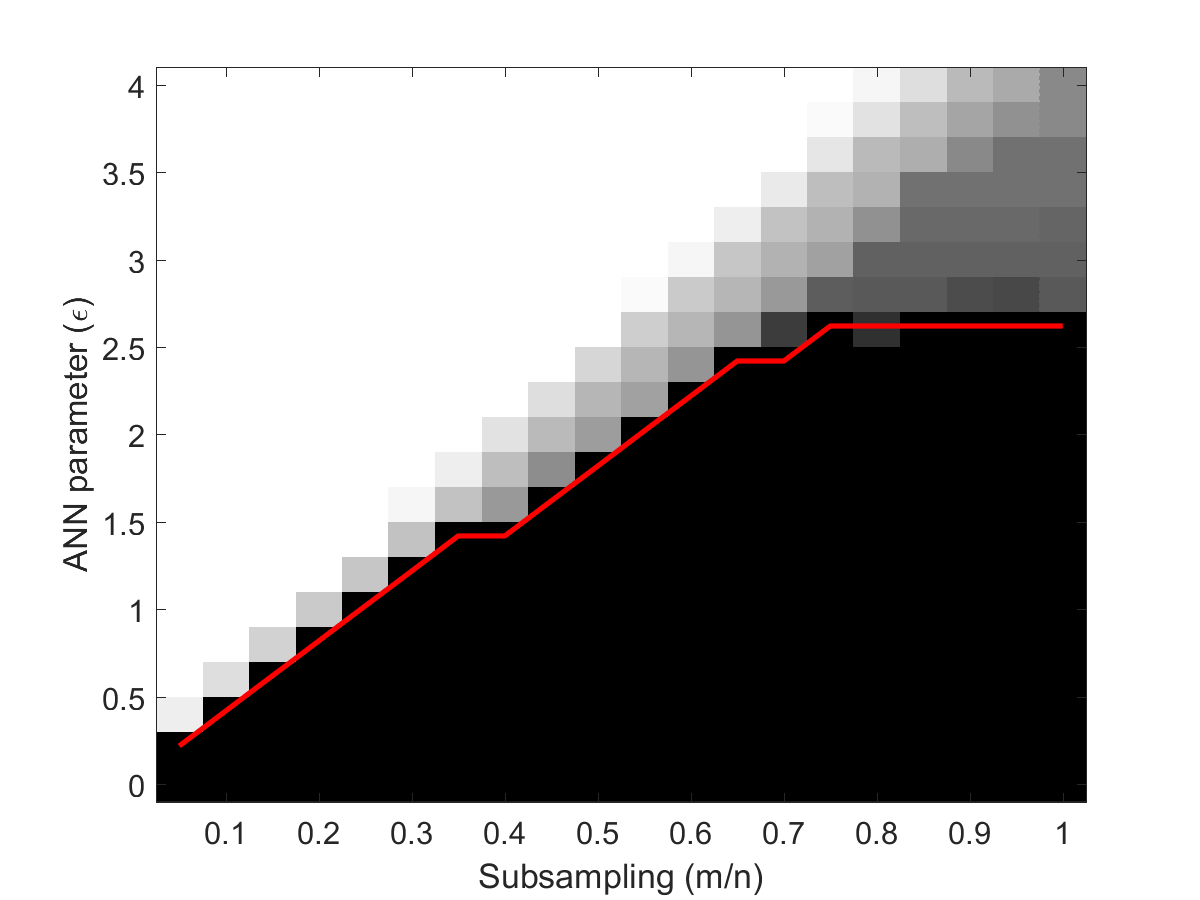} }
		\quad
		\subfloat
		{\includegraphics[width=.225\textwidth]{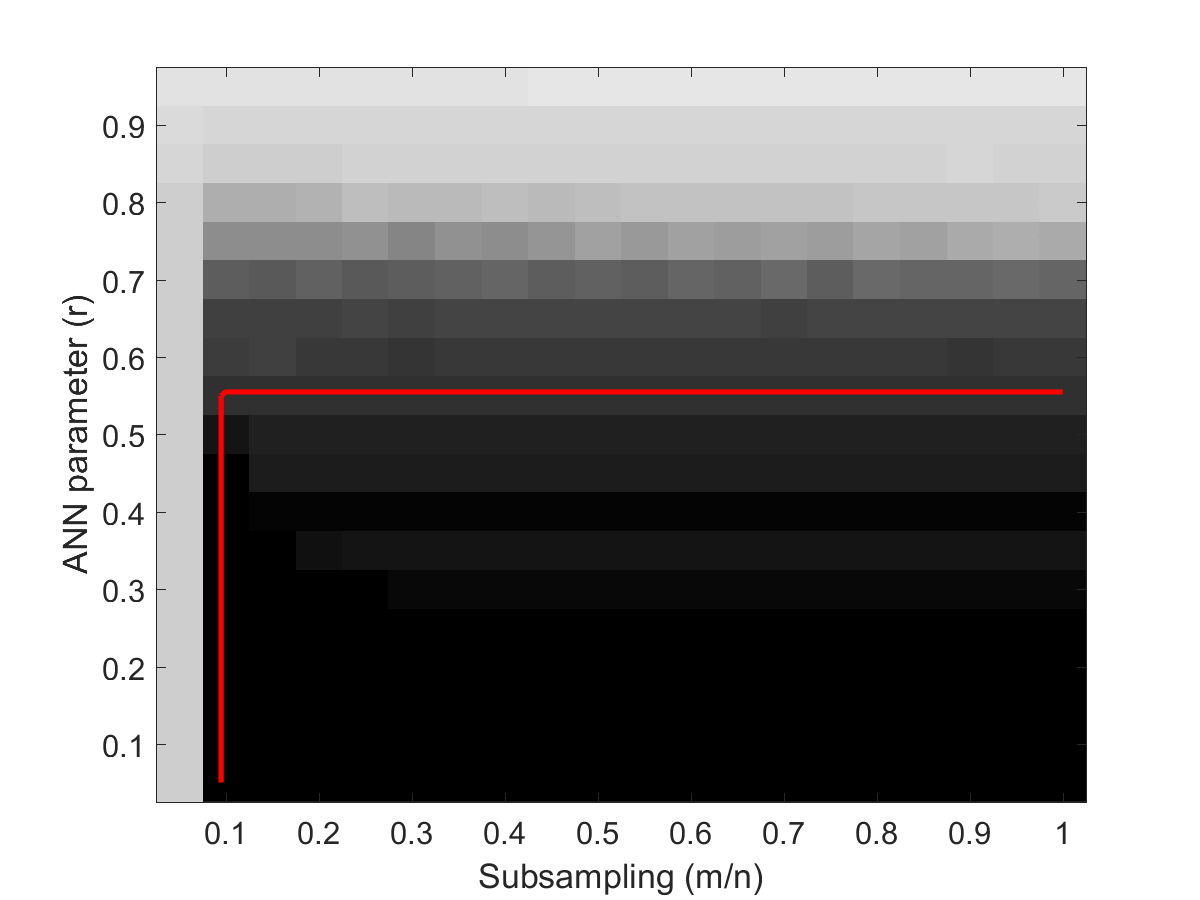} }
		\quad		
		\subfloat
		{\includegraphics[width=.225\textwidth]{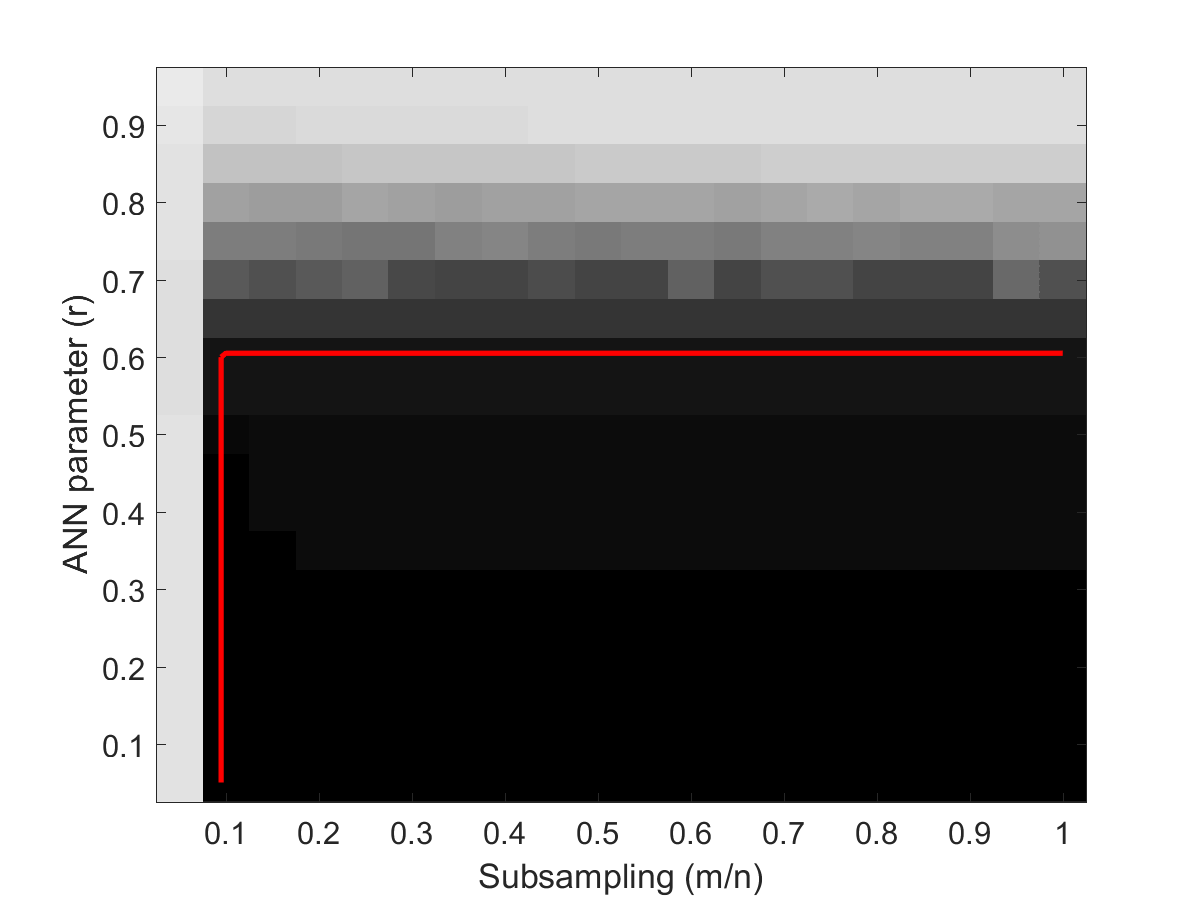} }		
		\quad
		\subfloat
		{\includegraphics[width=.225\textwidth]{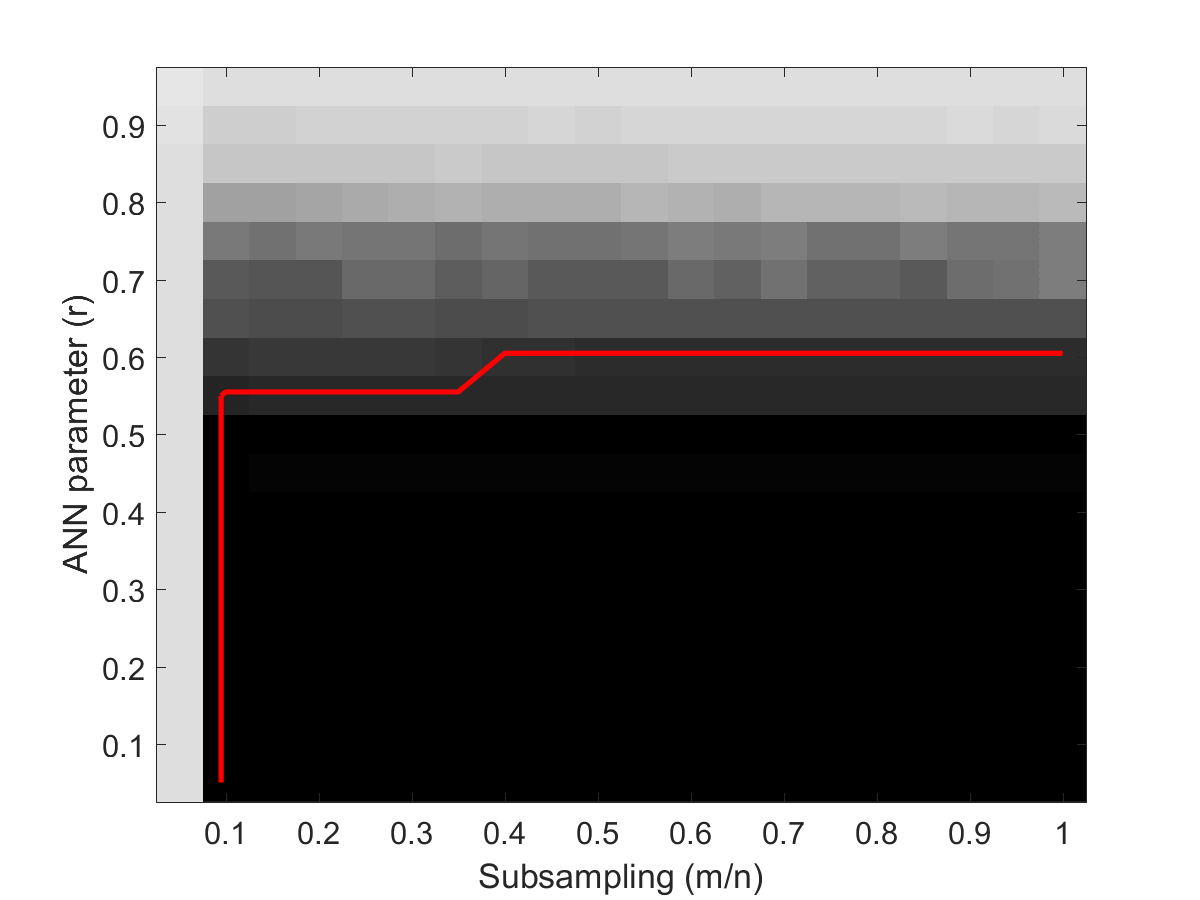} }
		\quad		
		\subfloat
		{\includegraphics[width=.225\textwidth]{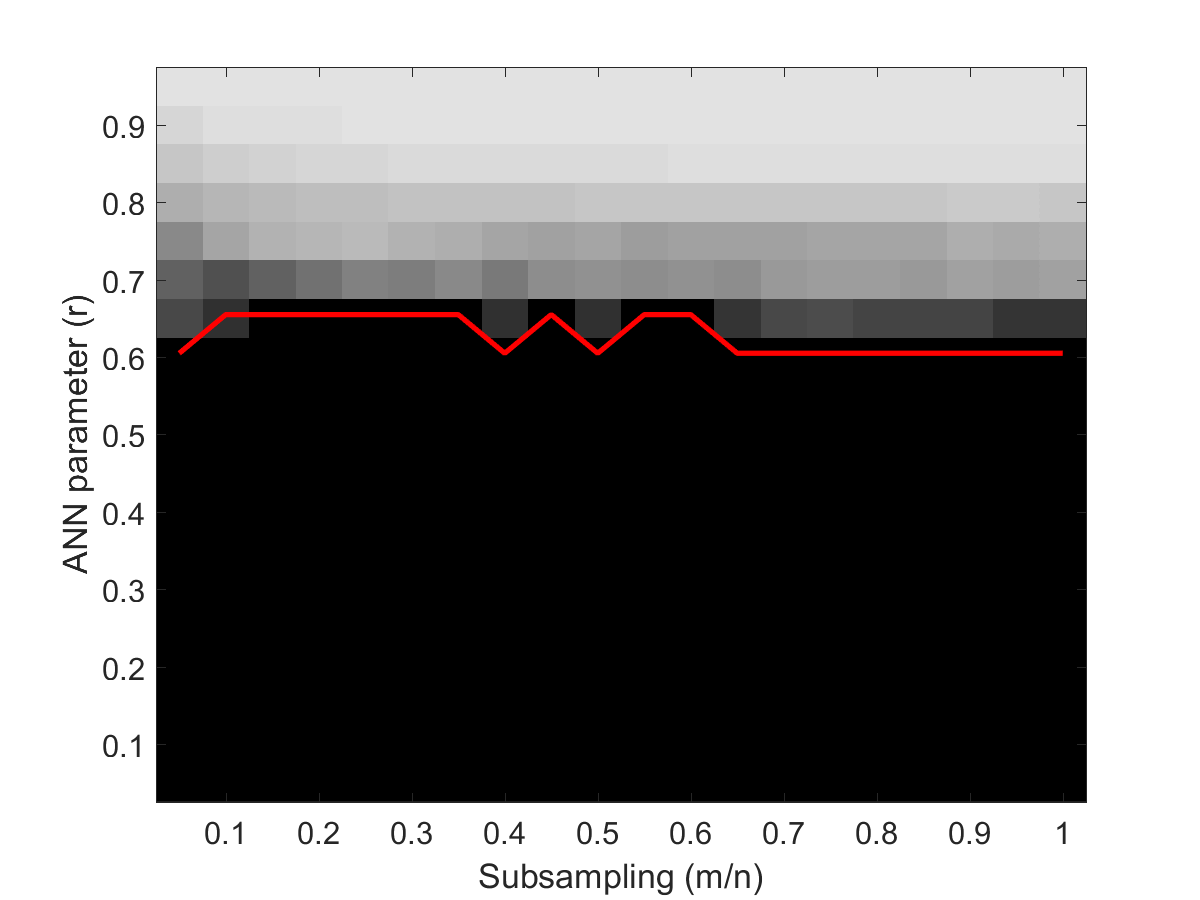} }

		\caption{Recovery phase transitions for IPG with approximate projection (i.e. ANN search). Image intensities correspond to the normalized solution MSE for a search parameter and a given subsampling ratio (ranging between 5-100$\%$).
			 Intensities in all plots are identically set with a logarithmic scale: black pixels correspond to accurate points with MSE $\leq 10^{-6}$, white pixels represent points with MSE $\geq 1$, and the region below the red curve is defined as the exact recovery region with MSE $\leq10^{-4}$. 			
			The columns from left to right correspond to the phase transitions of S-Manifold, Swiss roll, Oscillating wave and MR Fingerprints datasets. The top and the bottom rows correspond to two cover tree based ANN searches namely, the $(1+\epsilon)$-ANN and the PFP-ANN with decay parameter $r$.   \label{fig:PT}}
	\end{minipage}
\end{figure*}
\else
\begin{figure*}
	\centering
	\begin{minipage}{\textwidth}
		\centering
		\subfloat
		{\includegraphics[width=.35\textwidth]{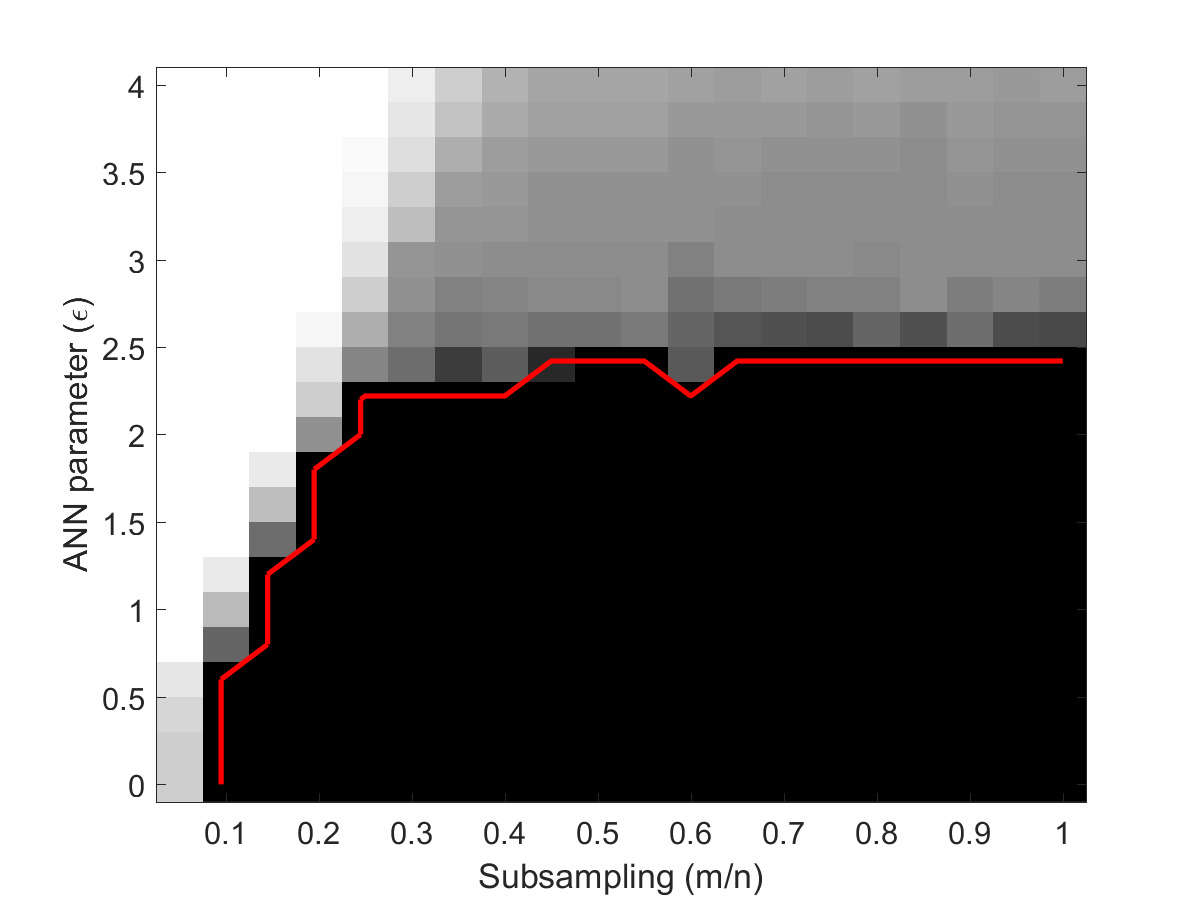} }
		\quad
		\subfloat
		{\includegraphics[width=.35\textwidth]{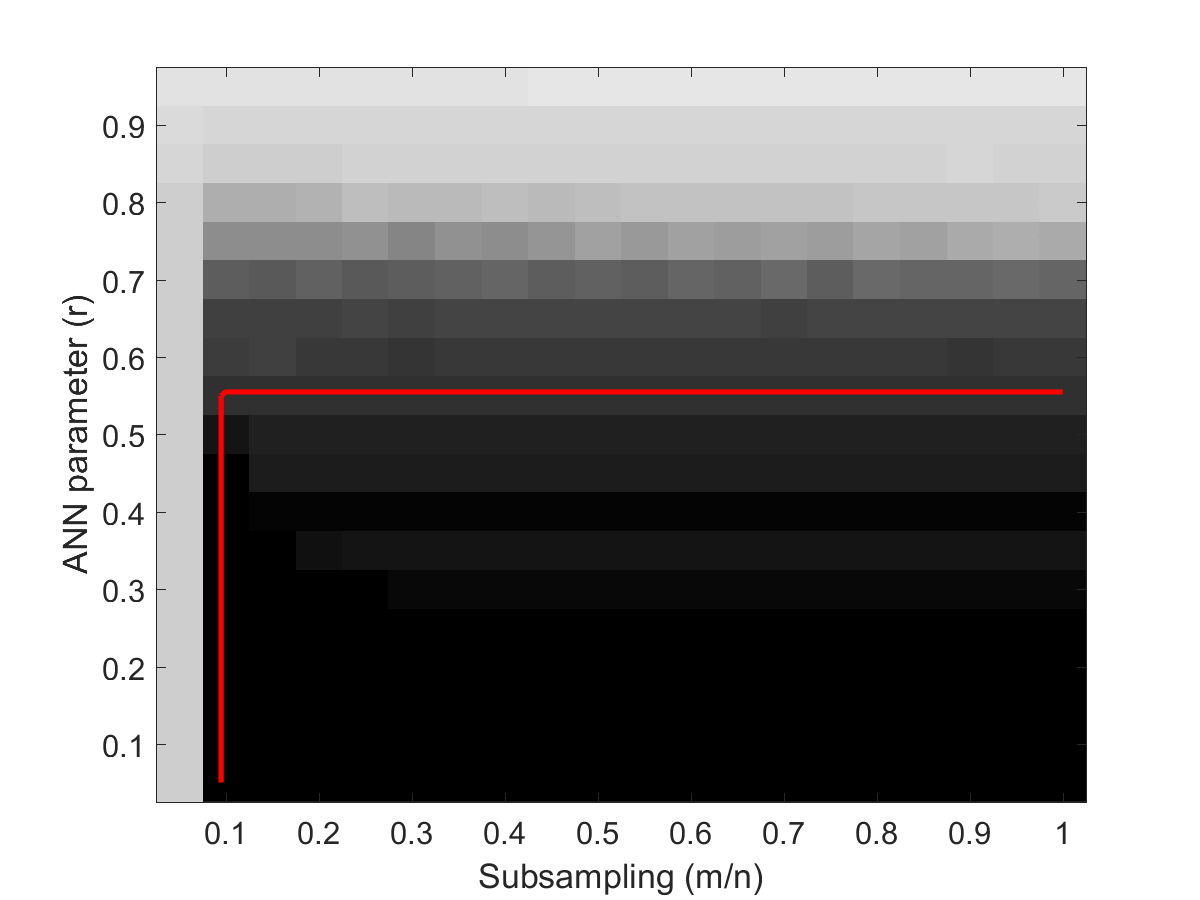} }
		\\
		\subfloat
		{\includegraphics[width=.35\textwidth]{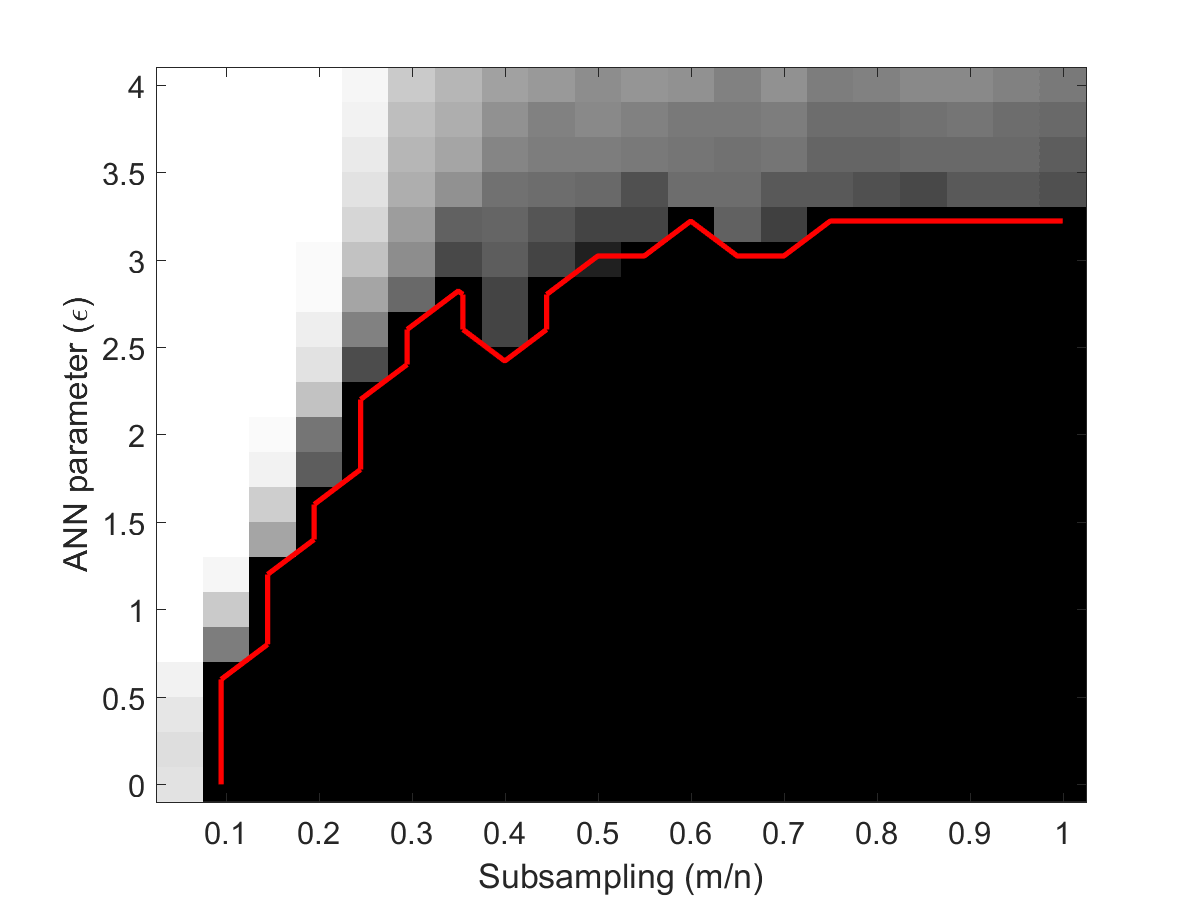} }		
		\quad		
		\subfloat
		{\includegraphics[width=.35\textwidth]{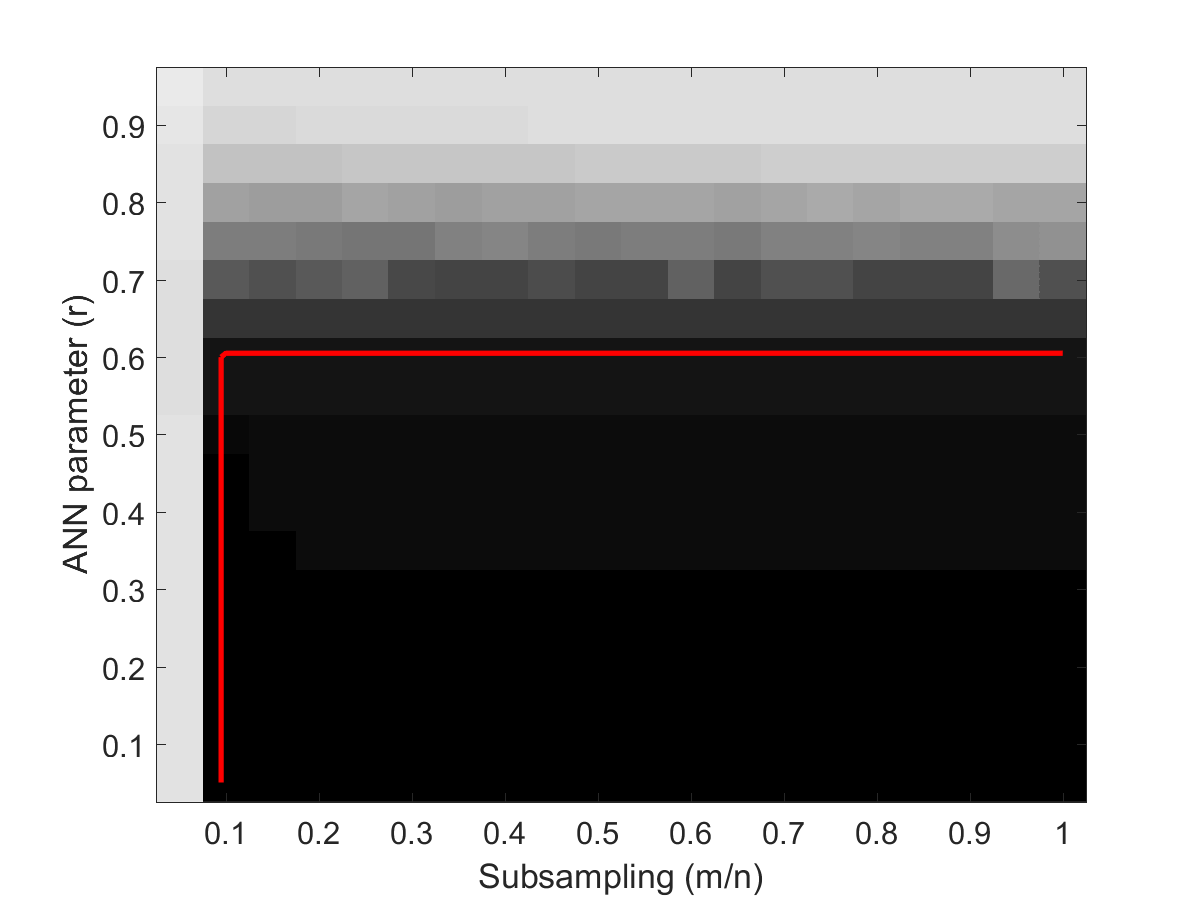} }
		\\
		\subfloat
		{\includegraphics[width=.35\textwidth]{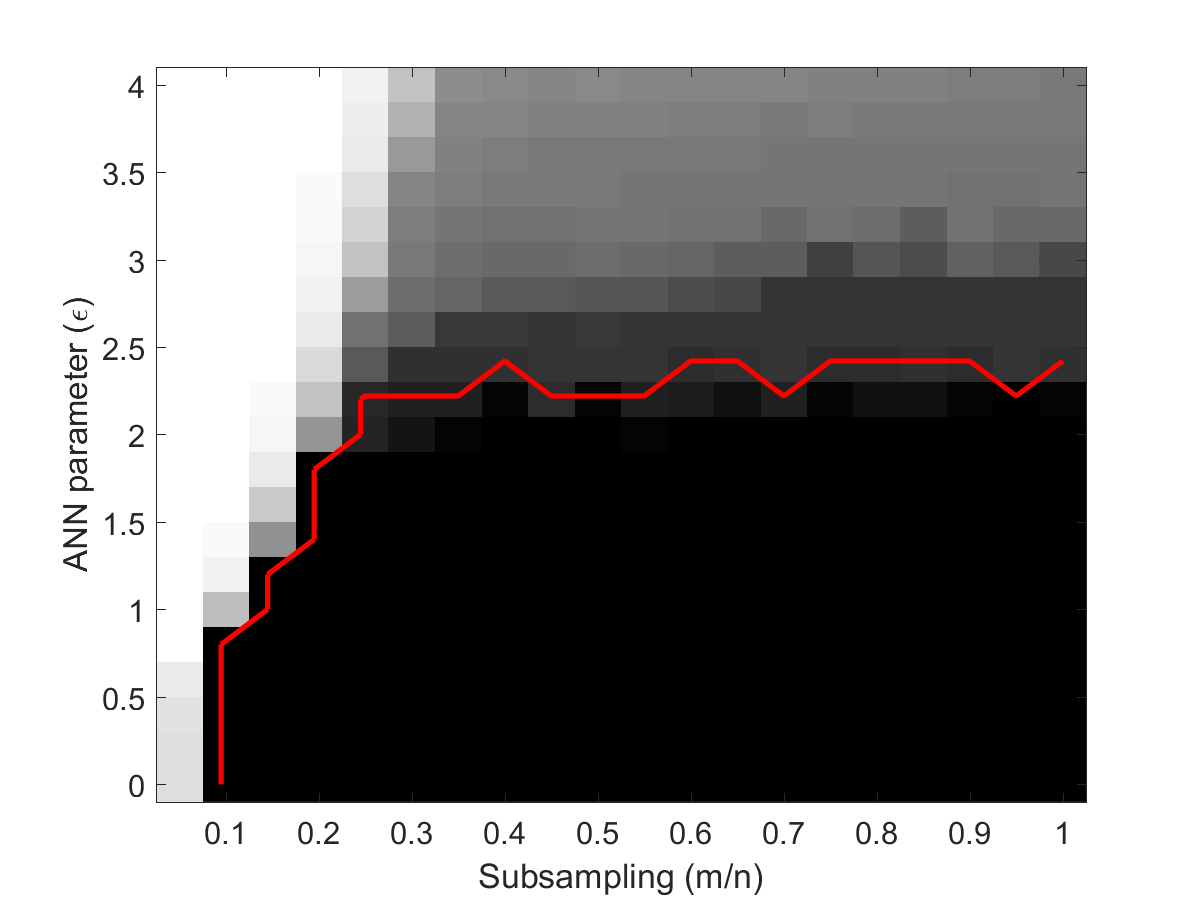} }		
		\quad
		\subfloat
		{\includegraphics[width=.35\textwidth]{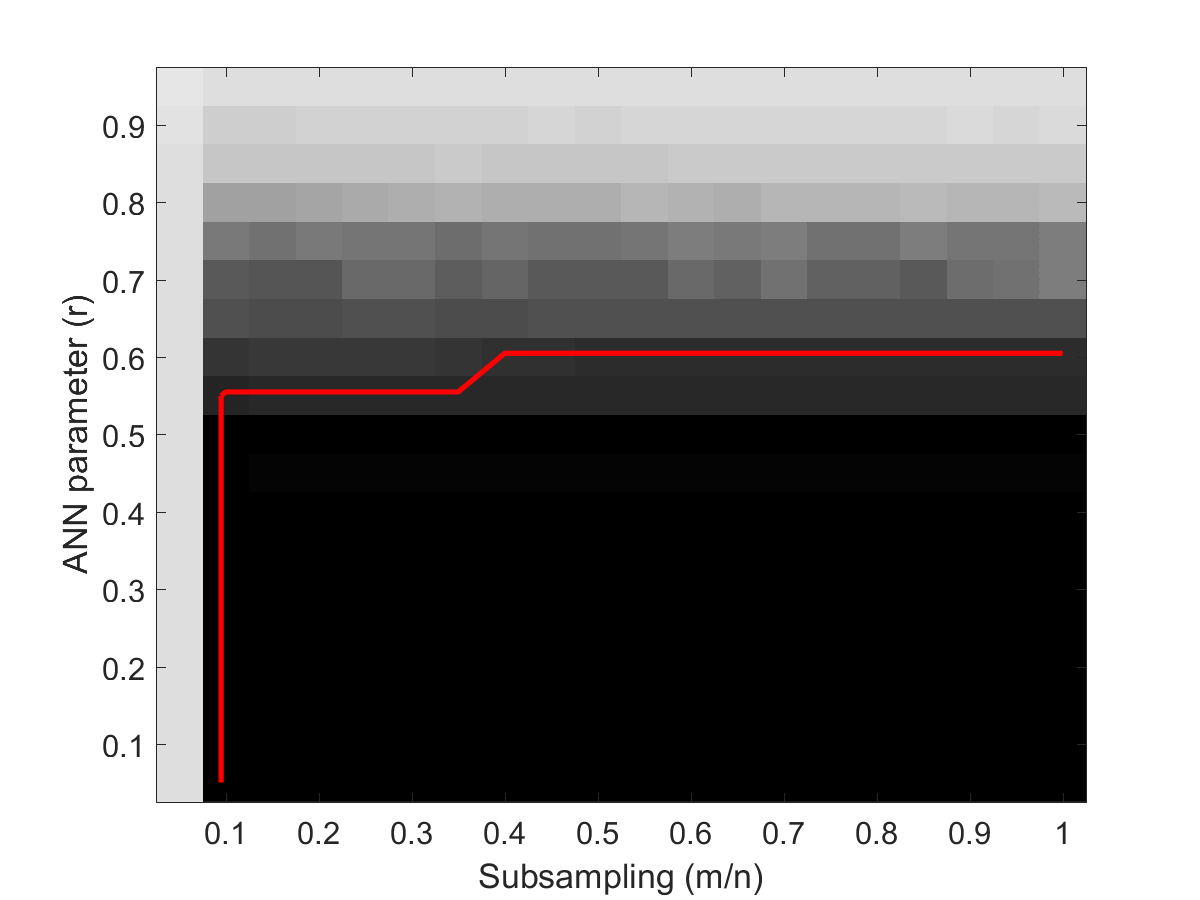} }
		\\
		\subfloat
		{\includegraphics[width=.35\textwidth]{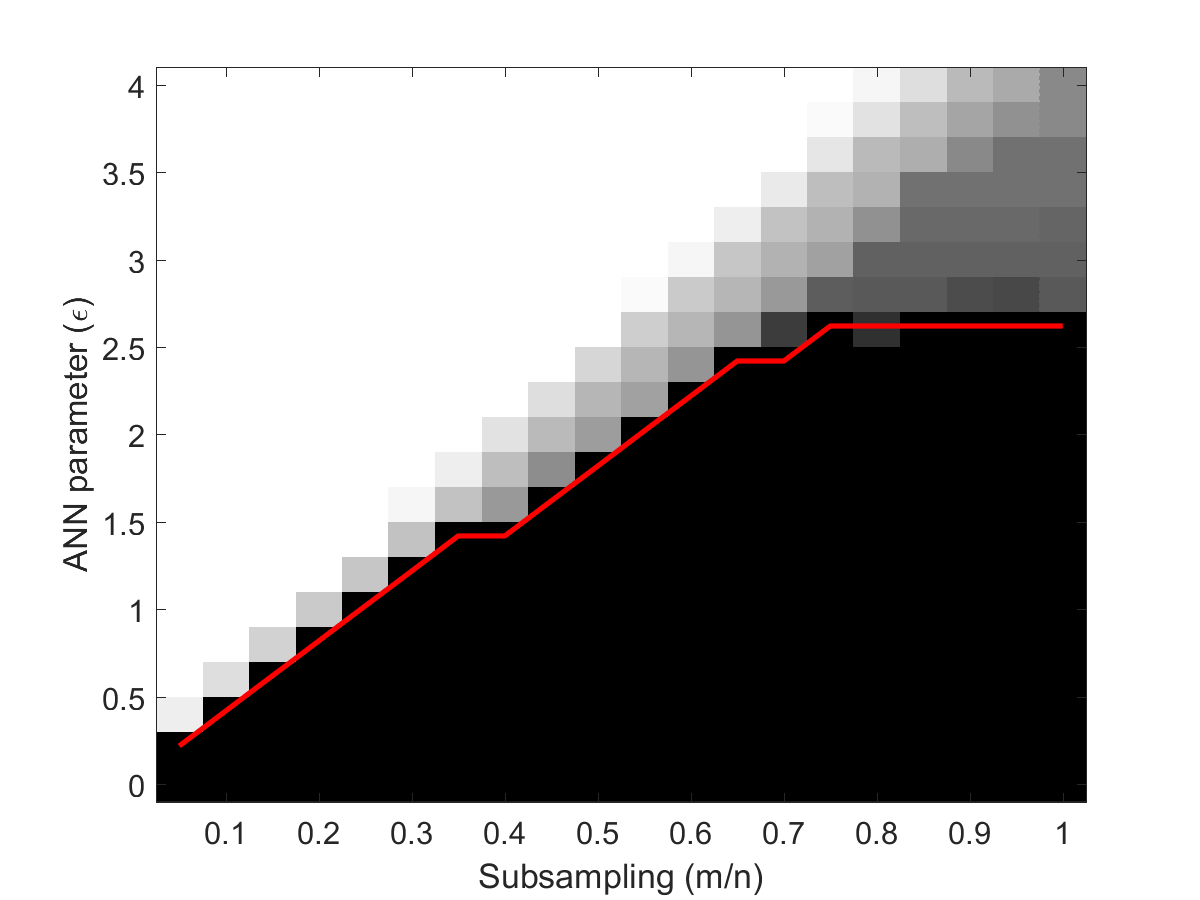} }
		\quad		
		\subfloat
		{\includegraphics[width=.35\textwidth]{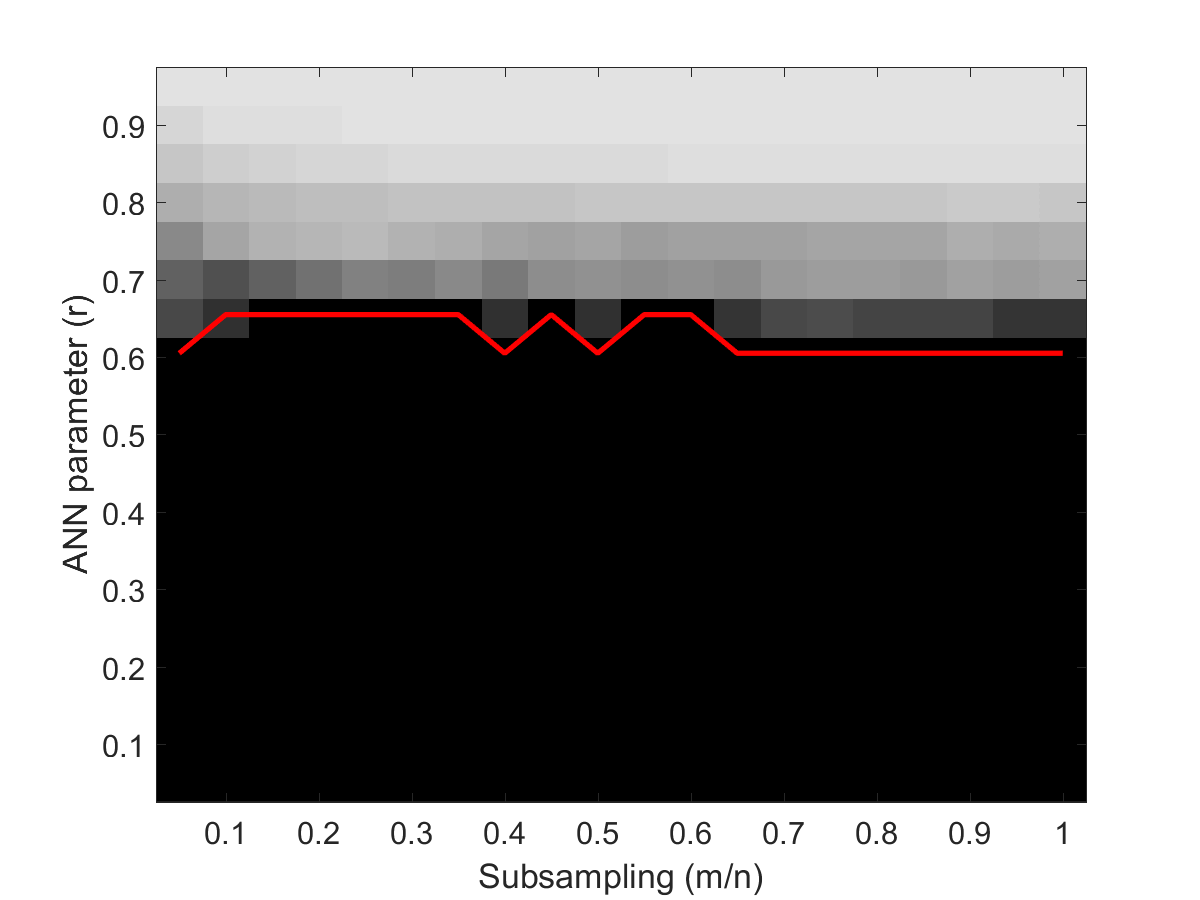} }

		\caption{Recovery phase transitions for IPG with approximate projection (i.e. ANN search). Image intensities correspond to the normalized solution MSE for a search parameter and a given subsampling ratio (ranging between 5-100$\%$).
			Intensities in all plots are identically set with a logarithmic scale: black pixels correspond to accurate points with MSE $\leq 10^{-6}$, white pixels represent points with MSE $\geq 1$, and the region below the red curve is defined as the exact recovery region with MSE $\leq10^{-4}$. 			
			Rows from top to bottom 
			 correspond to the phase transitions of S-Manifold, Swiss roll, Oscillating wave and MR Fingerprints datasets. The left and the right columns correspond to two cover tree based ANN searches namely, the $(1+\epsilon)$-ANN and the PFP-ANN with decay parameter $r$.   \label{fig:PT}}
	\end{minipage}
\end{figure*}
\fi

%
%
%

Along with a brute-force exact search,  three cover tree based ANN search strategies are investigated as described in the previous section:
\begin{itemize}
	\item FP-ANN for precision parameters $\nu_p= \{0.1, 0.05, .01, 0.001\}$.
	\item PFP-ANN for varying precision errors $\nu_p^k=r^k$ decaying at rates $r=\{0.05, 0.1,0.15,\ldots,0.95\}$.
	\item $(1+\epsilon)$-ANN for near optimality parameters $\epsilon=\{0, 0.2,0.4,\ldots,4\}$. 
	The case $\epsilon=0$ corresponds to an exact NN search, however by using the branch-and-bound algorithm on the cover tree proposed in~\cite{beygelzimer2006cover}.
\end{itemize}

\subsubsection*{Gaussian CS sampling}
From each dataset we select $J=50$ points at random and populate our signal matrix $X\in \RR^{\n\times J}$. We then subsample the signal using the linear noiseless model discussed  in \eqref{eq:datadrivenCS}, where the sampling matrix $A\in \RR^{m\times \n J}$ is drawn at random from the i.i.d. 
normal distribution. We denote by $\frac{m}{n}\leq 1$ (where, $n=\n J$) as the subsampling ratio used in each experiment.
 
Throughout we set the maximum number of IPG iterations to $30$. The step size is set to $\mu = 1/m\approx 1/\MM$ which is a theoretical value satisfying the restricted Lipschitz smoothness condition for the i.i.d. Normal sampling ensembles in our theorems and related works on iterative hard thresholding algorithms e.g. see~\cite{IHTCS,AIHT,MIP}.

Figure~\ref{fig:Decays} shows the normalized solution MSE measured by $\frac{\norm{x^k-x^\gt}}{\norm{x^\gt}}$ at each iteration of the exact and inexact IPG algorithms, and
for a given random realization of the sampling matrix $A$ and selected signals $X$. 
For the FP-ANN IPG the convergence rate is unchanged from the exact IPG algorithm but the reconstruction accuracy depends on the chosen precision parameter and for lower precisions the algorithm
stops at an earlier iteration with reduced accuracy, but with the benefit of requiring a smaller search tree. 

The PFP-ANN IPG ultimately achieves the same accuracy of the exact algorithm. 
Refining the approximations at a slow rate slows down the convergence of the algorithm (i.e. the staircase effect visible in the plots correspond to  $r=\{0.7,0.9\}$), whereas choosing too fast error decays, e.g. $r=0.1$, does not improve the convergence rate beyond the exact algorithm and thus potentially leads to computational inefficiency. The $(1+\epsilon)$-ANN IPG algorithm can also achieve the precision of an exact recovery for moderately chosen approximation parameters. The case $\epsilon=0$ (unsurprisingly) iterates the same steps as for the IPG with brute-force search. Increasing $\epsilon$ slows down the convergence and for a very large parameter, e.g. $\epsilon=\{3,4\}$, the algorithm diverges.

Figure~\ref{fig:PT} illustrates the recovery phase transitions for the inexact IPG using the PFP-ANN and $(1+\epsilon)$-ANN searches. 
The normalized MSE is averaged over $10$ random realizations of the sampling matrix $A$ and $20$ randomly subselected signal matrices $X$ for a given $A$. 
In each image the area below the red curve has the solution MSE less than $10^{-4}$ and is chosen as the recovery region. We can observe that the PFP-ANN oracle results in a  recovery region which is almost invariant to the chosen decay parameter $r$ (except for the slow converging case $r \gtrsim 0.6$, due to the limit on the maximum number of iterations). 

In the case of the $1+\epsilon$ oracle we see a different behaviour; smaller values of $\epsilon$ allow for a larger recovery region and larger approximations are restricted to work only 
in high sampling regimes. This observation is in agreement with our theoretical  bounds on recovery and it shows that the $(1+\epsilon)$-approximate oracles are sensitive to the compression ratio, even though an exact (or a better-chosen approximate) 
IPG might still report recovery in the same sampling regime.

Finally in Table~\ref{tab:comp} we report the total cost of projections for each iterative scheme. The cost is measured as the total number of pairwise distances calculated for performing the NN or ANN searches, and it is averaged over the same  trials as previously described\footnote{In our evaluations, we exclude the computation costs of the gradient updates, i.e. the forward and backward operators, which can become dominant when datasets are not very large and the sampling matrix is dense e.g. a Gaussian matrix. For structured embedding matrices such as the fast Johnson-Lindenstrauss transform~\cite{FJLT1} or randomized orthoprojectors e.g. in MRI applications the cost of gradient updates becomes a tiny fraction of the search step, particularly when dealing with a large size dataset.}. For a better evaluation we set the algorithm to terminate earlier (than 30 iterations) when the objective function does not progress more than a tolerance level $tol = 10^{-8}$.  
 For each scheme the reported parameter achieves an average normalized solution MSE $\leq10^{-4}$ in the smallest amount of computations. For comparison we also include the cost of exact IPG implemented with the brute-force and exact ($\epsilon=0$) cover tree NN searches.  When using a brute-force NN search the cost per iteration is fixed and it is equal to the whole dataset population; as a result the corresponding exact IPG reports the highest computation. Replacing the brute-force search with a cover tree based exact NN search significantly reduces the computations. This is related to the low dimensionality of the manifolds in our experiments for which a cover tree search, even for performing an exact NN,  turns out to require many fewer pairwise distances evaluations.  
  Remarkably, the approximate algorithm $(1+\epsilon)$-ANN IPG 
  consistently outperforms all other schemes by reporting 4-10 times acceleration compared to the exact algorithm with $\epsilon=0$, and about (or sometimes more than) 2 orders of magnitude acceleration compared to the IPG with an exact brute-force search; in fact for larger  datasets the gap becomes wider as the $(1+\epsilon)$-ANN complexity stays relatively invariant to the population size.   
The FP-ANN IPG reports similar computations as for the exact tree search ($\epsilon=0$) algorithm because in order to achieve the desired accuracy the (exact) search is performed up to a very fine level of the tree. 
  A gradual progress along the tree levels by the PFP-ANN IPG however improves the search time and reports a comparable computation cost to the $(1+\epsilon)$-ANN. 
Also it can be observed that by taking more samples the overall projection cost reduces which is related to the fast convergence (i.e. less iterations) of the algorithm once more measurements are available.

%
%


\ifCLASSOPTIONtwocolumn
\begin{table*}[t!]
	\vspace{0cm}
	\centering
	\scalebox{.95}{%
		\begin{tabular}{lccccccccccccccc}
			\toprule[0.2em]
			& \multicolumn{14}{c}{ Total NN/ANN cost $(\times 10^4)$ }   \\
			\midrule[0.05em]
			Subsampling ratio ($\frac{m}{n}$)  & \multicolumn{5}{c}{ $10\%$ } &  \multicolumn{4}{c}{ $20\%$ } &  \multicolumn{5}{c}{ $30\%$ }  \\
			\midrule[0.05em]
			Datasets   & SM & SR & OW & MRF & & SM & SR & OW & MRF & & SM &  SR & OW & MRF  \\
			\midrule[0.2em]
			Brute-force NN& 194.23 & 193.67 & 215.10 & 923.34 &&  130.80 & 127.19 & 140.89 & 744.23 & & 113.55 &  109.34 & 123.06 & 699.48\\
			\midrule[.05em]
			CT's exact NN $(\epsilon=0)$  & 8.11 & 8.90 & 15.47 & 33.05 & & 4.90 & 5.19 & 8.99 & 24.74 & &  3.87 & 4.08 & 7.19 & 20.91\\
			\midrule[.05em]
			FP-ANN  & 8.11 & 8.90 & 15.47 & - & & 4.90 & 5.19 & 9.00 & - & & 3.88 & 4.07 & 7.21 & - \\
			Parameter $\nup$  & 1E-3 & 1E-3 & 1E-3 &  & & 1E-3 & 1E-3 & 1E-3 & & & 1E-3 & 1E-3 & 1E-3 &  \\
			\midrule[.05em]
			PFP-ANN & 2.94 & 3.50 & 7.10 & 3.41 & & 1.96 &  2.41 & 3.94 & 2.84 & & 1.78 &  1.99 & 3.38 & 2.52\\
			Parameter $r$ & 4E-1 & 5E-1 & 5E-1 & 4E-1 & & 3E-1 &  3E-1 & 4E-1 & 4E-1 & & 4E-1 &  3E-1 & 4E-1 & 2E-1\\
			\midrule[.05em]
			$(1+\epsilon)$-ANN &  \textbf{2.36} & \textbf{2.77} & \textbf{4.54} & \textbf{2.78} & & \textbf{1.54} & \textbf{1.86} & \textbf{2.91} & \textbf{2.21} & & \textbf{1.31} & \textbf{1.60} & \textbf{2.46} & \textbf{1.92}\\	
			Parameter $\epsilon$ &   4E-1 & 4E-1 & 4E-1 & 4E-1 & & 4E-1 & 4E-1 & 4E-1 & 4E-1 & & 4E-1 & 4E-1 & 6E-1 & 4E-1\\							
			\bottomrule[0.2em]\\
		\end{tabular}}
		\caption{Average computational complexity of the exact/inexact IPG measured by the total number of pairwise distances (in the ambient dimension) calculated within the NN/ANN steps to achieve an average solution MSE $\leq 10^{-4}$ (algorithms with less accuracies are marked as '-'). For each ANN scheme the lowest cost and the associated parameter is reported.  
			SM, SR, OW and MRF abbreviate S-Manifold, Swiss roll, Oscillating wave and the MR Fingerprints datasets, respectively.}\label{tab:comp}
	\end{table*}
\else
\begin{table*}[t!]
	\vspace{0cm}
	\centering
	\scalebox{.83}{%
		\begin{tabular}{lccccccccccccccc}
			\toprule[0.2em]
			& \multicolumn{14}{c}{ Total NN/ANN cost $(\times 10^4)$ }   \\
			\midrule[0.05em]
			Subsampling ratio ($\frac{m}{n}$)  & \multicolumn{5}{c}{ $10\%$ } &  \multicolumn{4}{c}{ $20\%$ } &  \multicolumn{5}{c}{ $30\%$ }  \\
			\midrule[0.05em]
			Datasets   & SM & SR & OW & MRF & & SM & SR & OW & MRF & & SM &  SR & OW & MRF  \\
			\midrule[0.2em]
			Brute-force NN& 194.23 & 193.67 & 215.10 & 923.34 &&  130.80 & 127.19 & 140.89 & 744.23 & & 113.55 &  109.34 & 123.06 & 699.48\\
			\midrule[.05em]
			CT's exact NN $(\epsilon=0)$  & 8.11 & 8.90 & 15.47 & 33.05 & & 4.90 & 5.19 & 8.99 & 24.74 & &  3.87 & 4.08 & 7.19 & 20.91\\
			\midrule[.05em]
			FP-ANN  & 8.11 & 8.90 & 15.47 & - & & 4.90 & 5.19 & 9.00 & - & & 3.88 & 4.07 & 7.21 & - \\
			Parameter $\nup$  & 1E-3 & 1E-3 & 1E-3 &  & & 1E-3 & 1E-3 & 1E-3 & & & 1E-3 & 1E-3 & 1E-3 &  \\
			\midrule[.05em]
			PFP-ANN & 2.94 & 3.50 & 7.10 & 3.41 & & 1.96 &  2.41 & 3.94 & 2.84 & & 1.78 &  1.99 & 3.38 & 2.52\\
			Parameter $r$ & 4E-1 & 5E-1 & 5E-1 & 4E-1 & & 3E-1 &  3E-1 & 4E-1 & 4E-1 & & 4E-1 &  3E-1 & 4E-1 & 2E-1\\
			\midrule[.05em]
			$(1+\epsilon)$-ANN &  \textbf{2.36} & \textbf{2.77} & \textbf{4.54} & \textbf{2.78} & & \textbf{1.54} & \textbf{1.86} & \textbf{2.91} & \textbf{2.21} & & \textbf{1.31} & \textbf{1.60} & \textbf{2.46} & \textbf{1.92}\\	
			Parameter $\epsilon$ &   4E-1 & 4E-1 & 4E-1 & 4E-1 & & 4E-1 & 4E-1 & 4E-1 & 4E-1 & & 4E-1 & 4E-1 & 6E-1 & 4E-1\\							
			\bottomrule[0.2em]\\
		\end{tabular}}
		\caption{Average computational complexity of the exact/inexact IPG measured by the total number of pairwise distances (in the ambient dimension) calculated within the NN/ANN steps to achieve an average solution MSE $\leq 10^{-4}$ (algorithms with less accuracies are marked as '-'). For each ANN scheme the lowest cost and the associated parameter is reported.  
			SM, SR, OW and MRF abbreviate S-Manifold, Swiss roll, Oscillating wave and the MR Fingerprints datasets, respectively.}\label{tab:comp}
	\end{table*}
\fi	

\section{Conclusion and discussions}\label{sec:conclusion}
We studied the robustness of the iterative projected gradient  algorithm against inexact gradient and projection oracles and for solving inverse problems in compressed sensing. We considered fixed precision, progressive fixed precision and $(1+\epsilon)$-approximate oracles. A notion of model information preserving under a hybrid local-uniform embedding assumption is at the heart of our analysis. We showed that under the same assumptions, the algorithm with PFP approximate oracles achieves 
 the accuracy of the exact IPG.
For a certain rate of decay of the  approximation errors this scheme  can also maintain the  rate of linear convergence as for the exact algorithm. We  also conclude that choosing too fast decays does not help the convergence rate beyond the exact algorithm and therefore  can result in computational inefficiency. 
The $(1+\epsilon)$-approximate IPG can also achieves  
the accuracy of the exact algorithm, however
under a stronger embedding condition, slower rate of convergence and possibly more noise amplification  compared to the exact algorithm. We show that this approximation is sensitive to the CS subsampling regime i.e. for high compression ratios one can not afford too large approximation. 
We applied our results to a class of data driven compressed sensing problems, where we replaced exhaustive NN searches over large datasets with fast and approximate alternatives introduced by the cover tree structure. Our experiments indicate that the inexact IPG with  $(1+\epsilon)$-ANN searches (and also comparably the PFP type search) can significantly accelerate the CS reconstruction.  

Our results require a lower bound on the chosen step size which is a critical assumption for globally solving a nonconvex CS problem, see e.g. \cite{IHTCS,Blumen,AIHT}. With no lower bound on the step size only convergence to a local critical point is guaranteed (for the exact algorithm) see e.g.~\cite{Nesterov:nonconvex,Attouch}. Such assumption is restrictive for solving convex problems, where an upper bound on the step size is generally sufficient for the convergence. However recent studies~\cite{negahban:fast,Gabriel:FBlocal,Oymak:tradeoff} established fast \emph{linear} convergence for solving non strongly convex problems such as CS with the exact IPG, and by assuming a notion of \emph{local} (model restricted) strong convexity as well as choosing large enough step sizes. In our future work  
we would like to make more explicit connection between these results and our embedding assumptions and extend our approximation robustness study to the convex CS recovery settings with sharper bounds than e.g. \cite{BachinexactIPG}.

A limitation of our current method is the dependency on $\epsilon$ for the CS recovery using $(1+ \epsilon)$-ANN searches. In future we plan to investigate whether the ideas from~\cite{Hegde15} can be generalized to include our data-driven signal models.

Finally we did not provide much discussion or experiments on the applications of the approximate gradient updates within the IPG. We think such approximations might be related to the sketching techniques for solving large size inverse problems, see~\cite{Pilanci:IHS, GPIS}.
In our future work we would like to make explicit connections  in this regard as well.

\bibliographystyle{IEEEtran}
\bibliography{mybiblio}




\end{document}